\pdfoutput=1

\documentclass{article}

\usepackage{microtype}
\usepackage{graphicx}
\usepackage{booktabs}
\usepackage{array}
\usepackage{hyperref}
\let\outlineaddcontentsline\addcontentsline

\usepackage[preprint]{icml2026}
\usepackage{etoolbox}
\makeatletter
\patchcmd{\printAffiliationsAndNotice}{{\bf AUTHORERR: Missing \textbackslash{}icmlcorrespondingauthor.}}{}{}{%
  \PackageWarning{main}{could not remove the missing-corresponding-author error}}
\makeatother
\let\addcontentsline\outlineaddcontentsline

\usepackage{amsmath}
\usepackage{amssymb}
\usepackage{mathtools}
\usepackage{amsthm}
\usepackage{xcolor}
\usepackage[capitalize,noabbrev]{cleveref}
\usepackage{acro}
\usepackage{siunitx}
\theoremstyle{plain}
\newtheorem{proposition}{Proposition}[section]

\newcommand{\rev}[1]{#1}

\newcommand{\Mc}{\mathcal{M}}               %
\newcommand{\Amp}{\mathcal{A}}              %
\newcommand{\Tobs}{T_{\rm obs}}             %
\newcommand{\fw}{f_{w}}                     %
\newcommand{\fdot}{\dot f}
\newcommand{\snr}{\rho}
\newcommand{\R}{\mathbb{R}}
\newcommand{\E}{\mathbb{E}}
\newcommand{\Var}{\operatorname{Var}}
\newcommand{\MAD}{\operatorname{MAD}}
\newcommand{\Unif}{\mathcal{U}}
\newcommand{\Normal}{\mathcal{N}}
\newcommand{\CN}{\mathcal{CN}}
\newcommand{\dd}{\mathrm{d}}
\newcommand{\ii}{\mathrm{i}}
\newcommand{\T}{^{\top}}
\newcommand{\inner}[2]{\langle #1 \,|\, #2 \rangle}
\newcommand{\bins}{\,\text{bins}}
\newcommand{\vnet}{\hat v_{\eta}}           %

\DeclareSIUnit\solarmass{\ensuremath{M_\odot}}
\DeclareSIUnit\year{yr}
\DeclareSIUnit\parsec{pc}

\acsetup{make-links = false, first-style = long-short}

\DeclareAcronym{lisa}{short = LISA, long = Laser Interferometer Space Antenna}
\DeclareAcronym{gw}{short = GW, long = gravitational wave}
\DeclareAcronym{gb}{short = GB, long = galactic binary, long-plural-form = galactic binaries}
\DeclareAcronym{tdi}{short = TDI, long = time-delay interferometry}
\DeclareAcronym{snr}{short = SNR, long = signal-to-noise ratio}
\DeclareAcronym{psd}{short = PSD, long = power spectral density, long-plural-form = power spectral densities}
\DeclareAcronym{oms}{short = OMS, long = optical metrology system}
\DeclareAcronym{ssb}{short = SSB, long = solar-system barycentre}
\DeclareAcronym{sgwb}{short = SGWB, long = stochastic gravitational-wave background}
\DeclareAcronym{ddpc}{short = DDPC, long = Distributed Data Processing Centre}
\DeclareAcronym{stft}{short = STFT, long = short-time Fourier transform}

\DeclareAcronym{mcmc}{short = MCMC, long = Markov chain Monte Carlo}
\DeclareAcronym{rjmcmc}{short = RJMCMC, long = reversible-jump Markov chain Monte Carlo}
\DeclareAcronym{fmpe}{short = FMPE, long = flow-matching posterior estimation}
\DeclareAcronym{sbi}{short = SBI, long = simulation-based inference}
\DeclareAcronym{npe}{short = NPE, long = neural posterior estimation}
\DeclareAcronym{nsf}{short = NSF, long = neural spline flow}
\DeclareAcronym{maf}{short = MAF, long = masked autoregressive flow}
\DeclareAcronym{mle}{short = MLE, long = maximum-likelihood estimate}
\DeclareAcronym{ess}{short = ESS, long = effective sample size}
\DeclareAcronym{ecdf}{short = ECDF, long = empirical cumulative distribution function}
\DeclareAcronym{sbc}{short = SBC, long = simulation-based calibration}
\DeclareAcronym{tarp}{short = TARP, long = tests of accuracy with random points}
\DeclareAcronym{crps}{short = CRPS, long = continuous ranked probability score}

\DeclareAcronym{mmdit}{short = MMDiT, long = multimodal diffusion transformer}
\DeclareAcronym{rope}{short = RoPE, long = rotary position embedding}
\DeclareAcronym{mlp}{short = MLP, long = multilayer perceptron}
\DeclareAcronym{rms}{short = RMS, long = root mean square}

\DeclareAcronym{gpu}{short = GPU, long = graphics processing unit}
\DeclareAcronym{cpu}{short = CPU, long = central processing unit}

\newcommand{\RelatedWorkSourcesWord}{four}

\newcommand{\RelatedWorkTobs}{2}
\newcommand{\RelatedWorkDelmondTobs}{1}
\newcommand{\RelatedWorkPatch}{16}
\newcommand{\RelatedWorkFzeroLow}{0.1}
\newcommand{\RelatedWorkFzeroHigh}{16}
\newcommand{\RelatedWorkSnrLow}{7}
\newcommand{\RelatedWorkSnrHigh}{67}
\newcommand{\RelatedWorkVisibleShare}{0.84}
\newcommand{\RelatedWorkDelmondHardest}{6.3}
\newcommand{\RelatedWorkDelmondEasiest}{15.8}
\newcommand{\RelatedWorkCrowdingWindows}{4096}
\newcommand{\RelatedWorkClosestFull}{21}
\newcommand{\RelatedWorkCloserFull}{18}

\newcommand{\RelatedWorkClosestLowTail}{2.6}
\newcommand{\RelatedWorkCloserLowTail}{86}

\newcommand{\RelatedWorkCloserLowFreq}{58}

\newcommand{\RelatedWorkCloserCore}{34}
\newcommand{\RelatedWorkSubBinCore}{6}

\newcommand{\RelatedWorkCloserHighFreq}{19}

\newcommand{\RelatedWorkClosestHighTail}{41}
\newcommand{\RelatedWorkCloserHighTail}{9}

\newcommand{\ForwardModelVelocity}{29.8}

\newcommand{\ForwardModelArmGm}{2.5}

\newcommand{\ForwardModelFStar}{19.1}
\newcommand{\ForwardModelTobsYears}{2}

\newcommand{\ForwardModelWindowResponses}{4}
\newcommand{\ForwardModelInteriorResponses}{3}
\newcommand{\ForwardModelWindowStride}{2}
\newcommand{\ForwardModelAntennaHarmonics}{4}

\newcommand{\ForwardModelNarrowR}{64}

\newcommand{\ForwardModelFullR}{128}

\newcommand{\ForwardModelSpanMedian}{18}

\newcommand{\ForwardModelSpanBound}{2048}

\newcommand{\ForwardModelReferencePoints}{16384}

\newcommand{\ForwardModelNarrowTruncRelMedian}{5.7}

\newcommand{\ForwardModelNarrowTruncRelOffBin}{8.6}
\newcommand{\ForwardModelNarrowTruncOutsideMax}{1.00}

\newcommand{\ForwardModelFullTruncRelMedian}{3.9}

\newcommand{\ForwardModelFullTruncRhoDev}{4.5e-5}

\newcommand{\NoiseOms}{15}
\newcommand{\NoiseAcc}{3}

\newcommand{\NoiseBandLow}{0.1}
\newcommand{\NoiseBandHigh}{16}

\newcommand{\NoiseTOverALow}{8.3e-8}

\newcommand{\NoiseRobsonPower}{0.7}
\newcommand{\NoiseRobsonAmplitude}{0.4}
\newcommand{\NoiseC}{3.19}
\newcommand{\NoiseCError}{0.02}
\newcommand{\NoiseCGalactic}{3.35}
\newcommand{\NoiseCFlat}{0.4}
\newcommand{\NoiseCFlatBelow}{1.5}
\newcommand{\NoiseCRise}{26}
\newcommand{\NoiseCTop}{16}
\newcommand{\NoiseCSources}{8192}
\newcommand{\NoiseSqrtC}{1.8}

\newcommand{\NoiseGeneratorWindows}{3000}
\newcommand{\NoiseGeneratorBins}{1024}

\newcommand{\NoiseGeneratorDof}{4096}

\newcommand{\NoiseGeneratorKs}{0.83}

\newcommand{\NoiseStrubSnr}{7.951}
\newcommand{\NoiseStrubSnrTheirs}{7.93}

\newcommand{\PriorsNarrowR}{64}
\newcommand{\PriorsNarrowW}{256}
\newcommand{\PriorsNarrowLowMhz}{1}
\newcommand{\PriorsNarrowHighMhz}{3}
\newcommand{\PriorsNarrowWindows}{63021}
\newcommand{\PriorsNarrowCatalogueShare}{3.7}
\newcommand{\PriorsNarrowVisible}{15.8}

\newcommand{\PriorsNarrowWithVisible}{34.6}
\newcommand{\PriorsNarrowTrainingWithVisible}{77.0}

\newcommand{\PriorsNarrowDriftOver}{0}

\newcommand{\PriorsFullR}{128}
\newcommand{\PriorsFullW}{512}
\newcommand{\PriorsFullLowMhz}{0.1}
\newcommand{\PriorsFullHighMhz}{16}
\newcommand{\PriorsFullWindows}{501576}
\newcommand{\PriorsFullCatalogueShare}{100.0}
\newcommand{\PriorsFullVisible}{0.84}

\newcommand{\PriorsFullWithVisible}{1.5}
\newcommand{\PriorsFullTrainingWithVisible}{78.0}
\newcommand{\PriorsFullTrainingTopShare}{90.7}
\newcommand{\PriorsFullDriftOver}{0.067}
\newcommand{\PriorsFullDriftMedian}{3e-05}

\newcommand{\PriorsFullTopShare}{2.1}

\newcommand{\PriorsMeanF}{-3.74}
\newcommand{\PriorsStdF}{0.39}
\newcommand{\PriorsMeanFdot}{-20.57}

\newcommand{\PriorsMeanAmp}{-23.27}

\newcommand{\PriorsSlopeAmp}{0.48}
\newcommand{\PriorsInterceptAmp}{-21.47}
\newcommand{\PriorsResidualAmp}{0.25}

\newcommand{\PriorsCatalogueCutMhz}{0.1}
\newcommand{\PriorsVisibleSnr}{7}
\newcommand{\PriorsPriorWindows}{8192}
\newcommand{\PriorsCutWindows}{4096}
\newcommand{\PriorsCandidates}{64}

\newcommand{\PriorsNarrowStepMs}{64}
\newcommand{\PriorsNarrowDrawMs}{0.71}
\newcommand{\PriorsNarrowDrawShare}{1}

\newcommand{\PriorsFullStepMs}{73}
\newcommand{\PriorsFullDrawMs}{2}
\newcommand{\PriorsFullDrawShare}{3}

\newcommand{\PriorsNarrowStepChip}{A100-SXM4-80GB}
\newcommand{\PriorsFullStepChip}{A100 80GB PCIe}
\newcommand{\PriorsSources}{4}
\newcommand{\PriorsCoords}{8}
\newcommand{\PriorsStride}{2}

\newcommand{\PriorsUniformWindowsPercent}{10}

\newcommand{\PriorsBulgeWeight}{25}
\newcommand{\PriorsThickWeight}{16}
\newcommand{\PriorsThinWeight}{58}
\newcommand{\PriorsFloorWeight}{0.8}

\newcommand{\PriorsPlaneDeg}{10}
\newcommand{\PriorsPlaneShare}{95}
\newcommand{\PriorsPermutations}{24}
\newcommand{\PriorsCopies}{6144}
\newcommand{\PriorsBatch}{1024}
\newcommand{\PriorsDraws}{16}
\newcommand{\PriorsObservationsPerStep}{64}
\newcommand{\PriorsCovFF}{0.152}
\newcommand{\PriorsCovFdotFdot}{1.872}
\newcommand{\PriorsCovAmpAmp}{0.098}
\newcommand{\PriorsCovFFdot}{0.529}
\newcommand{\PriorsCovFAmp}{0.073}
\newcommand{\PriorsCovFdotAmp}{0.287}

\newcommand{\ChartCoords}{8}

\newcommand{\ChartFdotSlope}{3.48}
\newcommand{\ChartFdotStd}{0.18}
\newcommand{\ChartXfStd}{0.577}
\newcommand{\ChartSigmaL}{0.530}
\newcommand{\ChartSigmaB}{0.103}
\newcommand{\ChartSigmaLDeg}{30}
\newcommand{\ChartSigmaBDeg}{5.9}
\newcommand{\ChartSeamX}{5.9}
\newcommand{\ChartPoleX}{15.2}
\newcommand{\ChartEdgeDeg}{10}
\newcommand{\ChartSeamMass}{0.09}
\newcommand{\ChartPoleMass}{0.01}
\newcommand{\ChartUnitStdLow}{0.99}
\newcommand{\ChartUnitStdHigh}{1.03}
\newcommand{\ChartZStdLow}{0.14}
\newcommand{\ChartZStdHigh}{0.59}
\newcommand{\ChartMaxCorr}{0.031}
\newcommand{\ChartKappaNarrowLow}{1.00}
\newcommand{\ChartKappaNarrowHigh}{2.91}
\newcommand{\ChartKappaFullLow}{0.98}
\newcommand{\ChartKappaFullHigh}{2.90}

\newcommand{\ChartKappaLawRms}{3}
\newcommand{\ChartPriorDraws}{32768}
\newcommand{\ChartBinsPerUnitNarrow}{96}
\newcommand{\ChartBinsPerUnitFull}{192}
\newcommand{\ChartFThirtyTwoBinsFull}{2.3e-5}

\newcommand{\FlowMatchingDraws}{16}
\newcommand{\FlowMatchingCostDraws}{256}
\newcommand{\FlowMatchingWindows}{64}
\newcommand{\FlowMatchingBatch}{1024}

\newcommand{\FlowMatchingLossBins}{16}
\newcommand{\FlowMatchingLossDecades}{-6}

\newcommand{\FlowMatchingTimeDecades}{-6}
\newcommand{\FlowMatchingOdeSteps}{32}
\newcommand{\FlowMatchingOdeStepsMore}{128}
\newcommand{\FlowMatchingEvaluations}{64}
\newcommand{\FlowMatchingLastMid}{\tfrac{1}{64}}
\newcommand{\FlowMatchingCheckUniform}{1.000}
\newcommand{\FlowMatchingCheckNll}{1.000}
\newcommand{\FlowMatchingSnrLow}{7}
\newcommand{\FlowMatchingSnrQuantile}{99.99}
\newcommand{\FlowMatchingNarrowSnrHigh}{86}
\newcommand{\FlowMatchingFullSnrHigh}{67}
\newcommand{\FlowMatchingWidthWide}{1.00}
\newcommand{\FlowMatchingWidthFrame}{1.001}
\newcommand{\FlowMatchingFrameDecay}{0.001}

\newcommand{\FlowMatchingWidthNarrowWeak}{0.13}
\newcommand{\FlowMatchingWidthNarrowWeakMore}{0.49}
\newcommand{\FlowMatchingWidthFullWeak}{7e-2}
\newcommand{\FlowMatchingOdeStepsMost}{512}
\newcommand{\FlowMatchingWidthNarrowWeakMost}{0.96}
\newcommand{\FlowMatchingWidthFullWeakMost}{0.79}
\newcommand{\FlowMatchingCostNarrow}{0.22}
\newcommand{\FlowMatchingCostFull}{0.25}
\newcommand{\FlowMatchingCostSmallest}{0.08}
\newcommand{\FlowMatchingCostLargest}{0.92}
\newcommand{\FlowMatchingCostDrawsMore}{1024}
\newcommand{\FlowMatchingCostFullMore}{0.98}

\newcommand{\FlowMatchingNarrowBinsPerUnit}{96}
\newcommand{\FlowMatchingNarrowWeak}{1e-3}
\newcommand{\FlowMatchingNarrowLoud}{1e-4}
\newcommand{\FlowMatchingFullBinsPerUnit}{192}
\newcommand{\FlowMatchingFullWeak}{7e-4}
\newcommand{\FlowMatchingFullLoud}{7e-5}

\newcommand{\NetworkXsWidth}{256}
\newcommand{\NetworkXsHeads}{4}
\newcommand{\NetworkXsBlocks}{4}
\newcommand{\NetworkXsBackbone}{15466512}
\newcommand{\NetworkXsTotalNarrow}{16583200}
\newcommand{\NetworkXsTotalFull}{16593440}
\newcommand{\NetworkSWidth}{384}
\newcommand{\NetworkSHeads}{6}
\newcommand{\NetworkSBlocks}{6}
\newcommand{\NetworkSBackbone}{53674020}
\newcommand{\NetworkSTotalNarrow}{55938868}
\newcommand{\NetworkSTotalFull}{55954228}
\newcommand{\NetworkBWidth}{512}
\newcommand{\NetworkBHeads}{8}
\newcommand{\NetworkBBlocks}{8}
\newcommand{\NetworkBBackbone}{128974912}
\newcommand{\NetworkBTotalNarrow}{132781136}
\newcommand{\NetworkBTotalFull}{132801616}
\newcommand{\NetworkLWidth}{768}
\newcommand{\NetworkLHeads}{12}
\newcommand{\NetworkLBlocks}{12}
\newcommand{\NetworkLBackbone}{441188496}
\newcommand{\NetworkLTotalNarrow}{449257120}
\newcommand{\NetworkLTotalFull}{449287840}
\newcommand{\NetworkXlWidth}{1024}
\newcommand{\NetworkXlHeads}{16}
\newcommand{\NetworkXlBlocks}{16}
\newcommand{\NetworkXlBackbone}{1052770560}
\newcommand{\NetworkXlTotalNarrow}{1066674448}
\newcommand{\NetworkXlTotalFull}{1066715408}
\newcommand{\NetworkHeadDim}{64}
\newcommand{\NetworkRopePairs}{32}
\newcommand{\NetworkPatch}{16}
\newcommand{\NetworkTokenScalars}{144}
\newcommand{\NetworkAmplitudeScalars}{48}
\newcommand{\NetworkPhaseScalars}{96}
\newcommand{\NetworkSinksWord}{four}
\newcommand{\NetworkSourcesWord}{four}
\newcommand{\NetworkCoords}{8}
\newcommand{\NetworkCoordsWord}{eight}
\newcommand{\NetworkHeadOutputs}{16}
\newcommand{\NetworkExpand}{2}
\newcommand{\NetworkHarmonics}{6}
\newcommand{\NetworkHarmonicScalars}{20}
\newcommand{\NetworkSourceScalars}{52}
\newcommand{\NetworkFreqFinest}{1/64}
\newcommand{\NetworkTimeDecades}{6}
\newcommand{\NetworkSinusoidFreqs}{64}
\newcommand{\NetworkSinusoidDecades}{4}
\newcommand{\NetworkDraws}{16}

\newcommand{\NetworkBfBits}{8}
\newcommand{\NetworkBfUnity}{256}
\newcommand{\NetworkSnrTop}{67}
\newcommand{\NetworkSnrQuantile}{99.99}
\newcommand{\NetworkBinsNarrow}{256}
\newcommand{\NetworkTokensNarrow}{16}
\newcommand{\NetworkFreqPeriodsNarrow}{26}
\newcommand{\NetworkBinsFull}{512}
\newcommand{\NetworkTokensFull}{32}
\newcommand{\NetworkFreqPeriodsFull}{28}
\newcommand{\NetworkNarrowSavedMin}{10240}
\newcommand{\NetworkNarrowSavedMax}{40960}
\newcommand{\NetworkEncodeGflopNarrow}{0.85}
\newcommand{\NetworkDecodeGflopNarrow}{0.38}

\newcommand{\NetworkCacheMibNarrow}{0.3125}
\newcommand{\NetworkEncodeGflopFull}{1.5}
\newcommand{\NetworkDecodeGflopFull}{0.39}
\newcommand{\NetworkEncodeOverDecodeFull}{4}
\newcommand{\NetworkCacheMibFull}{0.5625}

\newcommand{\TrainingClipNorm}{1}
\newcommand{\TrainingMuonRate}{1e-4}
\newcommand{\TrainingAdamRate}{1e-4}
\newcommand{\TrainingMuonRms}{0.2}
\newcommand{\TrainingMomentum}{0.95}
\newcommand{\TrainingBetaOne}{0.9}
\newcommand{\TrainingBetaTwo}{0.95}
\newcommand{\TrainingWeightDecay}{0.1}
\newcommand{\TrainingNewtonSchulzSteps}{5}
\newcommand{\TrainingNewtonSchulzWord}{five}
\newcommand{\TrainingAdamEps}{1e-8}
\newcommand{\TrainingMuonFieldCount}{seven}
\newcommand{\TrainingMuonShareXS}{76}
\newcommand{\TrainingMuonShareB}{93}
\newcommand{\TrainingMuonShareXL}{98}
\newcommand{\TrainingBatch}{1024}
\newcommand{\TrainingObservations}{64}
\newcommand{\TrainingDraws}{16}

\newcommand{\TrainingWarmupPercent}{2}
\newcommand{\TrainingCooldownPercent}{20}

\newcommand{\TrainingStepsNarrow}{650000}

\newcommand{\TrainingStepsFull}{240000}

\newcommand{\TrainingComputeNarrowBHours}{12}

\newcommand{\TrainingComputeFullBHours}{5}

\newcommand{\TrainingBackboneSpeedupAMin}{10}
\newcommand{\TrainingBackboneSpeedupAMax}{43}

\newcommand{\TrainingAttentionGainPercent}{8}

\newcommand{\TrainingDrawCpuCores}{eight}
\newcommand{\TrainingDrawCpuFullSeconds}{1.6}

\newcommand{\EvaluationInjections}{128}

\newcommand{\EvaluationPosterior}{1024}
\newcommand{\EvaluationOdeSteps}{32}
\newcommand{\EvaluationPermutations}{24}

\newcommand{\EvaluationRandom}{200}
\newcommand{\EvaluationWindows}{328}
\newcommand{\EvaluationWindowSources}{1312}
\newcommand{\EvaluationDraws}{256}

\newcommand{\EvaluationBandOne}{15}
\newcommand{\EvaluationBandTwo}{40}
\newcommand{\EvaluationBandThree}{100}

\newcommand{\EvaluationIdealLo}{0.08}
\newcommand{\EvaluationIdealHi}{8.2e-3}
\newcommand{\EvaluationValidation}{1000}
\newcommand{\EvaluationValidationDraws}{127}
\newcommand{\EvaluationRankValues}{128}
\newcommand{\EvaluationGammaSims}{1000}
\newcommand{\EvaluationBandLevel}{95}

\newcommand{\EvaluationPowerInjections}{5\,000}
\newcommand{\EvaluationPowerDraws}{126}
\newcommand{\EvaluationPowerChunk}{100}
\newcommand{\EvaluationPowerAlpha}{5}
\newcommand{\EvaluationPowerMixShare}{10}
\newcommand{\EvaluationPowerNullCoord}{0.03--0.08}
\newcommand{\EvaluationPowerNullOverlap}{0.05}
\newcommand{\EvaluationPowerNullSnr}{0.07}
\newcommand{\EvaluationPowerNullLike}{0.02}

\newcommand{\EvaluationPowerPhaseCoord}{0.03--0.08}
\newcommand{\EvaluationPowerPhaseOverlap}{1.00}
\newcommand{\EvaluationPowerPhaseSnr}{0.17}
\newcommand{\EvaluationPowerPhaseLike}{1.00}

\newcommand{\EvaluationPowerSkyCoord}{0.03--0.08}
\newcommand{\EvaluationPowerSkyOverlap}{1.00}
\newcommand{\EvaluationPowerSkySnr}{0.10}
\newcommand{\EvaluationPowerSkyLike}{1.00}

\newcommand{\EvaluationPowerShuffleCoord}{0.03--0.08}
\newcommand{\EvaluationPowerShuffleOverlap}{1.00}
\newcommand{\EvaluationPowerShuffleSnr}{1.00}
\newcommand{\EvaluationPowerShuffleLike}{1.00}

\newcommand{\EvaluationPowerMixedCoord}{0.06--0.14}
\newcommand{\EvaluationPowerMixedOverlap}{0.97}
\newcommand{\EvaluationPowerMixedSnr}{0.14}
\newcommand{\EvaluationPowerMixedLike}{0.98}

\newcommand{\EvaluationPowerNarrowCoord}{0.57--0.67}
\newcommand{\EvaluationPowerNarrowOverlap}{1.00}
\newcommand{\EvaluationPowerNarrowSnr}{0.71}
\newcommand{\EvaluationPowerNarrowLike}{1.00}

\newcommand{\EvaluationPowerWideCoord}{0.34--0.41}
\newcommand{\EvaluationPowerWideOverlap}{1.00}
\newcommand{\EvaluationPowerWideSnr}{0.58}
\newcommand{\EvaluationPowerWideLike}{1.00}

\newcommand{\EvaluationPowerShiftCoord}{0.03--0.99}
\newcommand{\EvaluationPowerShiftOverlap}{0.71}
\newcommand{\EvaluationPowerShiftSnr}{0.37}
\newcommand{\EvaluationPowerShiftLike}{1.00}

\newcommand{\EvaluationPowerTarpWidth}{0.98--1.00}
\newcommand{\EvaluationPowerTarpBlind}{0.08}

\newcommand{\EvaluationMockFisherIdeal}{3.60}
\newcommand{\EvaluationMockFisherCover}{0.66}
\newcommand{\EvaluationMockFisherCoord}{$<$0.001}
\newcommand{\EvaluationMockFisherTarp}{$<$0.001}
\newcommand{\EvaluationMockFisherOverlap}{$<$0.001}
\newcommand{\EvaluationMockFisherLike}{$<$0.001}

\newcommand{\EvaluationMockFisherMcmc}{0.54--1.23}
\newcommand{\EvaluationMockWideIdeal}{10.8}
\newcommand{\EvaluationMockWideCover}{0.99}
\newcommand{\EvaluationMockWideCoord}{$<$0.001}
\newcommand{\EvaluationMockWideTarp}{$<$0.001}
\newcommand{\EvaluationMockWideOverlap}{$<$0.001}
\newcommand{\EvaluationMockWideLike}{$<$0.001}

\newcommand{\EvaluationMockWideMcmc}{1.01--3.70}
\newcommand{\EvaluationMockNarrowIdeal}{1.80}
\newcommand{\EvaluationMockNarrowCover}{0.40}
\newcommand{\EvaluationMockNarrowCoord}{$<$0.001}
\newcommand{\EvaluationMockNarrowTarp}{$<$0.001}
\newcommand{\EvaluationMockNarrowOverlap}{$<$0.001}
\newcommand{\EvaluationMockNarrowLike}{$<$0.001}

\newcommand{\EvaluationMockNarrowMcmc}{0.29--0.62}
\newcommand{\EvaluationMockBiasedIdeal}{3.60}
\newcommand{\EvaluationMockBiasedCover}{0.47}
\newcommand{\EvaluationMockBiasedCoord}{$<$0.001}
\newcommand{\EvaluationMockBiasedTarp}{$<$0.001}
\newcommand{\EvaluationMockBiasedOverlap}{$<$0.001}
\newcommand{\EvaluationMockBiasedLike}{$<$0.001}

\newcommand{\EvaluationMockBiasedMcmc}{0.54--1.23}
\newcommand{\EvaluationMockPriorIdeal}{\rev{\num{3500}}}
\newcommand{\EvaluationMockPriorCover}{0.66}
\newcommand{\EvaluationMockPriorCoord}{0.030}
\newcommand{\EvaluationMockPriorTarp}{0.546}
\newcommand{\EvaluationMockPriorOverlap}{$<$0.001}
\newcommand{\EvaluationMockPriorLike}{$<$0.001}

\newcommand{\EvaluationMockPriorMcmc}{1.26--528}
\newcommand{\EvaluationMockPhasesIdeal}{3.60}
\newcommand{\EvaluationMockPhasesCover}{0.66}
\newcommand{\EvaluationMockPhasesCoord}{$<$0.001}
\newcommand{\EvaluationMockPhasesTarp}{$<$0.001}
\newcommand{\EvaluationMockPhasesOverlap}{$<$0.001}
\newcommand{\EvaluationMockPhasesLike}{$<$0.001}

\newcommand{\EvaluationMockPhasesMcmc}{0.54--2.65}
\newcommand{\EvaluationMockFloorIdeal}{9.44}
\newcommand{\EvaluationMockFloorCover}{0.68}
\newcommand{\EvaluationMockFloorCoord}{$<$0.001}
\newcommand{\EvaluationMockFloorTarp}{$<$0.001}
\newcommand{\EvaluationMockFloorOverlap}{$<$0.001}
\newcommand{\EvaluationMockFloorLike}{$<$0.001}

\newcommand{\EvaluationMockFloorMcmc}{0.55--1.25}
\newcommand{\EvaluationMockWide}{3}
\newcommand{\EvaluationMockNarrow}{0.5}

\newcommand{\EvaluationMockFloorSnr}{18.4}
\newcommand{\EvaluationMockInjections}{200}
\newcommand{\EvaluationMockInjectionsFull}{200}
\newcommand{\EvaluationMockScorecardSources}{1312}
\newcommand{\EvaluationMockReferenceSources}{187}
\newcommand{\EvaluationMockFisherIdealFull}{4.45}
\newcommand{\EvaluationMockPriorCoverFull}{0.66}
\newcommand{\EvaluationMockFloorCoverQuiet}{0.68}
\newcommand{\EvaluationMockFloorCoverLoud}{0.94}

\newcommand{\EvaluationFisherDraws}{32}

\newcommand{\EvaluationStartFactor}{4}

\newcommand{\EvaluationMockReferenceSnr}{7}
\newcommand{\EvaluationMockLoudSnr}{40}
\newcommand{\EvaluationMockLoudSources}{16}

\newcommand{\McmcWalkers}{64}
\newcommand{\McmcTemps}{six}
\newcommand{\McmcTempMax}{10}
\newcommand{\McmcBurn}{100\,000}
\newcommand{\McmcKeep}{300\,000}
\newcommand{\McmcThin}{100}
\newcommand{\McmcSeeds}{one}
\newcommand{\McmcInjections}{128}
\newcommand{\McmcChains}{128}

\newcommand{\McmcChainsAll}{768}
\newcommand{\McmcKeepDraws}{1\,024}

\newcommand{\McmcJointCores}{two}

\newcommand{\McmcFdotSigmas}{5}
\newcommand{\McmcFdotCutMass}{3e-07}

\newcommand{\McmcConvRuns}{16 and 48 and 80 and 112}

\newcommand{\McmcConvSeeds}{four}

\newcommand{\McmcFisherHighMedianLo}{0.89}
\newcommand{\McmcFisherHighMedianHi}{0.96}
\newcommand{\McmcFisherMidSnr}{15}

\newcommand{\McmcFisherFaintSnr}{5}

\newcommand{\McmcCostJointLo}{4.3}
\newcommand{\McmcCostJointHi}{9.5}
\newcommand{\McmcCostNarrowFourLo}{4.3}
\newcommand{\McmcCostNarrowFourHi}{5.4}
\newcommand{\McmcCostNarrowEightLo}{8.8}
\newcommand{\McmcCostNarrowEightHi}{9.5}
\newcommand{\McmcCostFullSixteenLo}{33}
\newcommand{\McmcCostFullSixteenHi}{40}
\newcommand{\McmcLabelFreeFoundBins}{0.5}

\newcommand{\McmcPriorStartLoudChainRef}{1.11}
\newcommand{\McmcPriorStartLoudFlowChain}{0.83}
\newcommand{\McmcPriorStartFaintChainPrior}{0.32}
\newcommand{\McmcPriorStartFaintRefPrior}{0.07}
\newcommand{\McmcPriorStartFaintFlowPrior}{0.52}

\newcommand{\McmcPriorStartMidFoundLastLo}{0.17}
\newcommand{\McmcPriorStartMidFoundLastHi}{0.29}
\newcommand{\McmcPriorStartMidFoundRefLo}{0.90}
\newcommand{\McmcPriorStartMidFoundRefHi}{1.00}
\newcommand{\McmcPriorStartMidFoundFlowLo}{0.10}
\newcommand{\McmcPriorStartMidFoundFlowHi}{0.98}
\newcommand{\McmcPriorStartChance}{0.04}

\newcommand{\McmcPriorStartSixteenLoudChainRefLo}{1.2}
\newcommand{\McmcPriorStartSixteenLoudChainRefHi}{1.7}
\newcommand{\McmcPriorStartSixteenLoudGrowing}{six}
\newcommand{\McmcPriorStartSixteenLoudN}{nine}
\newcommand{\McmcPriorStartSixteenLoudFlowRefLo}{0.55}
\newcommand{\McmcPriorStartSixteenLoudFlowRefHi}{1.65}

\newcommand{\RatesCatalogue}{9e6}
\newcommand{\RatesVisibleSnr}{7}

\newcommand{\RatesNarrowWindow}{256}
\newcommand{\RatesNarrowAllLo}{54}
\newcommand{\RatesNarrowAllHi}{3547}
\newcommand{\RatesNarrowVisibleLo}{14}

\newcommand{\RatesNarrowVisibleMax}{165}

\newcommand{\RatesFullWindow}{512}

\newcommand{\RatesFullAllHi}{344442}

\newcommand{\RatesFullVisibleMax}{329}
\newcommand{\RatesFullBusiestMhz}{1.70}

\newcommand{\NarrowBandSharePercent}{99.9}
\newcommand{\NarrowBandVisibleSnr}{7}
\newcommand{\NarrowBandTopR}{1024}
\newcommand{\NarrowBandLowTailLowMhz}{0.1}
\newcommand{\NarrowBandLowTailHighMhz}{0.76}
\newcommand{\NarrowBandLowTailCatalogue}{93}
\newcommand{\NarrowBandLowTailVisible}{0.025}
\newcommand{\NarrowBandLowTailR}{16}
\newcommand{\NarrowBandLowFreqLowMhz}{0.76}
\newcommand{\NarrowBandLowFreqHighMhz}{1.6}
\newcommand{\NarrowBandLowFreqCatalogue}{6.4}
\newcommand{\NarrowBandLowFreqVisible}{23}
\newcommand{\NarrowBandLowFreqR}{32}
\newcommand{\NarrowBandCoreLowMhz}{1.6}
\newcommand{\NarrowBandCoreHighMhz}{3.1}
\newcommand{\NarrowBandCoreCatalogue}{0.88}
\newcommand{\NarrowBandCoreVisible}{63}
\newcommand{\NarrowBandCoreR}{64}
\newcommand{\NarrowBandHighFreqLowMhz}{3.1}
\newcommand{\NarrowBandHighFreqHighMhz}{7.5}
\newcommand{\NarrowBandHighFreqCatalogue}{0.10}
\newcommand{\NarrowBandHighFreqVisible}{14}
\newcommand{\NarrowBandHighFreqR}{128}
\newcommand{\NarrowBandHighTailLowMhz}{7.5}
\newcommand{\NarrowBandHighTailHighMhz}{16}
\newcommand{\NarrowBandHighTailCatalogue}{0.0023}
\newcommand{\NarrowBandHighTailVisible}{0.32}
\newcommand{\NarrowBandHighTailR}{256}

\newcommand{\NarrowBandLowTailWindows}{652}
\newcommand{\NarrowBandLowTailAllLo}{2000}
\newcommand{\NarrowBandLowTailAllHi}{44000}
\newcommand{\NarrowBandLowTailVisibleLo}{0}
\newcommand{\NarrowBandLowTailVisibleHi}{0.70}

\newcommand{\NarrowBandLowFreqWindows}{428}
\newcommand{\NarrowBandLowFreqAllLo}{330}
\newcommand{\NarrowBandLowFreqAllHi}{4000}
\newcommand{\NarrowBandLowFreqVisibleLo}{1.0}
\newcommand{\NarrowBandLowFreqVisibleHi}{78}

\newcommand{\NarrowBandCoreWindows}{356}
\newcommand{\NarrowBandCoreAllLo}{49}
\newcommand{\NarrowBandCoreAllHi}{660}
\newcommand{\NarrowBandCoreVisibleLo}{43}
\newcommand{\NarrowBandCoreVisibleHi}{160}

\newcommand{\NarrowBandHighFreqWindows}{542}
\newcommand{\NarrowBandHighFreqAllLo}{1.1}
\newcommand{\NarrowBandHighFreqAllHi}{97}
\newcommand{\NarrowBandHighFreqVisibleLo}{1.1}
\newcommand{\NarrowBandHighFreqVisibleHi}{87}

\newcommand{\NarrowBandHighTailWindows}{526}
\newcommand{\NarrowBandHighTailAllLo}{0.018}
\newcommand{\NarrowBandHighTailAllHi}{2.2}
\newcommand{\NarrowBandHighTailVisibleLo}{0.018}
\newcommand{\NarrowBandHighTailVisibleHi}{2.1}

\newcommand{\NarrowBandFittedFour}{4925}
\newcommand{\NarrowBandFittedEight}{8842}
\newcommand{\NarrowBandFittedSixteen}{15653}
\newcommand{\NarrowBandVisibleTotal}{62497}
\newcommand{\NarrowBandNarrowVisible}{84}

\newcommand{\ResultsRunsRows}{narrow & 2 & \num{1.79} & 64 & \num{35840} & 32 & finished & \num{-1.362} (57) \\
narrow & 4 & \num{1.66} & 64 & \num{33280} & 32 & finished & \num{-1.135} (62) \\
narrow & 8 & \num{1.41} & 64 & \num{28160} & 32 & finished & \num{-0.917} (59) \\
full & 4 & \num{3.20} & 64 & \num{64000} & 64 & diverged, epoch 51 & \num{-1.053} (43) \\
full & 8 & \num{2.69} & 64 & \num{53760} & 64 & finished & \num{-1.105} (64) \\
full & 16 & \num{2.11} & 64 & \num{42240} & 63 & finished & \num{-0.874} (64) \\}
\newcommand{\ResultsNarrowCounts}{2, 4 and 8}
\newcommand{\ResultsFullCounts}{4, 8 and 16}
\newcommand{\ResultsEpochs}{64}
\newcommand{\ResultsSnapshotEvery}{4}
\newcommand{\ResultsForkEpochs}{4, 8, 16, 32}
\newcommand{\ResultsCooldownPercent}{20}

\newcommand{\ResultsFullFourDivergedEpoch}{51}

\newcommand{\ResultsTFourMaxGrad}{3e6}
\newcommand{\ResultsTFourGaussMaxGrad}{9e17}

\newcommand{\ResultsNarrowHours}{32}
\newcommand{\ResultsFullHours}{63--64}
\newcommand{\ResultsWidthRows}{narrow-2 & 7--15 & 59 & 1.04 & 1.00 & 1.04 & 1.14 & 1.13 & 1.02 & 0.92 & 0.69 & 0.94 \\
\footnotesize e32 (40) & 15--40 & 34 & 1.03 & 1.00 & 1.07 & 1.17 & 1.09 & 1.08 & 0.97 & 0.69 & 0.96 \\
 & $\ge40$ & 1 & 1.08 & 1.41 & 0.73 & 2.63 & 2.26 & 0.81 & 1.00 & 0.88 & 1.00 \\
\addlinespace
narrow-4 & 7--15 & 126 & 1.17 & 1.02 & 1.05 & 1.30 & 1.23 & 1.07 & 0.79 & 0.68 & 0.95 \\
\footnotesize main (62) & 15--40 & 56 & 1.23 & 1.04 & 1.13 & 1.32 & 1.29 & 1.16 & 0.91 & 0.72 & 0.95 \\
 & $\ge40$ & 5 & 1.56 & 1.27 & 1.40 & 2.77 & 2.73 & 1.42 & 1.00 & 0.75 & 0.97 \\
\addlinespace
narrow-8 & 7--15 & 234 & 2.11 & 1.08 & 1.38 & 1.66 & 1.46 & 1.51 & 0.66 & 0.68 & 0.95 \\
\footnotesize e32 (40) & 15--40 & 130 & 1.41 & 1.13 & 1.56 & 1.84 & 1.64 & 1.61 & 0.84 & 0.72 & 0.96 \\
 & $\ge40$ & 12 & 1.62 & 1.24 & 1.40 & 3.00 & 2.75 & 1.44 & 0.92 & 0.72 & 0.93 \\
\addlinespace
full-4 & 7--15 & 50 & 1.66 & 1.07 & 1.21 & 1.57 & 1.39 & 1.21 & 0.62 & 0.69 & 0.95 \\
\footnotesize t44 (64) & 15--40 & 87 & 1.22 & 1.19 & 1.17 & 1.84 & 1.65 & 1.22 & 0.77 & 0.72 & 0.95 \\
 & $\ge40$ & 66 & 6.61 & 5.75 & 1.56 & 5.00 & 3.70 & 1.60 & 0.82 & 0.69 & 0.94 \\
\addlinespace
full-8 & 7--15 & 87 & 1.69 & 1.12 & 1.41 & 1.90 & 1.66 & 1.44 & 0.67 & 0.70 & 0.95 \\
\footnotesize main (64) & 15--40 & 176 & 2.19 & 1.44 & 1.42 & 2.69 & 2.11 & 1.42 & 0.69 & 0.69 & 0.95 \\
 & $\ge40$ & 145 & 10.7 & 9.05 & 1.97 & 8.61 & 5.79 & 1.92 & 0.66 & 0.71 & 0.96 \\
\addlinespace
full-16 & 7--15 & 179 & 3.27 & 1.12 & 1.57 & 2.09 & 1.83 & 1.64 & 0.56 & 0.68 & 0.94 \\
\footnotesize e32 (40) & 15--40 & 357 & 2.37 & 1.37 & 1.67 & 2.90 & 2.26 & 1.65 & 0.68 & 0.69 & 0.95 \\
 & $\ge40$ & 288 & 23.6 & 12.8 & 2.87 & 14.8 & 8.95 & 2.79 & 0.50 & 0.68 & 0.95 \\}
\newcommand{\ResultsMcmcRhatMax}{1.073}
\newcommand{\ResultsMcmcRhatNarrowMax}{1.005}
\newcommand{\ResultsMcmcEssMin}{1097}

\newcommand{\ResultsCoverSixtyEightLo}{0.67}
\newcommand{\ResultsCoverSixtyEightHi}{0.68}

\newcommand{\ResultsNarrowTwoWidthLo}{1.00}
\newcommand{\ResultsNarrowTwoWidthHi}{1.17}

\newcommand{\ResultsNarrowTwoOff}{7}
\newcommand{\ResultsNarrowTwoMcmcOff}{5}

\newcommand{\ResultsNarrowEightFzeroLow}{2.1}

\newcommand{\ResultsNarrowEightFoundLow}{66}
\newcommand{\ResultsNarrowEightOff}{29}

\newcommand{\ResultsFullFourFzeroMid}{1.2}
\newcommand{\ResultsFullFourFzeroLoud}{6.6}
\newcommand{\ResultsFullFourFdotLoud}{5.7}

\newcommand{\ResultsFullFourFoundLoud}{82}

\newcommand{\ResultsFullFourOff}{28}
\newcommand{\ResultsFullFourMcmcOff}{2}

\newcommand{\ResultsFullEightFdotLoud}{9.0}

\newcommand{\ResultsFullEightOff}{40}
\newcommand{\ResultsFullEightMcmcOff}{2}

\newcommand{\ResultsFullSixteenFzeroMid}{2.4}
\newcommand{\ResultsFullSixteenFzeroLoud}{24}
\newcommand{\ResultsFullSixteenFdotLoud}{13}

\newcommand{\ResultsFullSixteenFoundLoud}{50}

\newcommand{\ResultsFullSixteenOff}{49}
\newcommand{\ResultsFullSixteenMcmcOff}{0}

\newcommand{\ResultsFaintWidthLo}{0.96}
\newcommand{\ResultsFaintWidthHi}{1.08}
\newcommand{\ResultsNarrowEightFaintFzero}{5.6}
\newcommand{\ResultsFullSixteenFaintFzero}{2.0}
\newcommand{\ResultsFullSixteenFaintSkyLo}{0.81}
\newcommand{\ResultsFullSixteenFaintSkyHi}{0.85}

\newcommand{\ResultsForkRows}{narrow-2 & \num{-1.253} / 1.17 (5) & \num{-1.305} / 1.08 (10) & \num{-1.363} / 1.11 (19) & \num{-1.379} / 1.03 (40) & \num{-1.362} / 1.12 (57) \\
narrow-4 & \num{-0.899} / 1.45 (5) & \num{-0.965} / 1.18 (10) & \num{-1.019} / 1.17 (18) & \num{-1.092} / 1.09 (40) & \num{-1.135} / 1.23 (62) \\
narrow-8 & \num{-0.714} / 3.99 (5) & \num{-0.890} / 1.82 (10) & \num{-0.909} / 1.67 (20) & \num{-0.974} / 1.41 (40) & \num{-0.917} / 1.50 (59) \\
full-4 & \num{-1.020} / 2.56 (5) & \num{-1.135} / 1.97 (10) & \num{-1.110} / 1.92 (18) & \num{-1.200} / 1.48 (40) & n/c / 1.22 (64) \\
full-8 & \num{-0.844} / 5.17 (5) & \num{-0.948} / 3.50 (10) & \num{-0.989} / 2.80 (20) & \num{-1.083} / 2.41 (38) & \num{-1.105} / 2.19 (64) \\
full-16 & \num{-0.630} / 8.57 (5) & \num{-0.737} / 4.82 (10) & \num{-0.826} / 3.43 (20) & \num{-0.887} / 2.37 (40) & \num{-0.874} / 2.15 (64) \\}
\newcommand{\ResultsNarrowTwoForkGain}{0.13}
\newcommand{\ResultsFullFourForkGain}{0.18}
\newcommand{\ResultsFullFourMainLoud}{17}
\newcommand{\ResultsFullFourForkLoud}{9.7}
\newcommand{\ResultsFullFourContLoud}{6.6}
\newcommand{\ResultsNarrowFourSepClose}{3.1}
\newcommand{\ResultsNarrowFourSepCloseN}{18}

\newcommand{\ResultsNarrowFourSepFar}{1.1}

\newcommand{\ResultsNarrowEightSepClose}{3.3}
\newcommand{\ResultsNarrowEightSepCloseN}{80}

\newcommand{\ResultsNarrowEightSepFar}{1.3}

\newcommand{\ResultsSepCloseBins}{3}
\newcommand{\ResultsSepFarBins}{10}
\newcommand{\ResultsFullSplitMhz}{10}

\newcommand{\ResultsFullSixteenLoudHighShare}{43}
\newcommand{\ResultsFullSixteenUnfoundHighShare}{80}

\newcommand{\ResultsValidationRows}{narrow-2 & all & 2000 & 0.110 & 0.835 & 0.764 & $<$0.001 & 0.042 & $<$0.001 & +0.20 & +0.31 & 0.29 \\
\quad Gaussian mock & all & 2000 & $<$0.001 & $<$0.001 & $<$0.001 & $<$0.001 & $<$0.001 & $<$0.001 & +0.05 & +0.14 & 0.51 \\
\footnotesize e32 (40) & 15--40 & 261 & 0.006 & 0.737 & 0.765 & $<$0.001 & 0.526 & $<$0.001 & +0.14 & +0.28 & 0.33 \\
 & 40--100 & 18 & 0.072 & 0.227 & 0.032 & $<$0.001 & 0.558 & $<$0.001 & +0.37 & +0.42 & 0.09 \\
\addlinespace
narrow-4 & all & 4000 & 0.032 & 0.815 & 0.288 & $<$0.001 & $<$0.001 & $<$0.001 & +0.20 & +0.43 & 0.10 \\
\quad Gaussian mock & all & 4000 & $<$0.001 & $<$0.001 & $<$0.001 & $<$0.001 & $<$0.001 & $<$0.001 & +0.05 & +0.22 & 0.41 \\
\footnotesize main (62) & 15--40 & 537 & 0.004 & 0.856 & 0.025 & $<$0.001 & 0.334 & $<$0.001 & +0.19 & +0.43 & 0.10 \\
 & 40--100 & 38 & 0.100 & 0.609 & 0.990 & $<$0.001 & 0.220 & $<$0.001 & +0.41 & +0.48 & 0.04 \\
\addlinespace
narrow-8 & all & 8000 & 0.001 & 0.135 & 0.033 & $<$0.001 & $<$0.001 & $<$0.001 & +0.15 & +0.47 & 0.03 \\
\quad Gaussian mock & all & 8000 & $<$0.001 & $<$0.001 & $<$0.001 & $<$0.001 & $<$0.001 & $<$0.001 & +0.05 & +0.33 & 0.26 \\
\footnotesize e32 (40) & 15--40 & 1060 & 0.012 & 0.050 & 0.031 & $<$0.001 & 0.007 & $<$0.001 & +0.12 & +0.48 & 0.02 \\
 & 40--100 & 81 & 0.059 & 0.387 & 0.109 & $<$0.001 & 0.001 & $<$0.001 & +0.32 & +0.50 & 0.00 \\
\addlinespace
full-4 & all & 4000 & 0.016 & 0.013 & 0.531 & $<$0.001 & 0.156 & $<$0.001 & +0.20 & +0.40 & 0.11 \\
\quad Gaussian mock & all & 4000 & $<$0.001 & $<$0.001 & $<$0.001 & $<$0.001 & $<$0.001 & $<$0.001 & +0.05 & +0.23 & 0.40 \\
\footnotesize t44 (64) & 15--40 & 802 & 0.077 & 0.615 & 0.896 & $<$0.001 & 0.561 & $<$0.001 & +0.17 & +0.39 & 0.13 \\
 & 40--100 & 200 & 0.058 & 0.835 & 0.354 & $<$0.001 & 0.642 & $<$0.001 & +0.31 & +0.42 & 0.09 \\
\addlinespace
full-8 & all & 8000 & 0.135 & 0.206 & 0.036 & $<$0.001 & $<$0.001 & $<$0.001 & +0.20 & +0.47 & 0.02 \\
\quad Gaussian mock & all & 8000 & $<$0.001 & $<$0.001 & $<$0.001 & $<$0.001 & $<$0.001 & $<$0.001 & +0.04 & +0.31 & 0.28 \\
\footnotesize main (64) & 15--40 & 1604 & 0.067 & 0.779 & 0.035 & $<$0.001 & 0.387 & $<$0.001 & +0.19 & +0.48 & 0.02 \\
 & 40--100 & 376 & 0.014 & 0.631 & 0.421 & $<$0.001 & $<$0.001 & $<$0.001 & +0.36 & +0.49 & 0.01 \\
\addlinespace
full-16 & all & 16000 & 0.151 & 0.001 & 0.014 & $<$0.001 & $<$0.001 & $<$0.001 & +0.15 & +0.48 & 0.02 \\
\quad Gaussian mock & all & 16000 & $<$0.001 & $<$0.001 & $<$0.001 & $<$0.001 & $<$0.001 & $<$0.001 & +0.04 & +0.41 & 0.10 \\
\footnotesize e32 (40) & 15--40 & 3228 & 0.045 & 0.498 & 0.024 & $<$0.001 & $<$0.001 & $<$0.001 & +0.13 & +0.50 & 0.00 \\
 & 40--100 & 758 & 0.052 & 0.046 & 0.078 & $<$0.001 & $<$0.001 & $<$0.001 & +0.29 & +0.50 & 0.00 \\}
\newcommand{\ResultsValidationInjections}{1000}
\newcommand{\ResultsValidationDraws}{127}

\newcommand{\ResultsMockOverlapOffsetLo}{+0.04}
\newcommand{\ResultsMockOverlapOffsetHi}{+0.05}
\newcommand{\ResultsFlowOverlapOffsetLo}{+0.15}
\newcommand{\ResultsFlowOverlapOffsetHi}{+0.20}
\newcommand{\ResultsLikeExcessLo}{+0.07}
\newcommand{\ResultsLikeExcessHi}{+0.20}
\newcommand{\ResultsOverlapFixFlowLo}{+0.23}
\newcommand{\ResultsOverlapFixFlowHi}{+0.30}

\newcommand{\ResultsOverlapFixMockFreqSkyLo}{$-$0.02}
\newcommand{\ResultsOverlapFixMockFreqSkyHi}{+0.00}
\newcommand{\ResultsOverlapFixMockPhasesLo}{+0.04}
\newcommand{\ResultsOverlapFixMockPhasesHi}{+0.10}
\newcommand{\ResultsOverlapFreqPhasesFlowLo}{+0.21}
\newcommand{\ResultsOverlapFreqPhasesFlowHi}{+0.24}
\newcommand{\ResultsOverlapModuliFaintFlowLo}{$-$0.06}
\newcommand{\ResultsOverlapModuliFaintFlowHi}{$-$0.01}
\newcommand{\ResultsOverlapModuliFaintMockLo}{$-$0.01}
\newcommand{\ResultsOverlapModuliFaintMockHi}{+0.04}
\newcommand{\ResultsOverlapAbsZFlowLo}{+0.00}
\newcommand{\ResultsOverlapAbsZFlowHi}{+0.02}
\newcommand{\ResultsOverlapOwnSnrFlowLo}{+0.00}
\newcommand{\ResultsOverlapOwnSnrFlowHi}{+0.05}
\newcommand{\ResultsInvisibleSnr}{3}

\newcommand{\ResultsNarrowEightFaintFlowPrior}{0.87}
\newcommand{\ResultsNarrowEightFaintMcmcPrior}{0.19}

\newcommand{\ResultsFullSixteenFaintFlowPrior}{0.99}
\newcommand{\ResultsFullSixteenFaintMcmcPrior}{0.60}
\newcommand{\ResultsFaintFlowPriorLo}{0.84}
\newcommand{\ResultsFaintFlowPriorHi}{0.99}

\newcommand{\ResultsFullSixteenContractLow}{15}

\newcommand{\ResultsFullSixteenContractHigh}{2}
\newcommand{\ResultsNarrowFourFpBfMid}{1.23}
\newcommand{\ResultsNarrowFourFpThirtyTwoMid}{1.22}

\newcommand{\ResultsFullEightFpBfMid}{2.58}
\newcommand{\ResultsFullEightFpThirtyTwoMid}{2.14}

\newcommand{\ResultsFullEightFpBfLoud}{3.48}
\newcommand{\ResultsFullEightFpThirtyTwoLoud}{3.38}

\newcommand{\ResultsSamplerNarrowMax}{128}
\newcommand{\ResultsSamplerNarrowEight}{0.86}
\newcommand{\ResultsSamplerNarrowSixteen}{0.94}
\newcommand{\ResultsSamplerNarrowThirtyTwo}{0.98}
\newcommand{\ResultsSamplerNarrowSixtyFour}{1.00}

\newcommand{\ResultsSamplerFullEight}{0.97}

\newcommand{\ResultsSamplerGridFrom}{16}
\newcommand{\ResultsSamplerGridLo}{1.00}
\newcommand{\ResultsSamplerGridHi}{1.03}
\newcommand{\ResultsSamplerEulerNarrowThirtyTwo}{0.78}

\newcommand{\ResultsSamplerEulerFullThirtyTwo}{0.92}

\newcommand{\ResultsRobustInjections}{64}
\newcommand{\ResultsRobustDraws}{128}
\newcommand{\ResultsRobustNoiseFaintWidth}{0.61}

\newcommand{\ResultsRobustNoiseFaintWithin}{38}
\newcommand{\ResultsRobustBaseFaintWithin}{60}

\newcommand{\ResultsRobustBaseLoudWithin}{72}
\newcommand{\ResultsRobustNoiseSlotSnr}{8.5}
\newcommand{\ResultsRobustBaseSlotSnr}{4.8}

\newcommand{\ResultsRobustOtherWithinLo}{56}
\newcommand{\ResultsRobustOtherWithinHi}{79}

\newcommand{\ResultsRobustFullNoiseLoudWithin}{43}
\newcommand{\ResultsRobustFullBaseLoudWithin}{70}

\newcommand{\ResultsRobustFullOtherWithinLo}{57}
\newcommand{\ResultsRobustFullOtherWithinHi}{72}
\newcommand{\ResultsJsBins}{32}
\newcommand{\ResultsJsDraws}{1024}
\newcommand{\ResultsJsFloorMedian}{0.011}
\newcommand{\ResultsJsFloorHigh}{0.016}
\newcommand{\ResultsJsRows}{narrow-2 & 7--15 & 59 & 0.058 & 0.015 & 0.020 & 0.038 & 0.031 & 0.020 & 0.042 & 0.027 & 0.035 \\
\footnotesize e32 (40) & 15--40 & 34 & 0.029 & 0.015 & 0.024 & 0.035 & 0.026 & 0.018 & 0.038 & 0.018 & 0.029 \\
 & $\ge40$ & 1 & 0.050 & 0.057 & 0.089 & 0.38 & 0.19 & 0.028 & 0.11 & 0.032 & 0.12 \\
\addlinespace
narrow-4 & 7--15 & 126 & 0.10 & 0.019 & 0.029 & 0.067 & 0.051 & 0.027 & 0.055 & 0.034 & 0.052 \\
\footnotesize main (62) & 15--40 & 56 & 0.066 & 0.020 & 0.044 & 0.071 & 0.052 & 0.028 & 0.053 & 0.033 & 0.052 \\
 & $\ge40$ & 5 & 0.094 & 0.048 & 0.087 & 0.42 & 0.33 & 0.051 & 0.11 & 0.079 & 0.15 \\
\addlinespace
narrow-8 & 7--15 & 234 & 0.26 & 0.026 & 0.091 & 0.15 & 0.093 & 0.027 & 0.12 & 0.084 & 0.13 \\
\footnotesize e32 (40) & 15--40 & 130 & 0.18 & 0.029 & 0.11 & 0.16 & 0.13 & 0.029 & 0.12 & 0.078 & 0.12 \\
 & $\ge40$ & 12 & 0.31 & 0.078 & 0.13 & 0.32 & 0.36 & 0.023 & 0.13 & 0.056 & 0.16 \\
\addlinespace
full-4 & 7--15 & 50 & 0.14 & 0.025 & 0.059 & 0.12 & 0.080 & 0.038 & 0.081 & 0.055 & 0.074 \\
\footnotesize t44 (64) & 15--40 & 87 & 0.11 & 0.067 & 0.051 & 0.20 & 0.13 & 0.061 & 0.064 & 0.068 & 0.11 \\
 & $\ge40$ & 66 & 0.60 & 0.64 & 0.17 & 0.54 & 0.53 & 0.13 & 0.18 & 0.24 & 0.37 \\
\addlinespace
full-8 & 7--15 & 87 & 0.27 & 0.029 & 0.079 & 0.21 & 0.14 & 0.073 & 0.11 & 0.12 & 0.16 \\
\footnotesize main (64) & 15--40 & 176 & 0.32 & 0.13 & 0.15 & 0.38 & 0.31 & 0.071 & 0.13 & 0.17 & 0.23 \\
 & $\ge40$ & 145 & 0.67 & 0.65 & 0.24 & 0.62 & 0.57 & 0.10 & 0.21 & 0.30 & 0.42 \\
\addlinespace
full-16 & 7--15 & 179 & 0.39 & 0.038 & 0.17 & 0.26 & 0.21 & 0.079 & 0.21 & 0.18 & 0.23 \\
\footnotesize e32 (40) & 15--40 & 357 & 0.35 & 0.15 & 0.22 & 0.40 & 0.31 & 0.12 & 0.22 & 0.22 & 0.29 \\
 & $\ge40$ & 288 & 0.81 & 0.70 & 0.36 & 0.76 & 0.69 & 0.18 & 0.39 & 0.35 & 0.55 \\}
\newcommand{\ResultsJsFaintLo}{0.011}
\newcommand{\ResultsJsFaintHi}{0.042}
\newcommand{\ResultsJsFaintFullFzero}{0.015}
\newcommand{\ResultsJsFaintNarrowTwoFzero}{0.042}
\newcommand{\ResultsJsFaintNarrowFourFzero}{0.075}
\newcommand{\ResultsJsFaintNarrowEightFzero}{0.37}
\newcommand{\ResultsJsFaintFullSixteenFzero}{0.20}

\newcommand{\ResultsJsNarrowTwoMean}{0.033}

\newcommand{\ResultsJsFullFourLoudFzero}{0.60}
\newcommand{\ResultsJsFullFourLoudFdot}{0.64}
\newcommand{\ResultsJsFullFourLoudSky}{0.54}

\newcommand{\ResultsJsFullEightLoudFzero}{0.67}
\newcommand{\ResultsJsFullEightLoudFdot}{0.65}
\newcommand{\ResultsJsFullEightLoudSky}{0.62}

\newcommand{\ResultsJsFullSixteenLoudFzero}{0.81}
\newcommand{\ResultsJsFullSixteenLoudFdot}{0.70}
\newcommand{\ResultsJsFullSixteenLoudSky}{0.76}

\newcommand{\ResultsJsGaussWidthOneHalf}{0.053}
\newcommand{\ResultsJsGaussWidthTwo}{0.13}

\newcommand{\ResultsJsGaussWidthTen}{0.63}
\newcommand{\ResultsJsGaussOffsetHalf}{0.044}
\newcommand{\ResultsJsGaussOffsetOne}{0.16}

\newcommand{\ResultsSummaryRows}{n2-narrow & 2 & e32 (40) & 0.035 & 0.029 & 0.685 & 0.110 & 0.704 & 0.105 \\
n4-narrow & 4 & main (62) & 0.052 & 0.060 & 0.688 & 0.032 & 0.700 & 0.073 \\
n8-narrow & 8 & e32 (40) & 0.13 & 0.12 & 0.685 & 0.001 & 0.708 & 0.010 \\
n4-full & 4 & t44 (64) & 0.074 & 0.23 & 0.691 & 0.016 & 0.694 & 0.078 \\
n8-full & 8 & main (64) & 0.16 & 0.33 & 0.691 & 0.135 & 0.700 & 0.259 \\
n16-full & 16 & e32 (40) & 0.23 & 0.40 & 0.689 & 0.151 & 0.715 & $<$0.001 \\}
\newcommand{\ResultsSummaryFloor}{0.011}

\newcommand{\AmortisationNarrowShortEpochs}{5}
\newcommand{\AmortisationNarrowShortHours}{2.5}
\newcommand{\AmortisationNarrowLongHours}{32}
\newcommand{\AmortisationNarrowShortRatio}{1.2--4.0}
\newcommand{\AmortisationNarrowLongRatio}{1.1--1.5}
\newcommand{\AmortisationNarrowFlowSeconds}{0.22}
\newcommand{\AmortisationNarrowPasses}{658}
\newcommand{\AmortisationNarrowPassMinutes}{3.0}
\newcommand{\AmortisationNarrowHourWindows}{32}

\newcommand{\AmortisationFullShortHours}{5.0}
\newcommand{\AmortisationFullLongHours}{64}
\newcommand{\AmortisationFullShortRatio}{2.6--8.6}
\newcommand{\AmortisationFullLongRatio}{1.2--2.2}
\newcommand{\AmortisationFullFlowSeconds}{0.25}
\newcommand{\AmortisationFullPasses}{2614}
\newcommand{\AmortisationFullPassMinutes}{1.5}
\newcommand{\AmortisationFullHourWindows}{64}

\newcommand{\AmortisationInteriorResponses}{3}
\newcommand{\AmortisationDraws}{256}

\newcommand{\PriorCornerDraws}{200000}
\newcommand{\PriorCornerInner}{39}
\newcommand{\PriorCornerOuter}{86}

\icmltitlerunning{Joint Flow-Matching Posteriors for LISA Galactic Binaries}

\begin{document}

\twocolumn[
  \icmltitle{\rev{Joint Flow-Matching Posteriors for Overlapping LISA Galactic Binaries}}

  \begin{icmlauthorlist}
    \icmlauthor{Davide Sartor}{umass}
    \icmlauthor{Giorgio Mentasti}{apc}
  \end{icmlauthorlist}

  \icmlaffiliation{umass}{University of Massachusetts Amherst, Amherst, MA, USA}
  \icmlaffiliation{apc}{Universit\'e Paris Cit\'e, CNRS, Astroparticule et Cosmologie, F-75013 Paris, France}

  \icmlkeywords{flow matching, simulation-based inference, gravitational waves, LISA,
    simulation-based inference, label switching, transformers}

  \vskip 0.3in
]

\printAffiliationsAndNotice{}  %

\begin{abstract}
    \Ac{lisa} will observe millions of \acp{gb} at once, thousands of them resolvable, with signals that overlap in frequency. Separating them requires a global fit, a joint inference of all sources, which today takes \ac{mcmc} samplers hours to days on a cluster and is repeated for each new stretch of data. 
  We adapt recent advances in large-scale diffusion models, flow matching with a multimodal transformer, to learn the joint posterior once and then draw from it in a fraction of a second. 
  We evaluate our approach on a realistic simulator, with full instrument response and noise, priors fitted to a simulated Galactic catalogue, and sources sampled anywhere in the entire \qtyrange{\PriorsFullLowMhz}{\PriorsFullHighMhz}{\milli\hertz} frequency band.
  We demonstrate that the approach is scalable to up to \num{\NarrowBandFittedSixteen} sources fitted simultaneously, evaluating calibration and coverage w.r.t. the \ac{mcmc} baseline.
  The results show that this approach could be realistically scaled to tackle the full global fit problem.
\end{abstract}
\acresetall

\section{Introduction}
\label{sec:introduction}

\begin{figure}[t]
  \centering
  \includegraphics[width=\columnwidth]{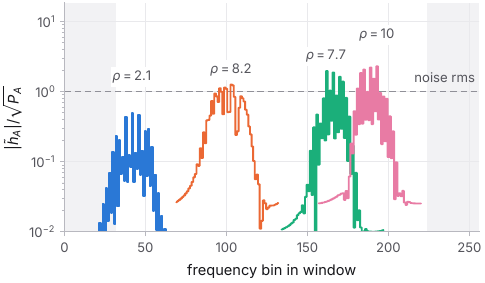}
  \caption{An example of four \ac{gb} signals in one narrow-band frequency window, in channel $A$ for
  two years of observation and normalised by the square root of the noise \acs{psd}, each labelled with
  its \acs{snr} from \cref{eq:gw-snr} \citep{cutler1994gravitational}.}
  \label{fig:forward-model-window}
\end{figure}

The space-based \ac{lisa}, due for launch in 2035, will open the millihertz band of gravitational
waves \citep{amaro2017lisa,colpi2024lisa}. Its most numerous sources are the compact binaries of
our own Galaxy: millions of quasi-monochromatic \acp{gb} \rev{from}
\qtyrange{\RelatedWorkFzeroLow}{\RelatedWorkFzeroHigh}{\milli\hertz}, of which thousands will be
resolved \citep{nelemans2001gravitational,korol2017prospects,karnesis2021characterization}. Below a
few millihertz their signals overlap (\cref{fig:forward-model-window}): the instrument's orbit and
the source's own chirp spread each signal over several frequency bins, neighbours share bins, and
their parameters must be inferred jointly (\cref{fig:related-crowding}).

\begin{table*}[t]
\caption{\Acp{gb} returned and cost of the global fits of the Sangria data set (top), of \acs{sbi}
for \acp{gb} (middle) and of this work (bottom).}
\label{tab:gw-globalfit}
\centering
\small
\newcommand{\tabgroup}[1]{\multicolumn{4}{@{}l}{\textsc{#1}}\\[1pt]}
\begin{tabular}{@{}llll@{}}
\toprule
pipeline & \acp{gb} returned & hardware, time & \rev{compute} \\
\midrule
GLASS \citep{littenberg2023prototype} & -- & \num{7488} \acsp{cpu}, \qty{5}{\day} & ${\sim}10^6$ \\
Erebor \citep{katz2024efficient} & \num{12000} (\num{8000} confident) & 4 A100, \qty{7}{\day} & ${\sim}700$ \acs{gpu} h \\
\citet{strub2024global} & \num{5100} (half a year) & \acs{cpu} cluster & \num{700} \\
Gee-Moo \citep{deng2025modular} & \num{9542} (\num{7620} identified) & \num{1000} cores, \qty{20}{\hour} & ${\sim}2\times10^4$ \\
\midrule
\tabgroup{SBI}
\citet{delmond2026neural} & 1; 2 as a proof of concept & 1 A100, ${\sim}1$ week; $<\qty{1}{\second}$ per posterior & ${\sim}170$ \acs{gpu} h \\
\citet{mao2025robust} & 1, with 3 free parameters & 1 L40S, \numrange{3}{5.5}\,h & \numrange{3}{5.5} \acs{gpu} h \\
SlotFlow \citep{houba2025slotflow} & up to 10 sinusoids, no response & 4 GH200, \qty{190}{\hour}; \qty{13}{\milli\second} per draw & ${\sim}760$ \acs{gpu} h \\
\midrule
\tabgroup{ours}
this work, $S=16$ & \num{\NarrowBandFittedSixteen} (at most) & 1 A100, \ResultsFullHours\,h; \AmortisationFullFlowSeconds\,s per window & \ResultsFullHours{} \acs{gpu} h \\
\bottomrule
\end{tabular}
\end{table*}

The reference method is the global fit, a trans-dimensional \ac{mcmc} over every resolvable
source \citep{littenberg2020global,littenberg2023prototype,katz2024efficient,strub2024global,deng2025modular}.
It is the standard for what the data support, but a joint chain over one of our frequency
windows alone takes \qtyrange{\McmcCostJointLo}{\McmcCostJointHi}{\hour}, and the whole fit runs
again on every new stretch of data. \Ac{sbi} learns the posterior once and then returns
it in a fraction of a second. For \acp{gb} it has so far treated one source, two ordered sources
in a narrow band, or a factorised set of toy signals
\citep{korsakova2024neural,houba2025slotflow,delmond2026neural} (\cref{tab:related-closest}).

We bring \acs{sbi} to the global fit's unit of work: the joint posterior of the $S$
sources of a frequency window, $8S$ parameters under the full instrument response, for windows
anywhere in the band. A conditional flow-matching model
\citep{lipman2023flow,liu2023flow,albergo2023building,dax2023flow} draws it with no likelihood
evaluation.

\paragraph{Contributions.}
\begin{itemize}
  \item \emph{Joint posteriors of overlapping \acp{gb} from the flow alone}, with $S$ from two to
    sixteen sources per window under broad catalogue priors, for windows anywhere in the band
    from \qtyrange{\PriorsFullLowMhz}{\PriorsFullHighMhz}{\milli\hertz}
    (\cref{sec:gw-slicing,sec:narrow-band}). Runs with increasing $S$ measure how the joint
    posterior scales.
  \item \emph{Coordinates without the posterior's symmetries.} A flat chart of eight coordinates
    per source folds each source's equivalent copies into one mode, and slots ordered in frequency
    remove the labelling modes (\cref{sec:method-chart}).
  \item \emph{A one-way multi-stream transformer} that encodes the data once per window, so every
    draw and every sampler step reuses it (\cref{sec:method-network}).
  \item \emph{Subbands of fixed shapes.} Dividing the band where the width of the sources'
    responses doubles gives five subproblems (\cref{sec:narrow-band}).
  \item \emph{Validation on known answers.} A toy with an exact posterior shows that ranks of
    data-space quantities catch phase, sky and dependence errors that parameter ranks miss, and
    mock posteriors are run through the whole evaluation (\cref{sec:validation}).
\end{itemize}

\section{Background and Related Work}
\label{sec:background}

\begin{figure*}[t]
  \centering
  \includegraphics[width=\textwidth]{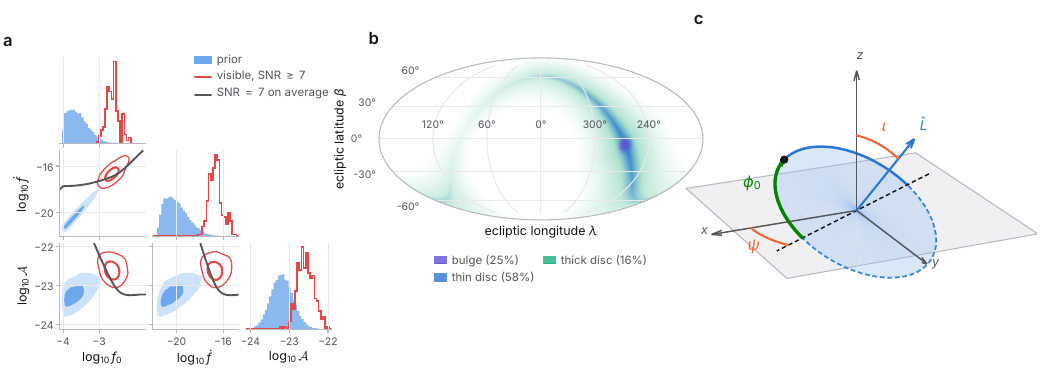}
  \caption{A visual representation of the \ac{gb} parameters and their priors. (a)~$(\log_{10}f_0, \log_{10}\fdot, \log_{10}\Amp)$ over the full band,
  \num{\ChartPriorDraws} sources in windows centred on the catalogue's $f_0$ marginal: regions
  enclosing \qty{\PriorCornerInner}{\percent} and \qty{\PriorCornerOuter}{\percent} of the prior
  (blue) and of the sources with $\snr\ge\PriorsVisibleSnr$ (red), and the line
  $\snr=\PriorsVisibleSnr$ (grey), which averages over the other parameters, so sources
  that reach it through a favourable sky or orientation also lie below it. (b)~The sky,
  \cref{eq:priors-sky}, in ecliptic coordinates $(\lambda, \beta)$: the bulge, the thick and the thin
  disc. (c)~The orientation: the orbit (blue, dashed
  below the sky plane) is inclined by $\iota$ to the sky plane $(x, y)$, with $z$ towards the observer; its line of nodes
  (black, dashed) is at $\psi$ from $x$, and $\phi_0$ is the phase along the orbit at the start of the observation.}
  \label{fig:priors}
\end{figure*}

\paragraph{The global fit.}
\label{sec:background-gw}\label{sec:gw-globalfit}
The \ac{lisa} data hold every source at once, so they are analysed by a global fit: a
reversible-jump \ac{mcmc} \citep{green1995reversible} over the sources of all kinds, their number
and the noise
\citep{cornish2005lisa,littenberg2020global,littenberg2023prototype,katz2024efficient,strub2024global,deng2025modular}.
On a simulated year of data, current pipelines return \numrange{5000}{12000} \acp{gb} for \num{700}
to about $10^6$ \ac{cpu} hours, and every new dataset or change of model pays that cost again
(\cref{tab:gw-globalfit}, \cref{app:related-globalfit}).

\paragraph{Simulation-based inference.}
\label{sec:inversion}\label{sec:background-generative}
Given data $d$, a prior $\pi(\theta)$ and a likelihood $p(d\mid\theta)$, the posterior is
$p(\theta\mid d)\propto p(d\mid\theta)\,\pi(\theta)$. Samplers such as \ac{mcmc} and nested
sampling \citep{skilling2006nested} evaluate the likelihood many times and start again for every
dataset. \Ac{sbi} \citep{cranmer2020frontier} instead trains a conditional density
estimator on simulated pairs $(\theta,d)$; in \ac{npe} the estimator models the posterior itself
\citep{papamakarios2016fast,greenberg2019automatic}. Training needs only a simulator, and the
posterior of a new dataset costs only network evaluations, but nothing guarantees the estimate. It
is checked with rank statistics over many simulations, \ac{sbc} \citep{talts2018validating} and
\ac{tarp} \citep{lemos2023sampling}, which are necessary conditions only (\cref{sec:validation}).

Flow matching, rectified flows and stochastic interpolants
\citep{lipman2023flow,liu2023flow,albergo2023building} train a generative model by regression: a
network learns the velocity that carries a simple base distribution to the data along prescribed
paths, and a sample solves an ordinary differential equation. Its quality depends on what the
network predicts, how the loss is weighted along the path and how the sampler spaces its steps
\citep{karras2022elucidating,kingma2023understanding}. For posteriors it entered \acs{sbi}
as \ac{fmpe} \citep{dax2023flow}. A posterior differs from an image in that its width
matters and can be orders of magnitude below the prior's, which is where our objective departs from
standard \ac{fmpe} (\cref{sec:method-fm}). Our network is a \ac{mmdit}
\citep{peebles2023scalable,esser2024scaling} with the data and the sources as its two streams
(\cref{sec:method-network}).

\paragraph{\Acs{sbi} for gravitational waves.}
\label{sec:background-gw-npe}\label{sec:background-gb}
\Acs{sbi} is established for the compact binaries of ground-based detectors
\citep{green2020gravitational,gabbard2022bayesian,dax2021real,dax2025real}, with importance sampling to
the exact posterior \citep{dax2023neural}, and for \ac{lisa}'s massive black-hole binaries
\citep{du2024advancing,liang2024rapid,spadaro2026accurate}. Overlapping signals have been analysed
jointly a pair at a time on the ground \citep{langendorff2023normalizing,papalini2025can}. For
\acp{gb}, normalising flows serve inside the global fit as priors and proposals fitted to one
source's chains \citep{korsakova2024neural}. As posterior estimators, \citet{delmond2026neural}
train a neural spline flow on one \ac{gb} under the full response and, as a proof of concept, on
two; \citet{mao2025robust} apply \ac{fmpe} to one source with three free parameters in gapped data;
and SlotFlow \citep{houba2025slotflow} infers up to ten sinusoids in white noise, with a posterior
that factorises over its slots. None returns a joint posterior over more than two sources under the
\ac{lisa} response (\cref{app:related}).

\section{The LISA Galactic-Binary Problem}
\label{sec:gw}

We address the \acp{gb} of the \ac{lisa} data alone, in stationary noise of known spectrum, a first
step towards a global fit by \acs{sbi} (\cref{sec:followups}).

\begin{figure}[t]
  \centering
  \includegraphics[width=\columnwidth]{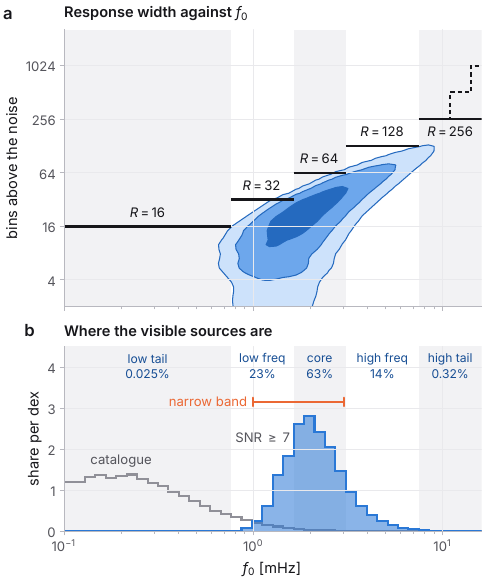}
  \caption{The band is divided where the response width doubles into five subbands (shaded
  and white strips; \cref{tab:narrow-band-bands}). (a)~Response width: sources with power above the noise, by $f_0$ and by the
  bins that hold all of it; contours hold \qtylist{50;90;99}{\percent} of them. Solid segments: each subband's $R$, which keeps every above-noise bin of
  \qty{\NarrowBandSharePercent}{\percent} of the sources at each $f_0$; in the high tail that $R$
  (dashed) climbs to \NarrowBandTopR{}. (b)~Where the sources are: share per dex of $f_0$ of the catalogue's sources (grey)
  and of the visible ones, $\snr\ge\NarrowBandVisibleSnr$ (blue), with each subband's share of the
  visible ones; the bracket marks the $f_0$ range of the narrow band (orange).}
  \label{fig:narrow-band}
\end{figure}

\subsection{Signal, Noise and Prior}
\label{sec:gw-data}\label{sec:gw-waveforms}\label{sec:gw-priors}

A \ac{gb} is a circular, slowly chirping binary of white dwarfs that emits a quasi-monochromatic
\ac{gw} at twice its orbital frequency \citep{nelemans2001gravitational,korol2017prospects}. Eight
parameters describe it,
\begin{equation}
  \theta = (f_0,\fdot,\Amp,\alpha,\delta,\psi,\iota,\phi_0):
  \label{eq:gw-parameters}
\end{equation}
frequency and frequency derivative at the start of the observation, strain amplitude, right
ascension and declination, polarisation angle, inclination and initial phase. Over the observation
time $\Tobs=\qty{\ForwardModelTobsYears}{\year}$ the detector's orbit modulates the signal, which
locates the source on the sky and spreads it over neighbouring frequency bins of width $1/\Tobs$.
JaxGB \citep{jaxgb} simulates the response in the noise-orthogonal \ac{tdi} channels $A$, $E$, $T$
\citep{tinto2021time,prince2002lisa} on the \ac{gpu}, so every training window is simulated on the
fly (\cref{app:forward-model}).

The noise is the stationary Gaussian instrument noise of the \ac{lisa} requirements
\citep{lisasst2018lisa,babak2021lisa}, with one-sided \ac{psd} $S_c(f)$ in channel $c$
(\cref{app:noise}). On the Fourier bins $f_k = k/\Tobs$, with $\tilde h_c(f;\theta_s)$ the response of
source $s$, the data of a band holding $N$ sources are
\begin{equation}
\begin{aligned}
  \tilde d_{k,c} &= \sum_{s=1}^{N}\tilde h_c(f_k;\theta_s) + \tilde n_{k,c},\\
  \tilde n_{k,c} &\sim \CN\!\bigl(0,\ P_c(f_k)\bigr),\qquad
  P_c(f) = \tfrac12\,S_c(f)\,\Tobs,
\end{aligned}
  \label{eq:noise-data}
\end{equation}
independent across bins and channels. A source's \ac{snr} is that of its response alone
\citep{cutler1994gravitational},
\begin{equation}
  \snr_s^2 = 4\sum_{c}\sum_{k}\frac{\bigl|\tilde h_c(f_k;\theta_s)\bigr|^2}{S_c(f_k)\,\Tobs},
  \label{eq:gw-snr}
\end{equation}
and we call a source \emph{visible} when $\snr\ge\RatesVisibleSnr$. The \ac{snr} accumulates over the
many bins a response spans, so per bin even a visible source stands little above the noise
(\cref{fig:forward-model-window}).

The $S_w$ sources of a window $w$ centred on $\fw$, $\Theta_w = (\theta_1,\dots,\theta_{S_w})$, are
independent draws from one per-source prior $\pi_w(\theta) = \pi(\theta\mid\fw)$,
\begin{equation}
  \pi_w(\Theta_w) = \prod_{s=1}^{S_w}\pi_w(\theta_s),
  \label{eq:priors-window-prior}
\end{equation}
fitted to a simulated Galactic catalogue of double white dwarfs
\citep{perego2025towards,breivik2020cosmic} (\cref{tab:priors-source}, \cref{fig:priors},
\cref{app:priors-sources}). It is closer to a real population than the flat or log-flat priors of
other analyses \citep{strub2022bayesian,deng2025modular,delmond2026neural}, and it is broad: most of
its sources lie below the noise.

\begin{table}[t]
\caption{The prior of one source in window $w$: a Gaussian in the logarithms of $(f_0,\fdot,\Amp)$,
with $f_0$ truncated to the window, a catalogue sky and an isotropic orientation.}
\label{tab:priors-source}
\centering
\footnotesize
\begin{tabular}{@{}lp{0.68\columnwidth}@{}}
\toprule
parameter & prior \\
\midrule
$(f_0, \fdot, \Amp)$ & $(\log_{10}f_0, \log_{10}\fdot, \log_{10}\Amp)\sim\Normal(\mu_\pi, \Sigma_\pi)$, \eqref{eq:priors-source} \\
$(\alpha, \delta)$ & bulge + disc, \eqref{eq:priors-sky} \\
$\psi$ & $\Unif(0, 2\pi)$ \\
$\iota\in[0,\pi]$ & $\cos\iota\sim\Unif(-1, 1)$ \\
$\phi_0$ & $\Unif(0, 2\pi)$ \\
\bottomrule
\end{tabular}
\end{table}

\subsection{Likelihood and Windows}
\label{sec:gw-slicing}\label{sec:gw-factorisation}\label{sec:gw-windows}

The noise is stationary, so the likelihood is diagonal in frequency, and a \ac{gb}'s response covers
only $R$ bins around $f_0\Tobs$ (\cref{sec:narrow-band}), so each bin involves only the sources whose
responses reach it. We therefore cut the band into windows of $K = \ForwardModelWindowResponses{}R$
bins. A source belongs to the window whose interior holds its $f_0$, and a guard band of $R/2$ bins
at each edge keeps its whole response inside (\cref{app:forward-model-windows}). With $d_w$ the data
of window $w$, the posterior of all the band's sources $\Theta$ factorises,
\begin{equation}
\begin{gathered}
  p(\Theta\mid d) = \textstyle\prod_{w} p(\Theta_w\mid d_w),\\
  p(\Theta_w\mid d_w) \propto p(d_w\mid\Theta_w)\,\pi_w(\Theta_w),
\end{gathered}
  \label{eq:gw-factorisation}
\end{equation}
with the Whittle likelihood
\begin{equation}
  \log p(d_w\mid\Theta_w) = -\sum_{k\in w}\sum_{c}\frac{\bigl|\tilde d_{k,c}-\sum_{s\in w}\tilde h_c(f_k;\theta_s)\bigr|^2}{P_c(f_k)}
  \label{eq:loglik}
\end{equation}
up to a constant. One network conditioned on $\fw$ then serves the whole band, window by window and in
parallel. It reads the window whitened, $y\in\R^{K\times6}$, the real and imaginary parts of
$\tilde d_{k,c}$ in the three channels divided by $\sqrt{P_c(f_k)/2}$, and its target is
$p(\Theta_w\mid y,\fw) = p(\Theta_w\mid d_w)$. In real data the wings of loud neighbours break the
factorisation, and a global fit updates overlapping segments in turn \citep{littenberg2020global}
(\cref{sec:followups}).

\section{Dividing the Band}
\label{sec:narrow-band}

Three facts about the population shape the method.

\subsection{Response Width and Subbands}
\label{sec:narrow-band-span}

\emph{Responses are narrow, and their width grows with frequency.} The median source needs only
\ForwardModelSpanMedian{} bins for all its above-noise power, against a hard bound of
$R=\ForwardModelSpanBound$ at the top of the band. We size $R$ as the smallest power of two that keeps
every above-noise bin of \qty{\NarrowBandSharePercent}{\percent} of the sources at each $f_0$; it
grows from $R=\NarrowBandLowTailR$ at the bottom of the band to \NarrowBandTopR{} at its top
(\cref{fig:narrow-band}a, \cref{app:forward-model-heterodyne}). A single $R$ must suit the top and is
far too wide below, so we cut the band where $R$ doubles into the five subbands of
\cref{tab:narrow-band-bands}. In the intended scheme one device trains each subband at its own $R$,
and the gradients of one network are summed across devices.

\begin{table*}[t]
\caption{Response width grows with frequency, while visible sources concentrate in the core. For each
subband: the $R$ that keeps every above-noise bin of \qty{\NarrowBandSharePercent}{\percent} of the
sources at each $f_0$; its share of the catalogue's sources and of the visible ones
($\snr\ge\NarrowBandVisibleSnr$); the mean number of sources, all and visible, in a window of
$\ForwardModelWindowResponses{}R$ bins (\cref{eq:gw-rates}); and the number of windows that tile it.}
\label{tab:narrow-band-bands}
\centering
\small
\setlength{\tabcolsep}{4pt}
\begin{tabular}{@{}lcrrrccr@{}}
\toprule
subband & $f_0$ [mHz] & $R$ & catal.\ [\%] & vis.\ [\%] & sources & visible & windows \\
\midrule
low tail & \numrange{\NarrowBandLowTailLowMhz}{\NarrowBandLowTailHighMhz} & \NarrowBandLowTailR & \NarrowBandLowTailCatalogue & \NarrowBandLowTailVisible & \numrange{\NarrowBandLowTailAllLo}{\NarrowBandLowTailAllHi} & \numrange{\NarrowBandLowTailVisibleLo}{\NarrowBandLowTailVisibleHi} & \num{\NarrowBandLowTailWindows} \\
low freq & \numrange{\NarrowBandLowFreqLowMhz}{\NarrowBandLowFreqHighMhz} & \NarrowBandLowFreqR & \NarrowBandLowFreqCatalogue & \NarrowBandLowFreqVisible & \numrange{\NarrowBandLowFreqAllLo}{\NarrowBandLowFreqAllHi} & \numrange{\NarrowBandLowFreqVisibleLo}{\NarrowBandLowFreqVisibleHi} & \num{\NarrowBandLowFreqWindows} \\
core & \numrange{\NarrowBandCoreLowMhz}{\NarrowBandCoreHighMhz} & \NarrowBandCoreR & \NarrowBandCoreCatalogue & \NarrowBandCoreVisible & \numrange{\NarrowBandCoreAllLo}{\NarrowBandCoreAllHi} & \numrange{\NarrowBandCoreVisibleLo}{\NarrowBandCoreVisibleHi} & \num{\NarrowBandCoreWindows} \\
high freq & \numrange{\NarrowBandHighFreqLowMhz}{\NarrowBandHighFreqHighMhz} & \NarrowBandHighFreqR & \NarrowBandHighFreqCatalogue & \NarrowBandHighFreqVisible & \numrange{\NarrowBandHighFreqAllLo}{\NarrowBandHighFreqAllHi} & \numrange{\NarrowBandHighFreqVisibleLo}{\NarrowBandHighFreqVisibleHi} & \num{\NarrowBandHighFreqWindows} \\
high tail & \numrange{\NarrowBandHighTailLowMhz}{\NarrowBandHighTailHighMhz} & \numrange{\NarrowBandHighTailR}{\NarrowBandTopR} & \NarrowBandHighTailCatalogue & \NarrowBandHighTailVisible & \numrange{\NarrowBandHighTailAllLo}{\NarrowBandHighTailAllHi} & \numrange{\NarrowBandHighTailVisibleLo}{\NarrowBandHighTailVisibleHi} & \num{\NarrowBandHighTailWindows} \\
\bottomrule
\end{tabular}
\end{table*}

\subsection{Visible Sources and Sources per Window}
\label{sec:gw-rates}\label{sec:narrow-band-prior}

\emph{Visible sources are rare and concentrated.} The catalogue puts
\qty{\NarrowBandLowTailCatalogue}{\percent} of its sources in the low tail, where they are faint;
the visible ones are its loud tail, and \qty{\NarrowBandCoreVisible}{\percent} of them fall in the
core (\cref{fig:narrow-band}b, \cref{fig:priors}a).

\emph{A window holds a known, small number of sources that matter.} Each of the
$N = \num{\RatesCatalogue}$ sources of the simulated Galaxy falls in a window with the probability
$p$ that the catalogue's marginal $\pi(f_0)$ puts there, so the window holds $n$ sources,
$n_{\rm vis}$ of them visible, with
\begin{equation}
\begin{gathered}
  n \sim \operatorname{Bin}\bigl(N, p\bigr),\qquad n_{\rm vis} \sim \operatorname{Bin}\bigl(N, p_{\rm vis}\bigr),\\
  p = \textstyle\int_{w} \pi(f_0)\,\dd f_0,\quad p_{\rm vis} = \textstyle\int_{w} \pi(f_0)\,P(\snr\ge\RatesVisibleSnr\mid f_0)\,\dd f_0.
\end{gathered}
  \label{eq:gw-rates}
\end{equation}
On average a window holds up to \num{\NarrowBandLowTailAllHi} sources at the bottom of the band,
almost none of them visible, and up to \num{\NarrowBandCoreVisibleHi} visible ones in its core
(\cref{tab:narrow-band-bands}, \cref{app:priors-counts}).

\subsection{Slots and the Trained Configurations}
\label{sec:narrow-band-bands}

Most sources of a window are below the noise, and for those the posterior is close to the prior. We
therefore fix the number of sources: every window holds exactly $S_w = S$ sources drawn from the
prior truncated to its interior, with $S$ set per run from 2 to 16, where a real analysis infers the
number with a trans-dimensional sampler \citep{littenberg2020global} and models the unresolved rest as
confusion noise \citep{karnesis2021characterization}. A slot the data do not resolve returns the
prior, and the flow works in a space of fixed dimension $8S$. A network of $S$ slots fits
\begin{equation}
  N_S = \int \frac{\dd f}{W(f)}\ \E\bigl[\min(n_{\rm vis},S)\mid f\bigr],\quad W = \ForwardModelWindowResponses{}R/\Tobs,
  \label{eq:narrow-band-fitted}
\end{equation}
visible sources over the divided band: \num{\NarrowBandFittedFour} at $S=4$,
\num{\NarrowBandFittedEight} at $S=8$ and \num{\NarrowBandFittedSixteen} at $S=16$, of
\num{\NarrowBandVisibleTotal} in the band, to be compared with the global fits of
\cref{tab:gw-globalfit}.

We train two configurations, each on one device. The \emph{narrow band},
\qtyrange{\PriorsNarrowLowMhz}{\PriorsNarrowHighMhz}{\milli\hertz} at $R=\ForwardModelNarrowR{}$, is
the core subproblem, holding \qty{\NarrowBandNarrowVisible}{\percent} of the visible sources where they
overlap most. The \emph{full band} covers the whole distribution at $R=\ForwardModelFullR{}$, the width
of the high-frequency subband: too narrow for the high tail, where
\qty{\NarrowBandHighTailVisible}{\percent} of the visible sources lose their outer bins
(\cref{sec:results-full}). Training all five subbands together is left to future work.

\section{Flow in Relative Coordinates}
\label{sec:flow}\label{sec:method}

We train one conditional generative model of the window posterior $p(\Theta_w\mid y,\fw)$ of
\cref{eq:gw-factorisation} over all windows of a band. A loud source's posterior is decades
narrower than its prior and a quiet source's is the prior, so no absolute scale suits both, and
the flow works in relative quantities throughout. Its coordinates are relative to the window and
its noise, and carry none of the posterior's symmetries (\cref{sec:method-chart}). Its variable
is relative to a Gaussian frame that the network reads off the data (\cref{sec:method-frame}).
Its loss measures each error relative to the spread the data leave (\cref{sec:method-fm}).
\Cref{sec:architecture} describes the network and its training.

\begin{figure*}[t]
  \centering
  \includegraphics[width=\textwidth]{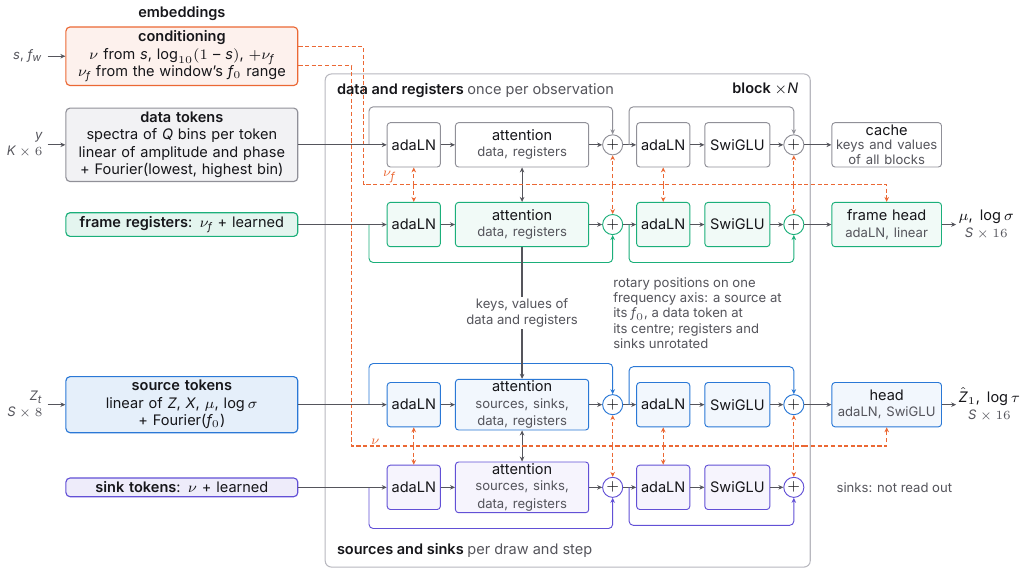}
  \caption{The network: four streams, each with its own normalisations, attention projections and
  feed-forward. The source stream (blue) has one token per source slot, in order of $f_0$, and the
  sink stream (violet) \NetworkSinksWord{} tokens; they attend jointly. The data stream (grey) has
  one token per \NetworkPatch{} frequency bins of the window, and the register stream (green) one
  frame register per slot, which the frame head reads; they attend jointly. Conditioning (orange, dashed) modulates every normalisation and residual gate:
  $\nu$ from the flow time and the window in the source stream and head, $\nu_f$ from the window
  alone in the data stream, the registers and the frame head. The data side never reads the
  sources, so its keys and values and the frame are computed once per observation and serve every
  draw and sampler step. Rotary positions place each source at its $f_0$ and each data token at its
  centre frequency.}
  \label{fig:network-architecture}
\end{figure*}

\subsection{Coordinates Relative to the Window}
\label{sec:method-chart}\label{sec:gw-chart}

The likelihood is invariant under $(\psi,\phi_0)\to(\psi+\pi/2,\phi_0+\pi)$, which gives each source
four copies in $(\psi,\phi_0)\in[0,2\pi)^2$ (\cref{eq:psi-phi-symmetry}), and under the $S!$
permutations of the sources. A posterior over the physical parameters therefore has $4^S\,S!$
equivalent modes, and a flow on them must put mass on every one. The velocity that flow matching
regresses is then an average over the copies for much of the path, which washes out the phase.

The flow draws $X\in\R^{8S}$, eight coordinates per source (\cref{eq:chart}). The frequency is
scaled to the window, $\log\fdot$ is standardised given $f_0$, and the sky is in galactic
coordinates scaled by the prior's widths. In place of amplitude, inclination, polarisation and
phase the source carries its two complex helicity amplitudes, in units of the noise at the window
centre and with the radius warped by $\operatorname{asinh}$,
\begin{equation}
\begin{aligned}
  \zeta_\pm &= \frac{\Amp}{u(\fw)}\,\frac{(1\mp\cos\iota)^2}{4}\,e^{\ii(\phi_0^{\rm mid}\pm2\psi)},\\
  z_\pm &= \operatorname{asinh}\lvert\zeta_\pm\rvert\;\frac{\zeta_\pm}{\lvert\zeta_\pm\rvert},
\end{aligned}
  \label{eq:chart-helicity}
\end{equation}
where $u(\fw)$ is the noise amplitude at the window centre (\cref{app:chart-unit}) and
$\phi_0^{\rm mid}$ the phase referred to mid-observation (\cref{eq:chart-phase}). The sources occupy
$S$ slots in order of $f_0$, the ordering cure for label switching
\citep{stephens2000dealing,jasra2005markov,buscicchio2019label} that the global fit also uses
\citep{littenberg2020global}.

The signal is linear in $\zeta_\pm$ (\cref{eq:forward-model-helicity}), so at fixed frequency, drift
and sky the likelihood is Gaussian in these four real numbers, and the four extrinsic parameters
reach the data only through them. The four phase copies of a source share $\phi_0\pm2\psi$ and so
map to one point, and the frequency ordering leaves each mode one copy instead of $S!$
(\cref{app:chart-slots}). Face-on sources, where much of the prior's mass lies, sit at a regular point of the
plane, where one of $z_\pm$ vanishes, and not at a coordinate pole. Every coordinate is unbounded and
without a periodic direction, so the flow's paths are straight lines (\cref{app:chart-flat}).
\Cref{app:chart-helicity} derives these properties.

\subsection{A Frame Relative to the Data}
\label{sec:method-frame}

Standard \ac{fmpe} fails on posteriors a small fraction of a bin wide in $f_0$, as a Gaussian model
of one coordinate shows (\cref{prop:fm-sensitivity}, \cref{fig:fm-sensitivity}). The velocity loss
pays an excess of order $s_0 w_m$ for a model $w_m$ wide against a prior of width $s_0$, whatever
the true width $w\ll w_m$, so it barely rewards the last decades of narrowing, and a sampler with
uniform steps on the prior's scale returns the narrowest posteriors collapsed.

The network first reads from the data alone a Gaussian \emph{frame} for each slot, a mean $\mu$
and a standard deviation $\sigma$ per coordinate (\cref{eq:fm-frame}). It is trained on the
Gaussian likelihood of the window's true sources $X_1$ in the flow coordinates, so that it learns
the mean and width of each coordinate's posterior marginal; a weight keeps loud sources, whose
frames are the narrowest, from dominating the batch (\cref{eq:fm-frame-loss}). The flow runs on
$Z=(X-\mu)/\sigma$, from a standard normal $Z_0$ to $Z_1=(X_1-\mu)/\sigma$ along the straight path
\begin{equation}
  Z_s = Z_0 + s\,(Z_1 - Z_0),\qquad s\in[0,1].
  \label{eq:fm-path}
\end{equation}
In the frame the posterior of $Z_1$ has a mean near zero and a width near one for every source, so
the flow spends its capacity on what a Gaussian per coordinate misses: correlations, tails and
multiple modes.

\subsection{A Loss Relative to the Posterior Width}
\label{sec:method-fm}

Given $Z_s$ and $s$ the network predicts, per coordinate, the mean $\hat Z_1$ and the log standard
deviation $\log\tau$ of a Gaussian over the endpoint $Z_1$. Its velocity is the one that reaches
the mean in the remaining path,
\begin{equation}
  \vnet = \frac{\hat Z_1 - Z_s}{1-s},
  \label{eq:fm-velocity}
\end{equation}
and the network minimises the sum of the frame's likelihood $\mathcal L_{\rm frame}$ and the
endpoint's negative log-likelihood, up to a constant and a factor of two,
\begin{equation}
  \mathcal L = \mathcal L_{\rm frame} + \E\,\frac{1}{8S}\sum_{k,i}
  \Bigl[\Bigl(\frac{Z_{1,ki}-\hat Z_{1,ki}}{\tau_{ki}}\Bigr)^2 + 2\log\tau_{ki}\Bigr],
  \label{eq:loss}
\end{equation}
with $k$ the slot and $i$ the coordinate, and the expectation over windows, draws of $Z_0$ and times uniform along the path. The two terms
share the network but not their gradients: the frame sees only the first, and the flow reads the
frame through a stop-gradient, so it cannot move the frame to ease its own task
(\cref{app:fm-frame}). At the optimal $\tau$ the second term is, for every point, coordinate and
source, a velocity loss divided by its own irreducible part,
\begin{equation}
  \frac{(\vnet - u)^2}{\Var(u\mid Z_s,y,\fw)},\qquad u = Z_1 - Z_0 ,
  \label{eq:fm-relative}
\end{equation}
because the factors of $1-s$ cancel (\cref{app:fm-loss}).

A source the data pin down therefore weighs as much as one whose posterior is its prior, with no
weights over time or coordinates. In the
Gaussian model the likelihood pays $2$ for every e-fold by which the model is too wide, whatever
the true width: $\Delta L_{\rm nll}\simeq\ln(w_m^2/w^2)+w^2/w_m^2-1$ depends on the widths only
through $w_m/w$. The frame brings the width to the order of one in $Z$, where the uniform sampler
is exact in the model. Draws use \FlowMatchingOdeSteps{} midpoint steps of $\vnet$ and are mapped
back by $X=\mu+\sigma Z$ (\cref{app:fm-sampling}). \Cref{app:fm} gives the derivations.

\section{Architecture and Training}
\label{sec:architecture}

One network gives both outputs of \cref{sec:flow}: the frame of a window from its data alone, and
the endpoint of every source given a point on the path. The second is a diffusion model's task,
denoising a few tokens conditioned on a long sequence, so we take the network from that literature.

\subsection{Why a Multimodal Diffusion Transformer}
\label{sec:method-network}\label{sec:gw-tokens}

Diffusion and flow models \citep{sohl2015deep,ho2020denoising,song2021score,lipman2023flow} have
converged on transformers. The diffusion transformer \citep{peebles2023scalable} carries the time
and the class by modulating every layer normalisation, with gated residual branches that start as
the identity. The \ac{mmdit} \citep{esser2024scaling}, built for text-to-image rectified flows
\citep{liu2023flow}, extends it to a condition that is itself a sequence: text and image tokens
keep their own weights, as two streams, and meet in one joint attention in every block. It scales
to eight billion parameters with steadily improving samples.

Our problem has the same shape: the condition is the data, a long sequence of spectral tokens, and
the sample is the set of sources, one token per source. Joint attention lets every source read the
data and the other sources in one operation, as a joint posterior of overlapping sources requires.
Every map is shared across source tokens, so the same network takes any number of sources,
including the varying number of the trans-dimensional global fit. The source side is also
permutation-equivariant, as in set transformers \citep{zaheer2017deep,lee2019set}: the slot order
of \cref{sec:method-chart} reaches the network only through each slot's frame.

\subsection{Four Streams and One-Way Attention}
\label{sec:method-streams}

\paragraph{Streams.}
The network has four streams, each with its own normalisations, attention projections and
feed-forward maps (\cref{fig:network-architecture}). The \emph{data} stream holds the tokens of the
window. The \emph{register} stream \citep{darcet2024vision} has one token per source slot, from
which a head reads the slot's frame $(\mu,\log\sigma)$ (\cref{app:network-frame}). The
\emph{source} stream has one token per slot, from which a head reads the endpoint $\hat Z_1$ and
$\log\tau$ (\cref{app:network-head}). The \emph{sink} stream has \NetworkSinksWord{} learned
tokens. Conditioning vectors modulate every normalisation and gate every residual branch: $\nu_f$,
from the window alone, in the data and register streams, and $\nu$, from the flow time and the
window, in the source and sink streams. The frame depends on the data alone, so it is read from the
data side; the endpoint depends on the flow time and the sources' current point, so it is read from
the source side.

\paragraph{One-way attention and caching.}
A window is a long sequence with few sources, and a posterior needs many draws of many sampler
steps each. In the \ac{mmdit} the text stream reads the image tokens and the time, so the whole
network runs at every step. Ours reads in one direction. With queries $q$, keys
$k$ and values $v$ of the data ($y$), register ($r$), source ($x$) and sink ($s$) streams, each
block computes
\begin{equation}
\begin{aligned}
  [o^y; o^r] &= \mathrm{Attn}\bigl([q^y; q^r]; [k^y; k^r], [v^y; v^r]\bigr),\\
  [o^x; o^s] &= \mathrm{Attn}\bigl([q^x; q^s]; [k^x; k^s; k^y; k^r], [v^x; v^s; v^y; v^r]\bigr),
\end{aligned}
  \label{eq:network-oneway}
\end{equation}
so the data and the registers attend only to each other, and the sources and the sinks attend to
everything (\cref{eq:network-block}). The data side's keys and values, and the frame, then depend
on $(y,\fw)$ alone: they are computed once per window and cached, like the prompt of a language
model, and every draw and sampler step runs only the source side. The data side, which holds most
of the tokens, is paid once per posterior, not once per network evaluation
(\cref{app:network-blocks}, \cref{tab:network-sizes}).

\paragraph{Sinks.}
With one-way attention, nothing on the data side reads the current sources. The sinks restore a
limited form of this: they read the data, the registers and the sources, and the sources read them
back, so a sink can gather the part of the data that matters for the current draw and pass it to
every source in the next block. They are optional and few, so they cost little and leave the cache
intact, and they give the sources' attention somewhere to go other than the data
\citep{xiao2024efficient,darcet2024vision}.

\subsection{Embeddings}
\label{sec:method-embeddings}

What each token holds decides what the attention can resolve: a posterior a small fraction of a bin
wide requires frequency at sub-bin resolution in every embedding (\cref{tab:network-embeddings},
\cref{app:network}).

\paragraph{Data tokens.}
The window is cut into patches of \NetworkPatch{} consecutive frequency bins, one token each. An
inverse Fourier transform over the patch turns its bins into as many time samples per channel, \rev{a}
\ac{stft} with the tokens as its bands (\cref{eq:network-stft}). It is critically sampled and
untapered, so it is invertible and keeps the noise white, and it puts each source parameter in a
simple form: the sub-bin frequency is a phase slope over the samples, the initial phase its
intercept, and the sky position a slow annual modulation. Loud sources are far above the unit
noise, so each complex sample enters as a compressed amplitude,
$\operatorname{asinh}\lvert\check y\rvert$, and a gated phase,
$\check y/\sqrt{1+\lvert\check y\rvert^2}$ (\cref{app:network-input}).

\paragraph{Source tokens.}
A source token is a linear map of the slot's point in the frame, $Z_s$, the same point in the flow
coordinates, $X=\mu+\sigma Z_s$, the higher harmonics of its two helicity phases, and its frame
$(\mu,\log\sigma)$ through a stop-gradient (\cref{app:network-sources}). It carries no slot label:
a source is told apart by its frame and its frequency.

\paragraph{Global frequency.}
Whitening and windowing remove the absolute frequency from the data, yet the response width, the
prior's drift and the amplitude a source needs to be seen all depend on it. It enters as sinusoids
with octave-spaced periods, from \NetworkFreqFinest{} of a bin to past twice the top of the band,
in three places: the edges of each data token, the $f_0$ of each source token, and the window edges
in $\nu_f$. The finest periods resolve a source's position inside a bin; the coarsest locate the
window in the band.

\paragraph{Rotary attention.}
Inside the attention, rotary embeddings \citep{su2024roformer,miyato2024gta} rotate the queries,
keys and values of the data and source tokens by their frequency, a data token's centre and a
source's $f_0$, and rotate each output back by the query's angle. A score then depends only on the
frequency separation of two tokens, so a source finds the data near its $f_0$, and its neighbours
by their distance in bins, with the same weights anywhere in the band. The rotation uses absolute
frequencies reduced modulo each period in double precision, so periods shorter than a token still
resolve frequency inside a patch (\cref{app:network-positions}). The registers and the sinks carry
no position.

\paragraph{Flow time.}
The time enters $\nu$ as sinusoids of $s$ and of $\log_{10}(1-s)$, clipped at \NetworkTimeDecades{}
decades, which resolves the end of the path, where the fine structure of a narrow posterior forms
(\cref{app:network-conditioning}).

\begin{figure*}[t]
\centering
\includegraphics[width=\textwidth]{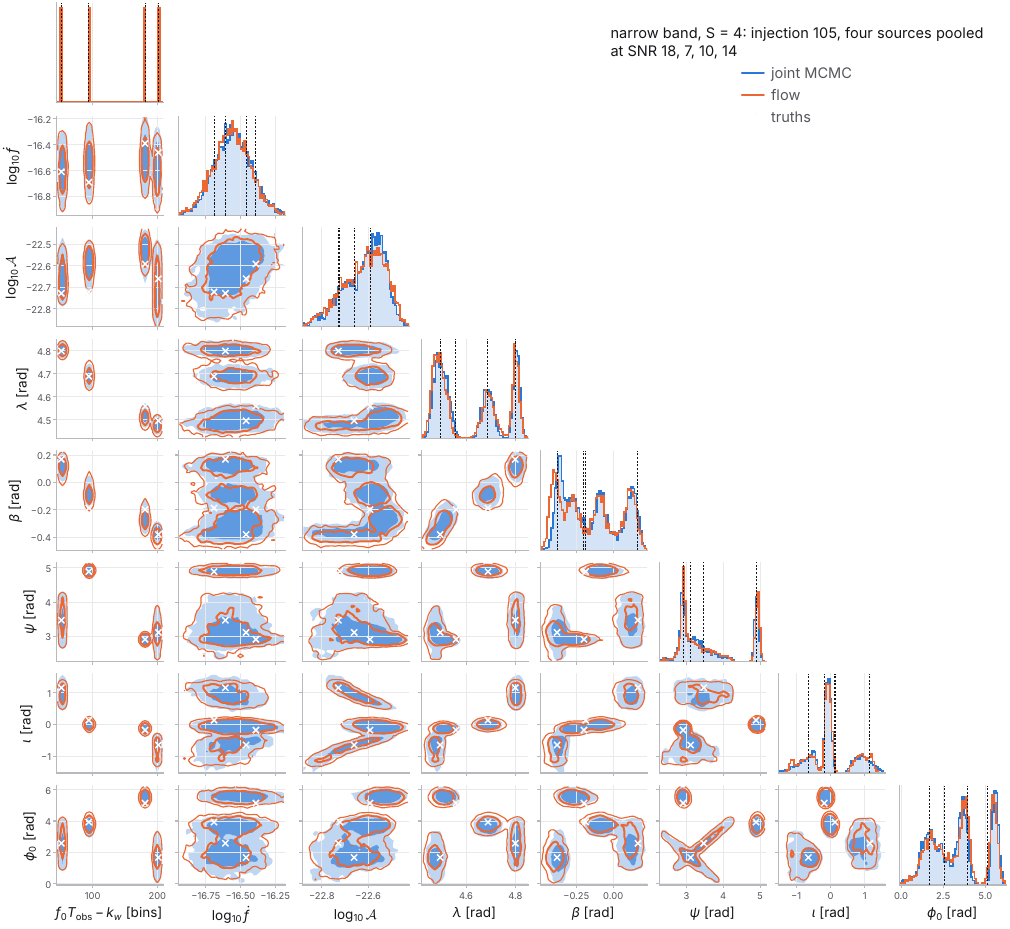}
\caption{Flow and joint \ac{mcmc} (filled) on one eval window of the narrow band, $S=4$:
\qty{68}{\percent} and \qty{95}{\percent} contours of the eight parameters with the four sources' draws
pooled, so each source is one cluster; crosses and dotted lines mark the truths. $f_0$ is in bins from
the window start, the sky in ecliptic coordinates, and $\psi$, $\phi_0$ are folded onto the copy nearest
the truth. The legend names the window and the sources' \acp{snr}.}
\label{fig:corner-narrow-4}
\end{figure*}

\begin{table*}[t]
\centering
\caption{Summary of the six models. Model: the 64-epoch main run, its fork cooled from epoch 32
(e32) or the continuation from epoch 44 (t44), with the epoch evaluated in brackets
(\cref{app:results-runs}). \rev{Every model is scored on the same \EvaluationInjections{} evaluation windows with
joint \acs{mcmc} and \ResultsValidationInjections{} prior windows of the \acs{sbc}.} JS: Jensen--Shannon divergence in bits between the flow's and the chains'
one-dimensional marginals, median over the sources at $\snr=7$--15 and $\snr\ge15$ of the mean over
the eight coordinates (\cref{tab:results-js}); \ResultsSummaryFloor{} from sampling noise alone.
\acs{sbc}: share of true coordinates inside the central \qty{68}{\percent} of the draws and the
smallest $\gamma$-test $p$-value over the eight coordinates, for the flow on the prior windows and
for the joint \acs{mcmc} on the evaluation windows (\cref{app:results-validation}).}
\label{tab:results-summary}
\small
\begin{tabular}{@{}lrlrrrrrr@{}}
\toprule
 & & & \multicolumn{2}{c}{JS} & \multicolumn{2}{c}{\acs{sbc} flow} & \multicolumn{2}{c}{\acs{sbc} \acs{mcmc}} \\
\cmidrule(lr){4-5}\cmidrule(lr){6-7}\cmidrule(l){8-9}
problem & $S$ & model & 7--15 & $\ge15$ & 68\% & $p_{\min}$ & 68\% & $p_{\min}$ \\
\midrule
\ResultsSummaryRows
\bottomrule
\end{tabular}
\end{table*}

\subsection{Training}
\label{sec:method-training}\label{sec:gw-training}

Every window is simulated afresh on the device, inside the compiled training step, so no window is
seen twice and the training set is the prior itself (\cref{app:priors-training}). Centres $\fw$
drawn from the catalogue's $f_0$ marginal would mostly hold no visible source, so we weight the
marginal by the chance that a source there is visible, and draw a share of the centres
log-uniformly over the band so that training also sees windows with nothing to resolve
(\cref{eq:priors-window}). The weights choose which windows training sees, not their sources, so
the posterior given a window is unchanged (\cref{app:priors-window}). The window receives $S$
sources from the prior, their responses and a noise draw, and is whitened. Unlike
\citet{delmond2026neural}, who precompute a bank of responses and keep sources within an \ac{snr}
range, we apply no cut: most of a window's sources lie below the noise, and for them the flow must
return the prior.

Each window in a batch is paired with several draws of time and base point, which share its cached
data side (\cref{app:training-batch}). The optimiser is Muon \citep{jordan2024muon,liu2025muon} for the hidden weight matrices and
AdamW for the rest, with a warmup, a constant learning rate and a cooldown to zero
\citep{hagele2024scaling}, and the model is evaluated at the checkpoint with the lowest loss on
held-out windows (\cref{app:training,app:evaluation-weights}).

\section{Experimental Results}
\label{sec:experiments}\label{sec:results}

We train the size-B network \rev{(\NetworkBBlocks{} blocks of width \NetworkBWidth{} with \NetworkBHeads{} heads, about 133\,M
parameters; \cref{tab:network-sizes})} on the two configurations of
\cref{sec:narrow-band-bands}, the narrow band with $S={}$\ResultsNarrowCounts{} sources per window
and the full band with $S={}$\ResultsFullCounts, one training run each (\cref{app:results-runs}).

\begin{figure*}[t]
  \centering
  \includegraphics[width=\textwidth]{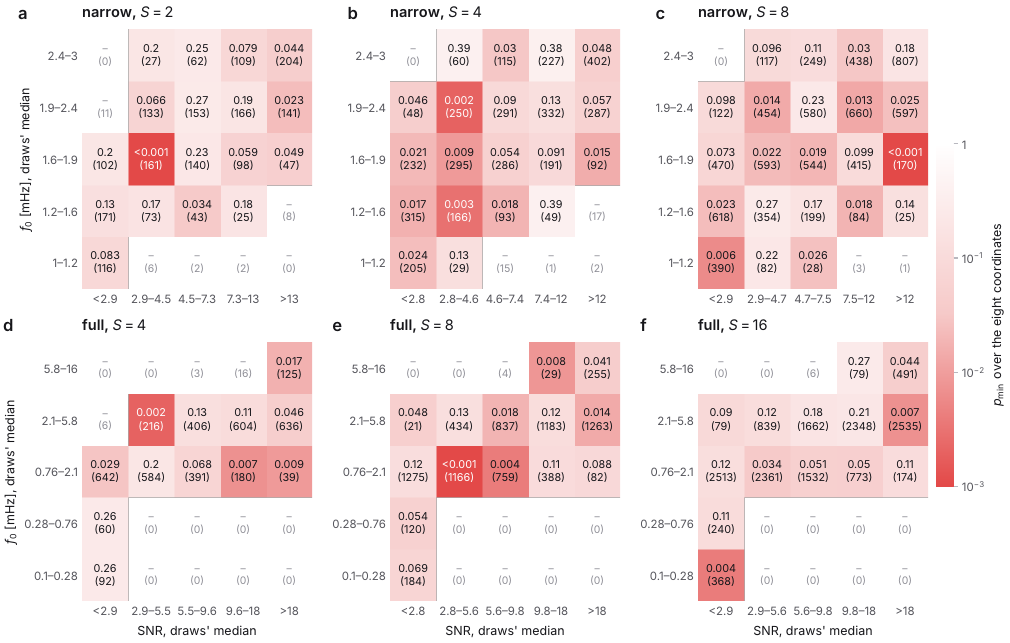}
  \caption{\acs{sbc} by frequency and \acs{snr} for the six models of \cref{tab:results-summary}, on the same
  \num{1000} prior windows: in each cell, the smallest $\gamma$-test $p$-value over the eight flow coordinates
  and, in brackets, the number of sources. Rows are five $f_0$ ranges, log-uniform over the band; columns are
  five \acs{snr} ranges holding about the same number of sources. Both are the draws' medians, so the binning
  depends on the data alone. Cells with fewer than 20 sources are left blank, and the test resolves $p$ down to
  \num{0.001}. With eight coordinates per cell, an exact posterior gives $p_{\min} < 0.01$ in up to about one cell in
  thirteen.}
  \label{fig:sbc-heatmap}
\end{figure*}

\subsection{Validating Without the Truth}
\label{sec:validation}

A joint posterior over $8S$ parameters has no closed form, so we score each model in two passes
(\cref{app:evaluation}). The \emph{validation pass} ranks the truth among
\ResultsValidationDraws{} draws for each of \ResultsValidationInjections{} windows simulated from the
prior, in the flow coordinates and in data space (\ac{sbc}, \ac{tarp}, and ranks of the overlap
and the log-likelihood). The \emph{reference pass} compares the flow with joint \ac{mcmc} on
\EvaluationInjections{} evaluation windows (\cref{app:mcmc}): the ratio of the flow's to the chains'
width, $1.4826\,\MAD$ of the draws, and the Jensen--Shannon (JS) divergence of the marginals. A
source is \emph{found} when the flow's $f_0$ width is below one bin, and an $f_0$ draw is
\emph{off-peak} when it lies more than five chain widths and half a bin from the chains' median.

Neither pass alone is sufficient. On mock posteriors with known defects the coordinate ranks accept
a posterior that returns the prior or shuffles parameters between draws, and coverage after
relabelling reads a width floor on loud sources as conservative (\cref{tab:evaluation-power,tab:evaluation-mocks}).
Data-space ranks catch the first two; only a width ratio against a reference shows the third. The
Gaussian mock is not a clean null, so a flow's data-space ranks are read against its offset.

The reference holds only where the chains have converged. For the narrow band with $S=8$ and the
full band with $S=16$ the chains started from Fisher draws never leave that start for faint
sources: below $\snr=\ResultsInvisibleSnr$ their $f_0$ widths are \ResultsNarrowEightFaintMcmcPrior{}
and \ResultsFullSixteenFaintMcmcPrior{} of the prior's, against \ResultsNarrowEightFaintFlowPrior{}
and \ResultsFullSixteenFaintFlowPrior{} for the flow. We therefore compare from $\snr=7$ up, where
prior-started chains confirm the reference for $S=8$; for the full band with $S=16$ they come out
wider, so its ratios up to $\snr\approx20$ are upper bounds on the flow's excess width
(\cref{app:results-mcmc}). The chains serve as a rough reference for the flow's widths, not as a posterior in their own
right, so we do not refine them further.

\subsection{Narrow Band}
\label{sec:results-main}\label{sec:results-narrow}

With two sources per window the flow's posteriors match the joint chains, and the match degrades
as sources are added (\cref{tab:results-summary,fig:results}).

At $\snr=7$--40 and $S=2$ every width ratio lies in
\ResultsNarrowTwoWidthLo--\ResultsNarrowTwoWidthHi, and the off-peak share of loud $f_0$ draws,
\qty{\ResultsNarrowTwoOff}{\percent}, is close to the chains' own \qty{\ResultsNarrowTwoMcmcOff}{\percent}.
With $S=8$ the $f_0$ width is \ResultsNarrowEightFzeroLow{} times the chains' at $\snr=7$--15,
\qty{\ResultsNarrowEightFoundLow}{\percent} of those sources are found and
\qty{\ResultsNarrowEightOff}{\percent} of the loud draws are off-peak. Below $\snr=7$ the flow
returns the prior, as do the chains. The sky is the weakest parameter, with an excess that grows
with the \ac{snr} at every $S$ (\cref{fig:results}b). The JS divergence ranks the models the same
way (\cref{app:results-js}).

\Cref{fig:corner-narrow-4} shows all four sources of one window with $S=4$, the one among those whose four
sources are visible where the flow overlaps the chains best: the flow places every source where the
chains have it, with their widths and correlations. The closest loud pair of the same model and the
windows of the other models, chosen by rule, are in \cref{app:results-corners}.

\subsection{Full Band}
\label{sec:results-full}

On the full band the flow is close to the chains for faint sources but does not yet resolve the
loud ones. At $\snr=15$--40 its $f_0$ widths are \ResultsFullFourFzeroMid{} to
\ResultsFullSixteenFzeroMid{} times the chains' for $S=4$ to 16; above $\snr=40$ they are
\ResultsFullFourFzeroLoud{} to \ResultsFullSixteenFzeroLoud{} times, and only
\qty{\ResultsFullSixteenFoundLoud}{\percent} of the loud sources are found for $S=16$
(\qty{\ResultsFullFourFoundLoud}{\percent} for $S=4$).

The misses are at high frequency. Above \qty{\ResultsFullSplitMhz}{\milli\hertz} lie
\qty{\ResultsFullSixteenLoudHighShare}{\percent} of the loud sources for $S=16$ but
\qty{\ResultsFullSixteenUnfoundHighShare}{\percent} of those not found, and the excess is mostly the
flow's refinement of its frame: read from the data alone, the frame is wide by design on loud
sources, and the flow narrows it \ResultsFullSixteenContractLow-fold below that frequency but only
\ResultsFullSixteenContractHigh-fold above (\cref{app:results-mcmc}). These sources have the widest
responses, some beyond the $R$ of this configuration (\cref{sec:narrow-band-bands}), and the prior
produces few of them, so training sees few examples\rev{; \cref{sec:followups} proposes how to give it more}.

\subsection{Calibration}
\label{sec:results-calibration}

Coverage is nominal in all six problems, yet the data disagree with every flow slightly. After
relabelling, every model covers \ResultsCoverSixtyEightLo--\ResultsCoverSixtyEightHi{} of the true
coordinates at the \qty{68}{\percent} level, even where the loud full-band $f_0$ widths are several
times the chains': as the mocks predicted, coverage does not see excess width.

The data-space ranks do (\cref{tab:results-validation}, \cref{fig:results-rank-ecdf}). With
\ResultsValidationInjections{} windows the $\gamma$ test \citep{sailynoja2022graphical} rejects even
the Gaussian mock, so we read the offset of the mean rank from one half, positive when the truth
fits the data better than the draws do. Every flow's log-likelihood offset exceeds its mock's by
\ResultsLikeExcessLo{} to \ResultsLikeExcessHi, and its overlap offset is \ResultsFlowOverlapOffsetLo{}
to \ResultsFlowOverlapOffsetHi{} against the mock's \ResultsMockOverlapOffsetLo{} to
\ResultsMockOverlapOffsetHi, the narrow band with $S=2$ included although its marginals have the
chains' widths. No single group of coordinates carries the offset: a draw with a displaced $f_0$
carries a phase that partly makes up for it, so each draw is a coherent fit displaced from the truth
along a direction that mixes the groups (\cref{app:results-validation}). Whether another chart or
loss removes the offset is open.

\rev{\paragraph{By frequency and \acs{snr}.}}\label{sec:results-sbc-heatmap}
Coverage and $p_{\min}$ in \cref{tab:results-summary} pool every source of a problem, but the posterior's
difficulty changes with frequency and loudness. \Cref{fig:sbc-heatmap} splits the \acs{sbc} of each model by
both, each source binned by its draws' median $f_0$ and \acs{snr}, which depend on the data alone: binning by
the true values would condition on the truth, and even exact posteriors would fail. \rev{Of the 100 cells
with enough sources, 14 have $p_{\min}<0.01$, against about 8 expected for exact posteriors, and 3 fall
below the test's resolution of 0.001. They follow no pattern in frequency, \acs{snr} or coordinate, so
binning reveals no region where the flow is miscalibrated beyond the pooled offset above.}
\subsection{Cost of a Posterior}
\label{sec:results-cost}

A posterior costs a fraction of a second. \FlowMatchingCostDraws{} draws of a narrow-band window take
\qty{\FlowMatchingCostNarrow}{\second} on one \ac{gpu} and of a full-band window
\qty{\FlowMatchingCostFull}{\second} (\cref{fig:fm-cost}), with the widths settled by 32 to 64
sampler steps (\cref{app:evaluation-sampler}). A joint \ac{mcmc} chain of one evaluation window takes
\qtyrange{\McmcCostJointLo}{\McmcCostJointHi}{\hour} on \McmcJointCores{} \ac{cpu} cores\rev{ in the narrow band, and about
\qtyrange{\McmcCostFullSixteenLo}{\McmcCostFullSixteenHi}{\hour} in the full band with $S=16$}
(\cref{app:mcmc-cost}).

\section{Conclusions and Discussion}
\label{sec:conclusions}

Conditional flow matching returns joint posteriors of overlapping \acp{gb} under the full
instrument response, with no likelihood evaluation and in a fraction of a second per window. The
chart, the frequency order and the frame turn a posterior that is multimodal by symmetry and
decades narrower than its prior into one that a flow on straight paths can learn, and the one-way
network makes each draw cost only the source side. On the core subproblem with two sources per
window the posteriors have the widths of a joint \ac{mcmc}; they widen as sources are added, and
over the whole band the loudest sources are not yet resolved.

\paragraph{Limitations.}
\label{sec:limitations}
The flow falls short of the chains where the posterior is narrowest. In the narrow band its $f_0$
widths reach \ResultsNarrowEightFzeroLow{} times the chains' with eight sources per window, and the
sky two to three times above $\snr=40$. In the full band the $f_0$ of sources above $\snr=40$ is
\ResultsFullFourFzeroLoud{} to \ResultsFullSixteenFzeroLoud{} times too wide, and with $S=16$ only
\qty{\ResultsFullSixteenFoundLoud}{\percent} of them are found. The misses lie above
\qty{\ResultsFullSplitMhz}{\milli\hertz}, where training sees few loud sources and the flow refines
its frame least (\cref{sec:results-full}). In every model the drawn signals fit the data slightly
worse than the true one, which the marginal widths do not show (\cref{sec:results-calibration}).

The evidence has limits of its own. Our tests are necessary conditions; only the reference chains
give widths, they exist for \EvaluationInjections{} windows per model, and for the faint sources of
the largest problems they are themselves unreliable (\cref{sec:validation}). Every result comes from
one training run per configuration, so the run-to-run spread is not measured; a second seed, the
narrow band with $S=2$ first, would show which differences are real (\cref{app:results-forks}).
Out of distribution, the flow tolerates sources past its prior but turns confident and wrong when
the noise is louder than its assumed \ac{psd} (\cref{app:evaluation-sampler}). Finally, the
simulator is simpler than real \ac{lisa} data, and it generates the training data, the test windows
and the reference chains' likelihood alike. Its simplifications therefore limit realism, not the
consistency of the comparison.

\paragraph{Proposed follow-ups.}
\label{sec:followups}
The first step is realistic orbits, with unequal and time-varying arms and \ac{tdi} generation 2
(\cref{app:forward-model-tdi}), and data gaps. We expect both to enter through the training
simulations, with no change to the cost of a draw or to how its posteriors are checked. A sampler
needs each effect written into its likelihood: gaps break the stationarity that makes
\cref{eq:loglik} diagonal, and whatever the likelihood leaves out biases the chains
\citep{baghi2019gravitational,dey2021effect}.

The second step is the window, which now has a fixed length and exactly $S$ binaries. A global fit
needs a number of sources that follows the density of the Galaxy, a length that follows the width
of the responses, and sources that straddle an edge or leak in from a neighbour; windows also
overlap, a stride of \ForwardModelWindowStride{} responses apart, so each source is inferred in
several and their posteriors must be combined \citep{littenberg2020global}. Beyond these come the
confusion foreground of the unresolved binaries
\citep{timpano2006characterizing,karnesis2021characterization} and the other source classes, noise
that is estimated with the sources and is non-stationary with glitches (\cref{app:noise}), the cubic
phase of $\ddot f = \tfrac{11}{3}\fdot^2/f_0$, which matters most at the top of the band, and
responses longer than $R$ bins, which the widest sources of the full band now lose
(\cref{app:forward-model-heterodyne}).

Three changes to the training recipe come next. \emph{Slot order.} Slots in order of $f_0$
(\cref{sec:method-chart}) give each slot a fixed share of the window, whether or not a source there
is visible. In order of \ac{snr}, the first slots would hold the visible sources and the last ones
the prior, which suits a varying number of sources, but two sources of similar \ac{snr} can swap
slots between noise draws, and the posterior would again hold both labellings.

\emph{Frame loss.} In the frame's weighted Gaussian likelihood (\cref{eq:fm-frame-loss}), one entry
whose frame collapses far below its error can fill a clipped step on its own
(\cref{app:fm-frame}). A Student-t loss bounds those steps and gave the best full-band model, but
was tried only on the full band with $S=4$.

\emph{Window prior.} Training centres most windows where visible sources are common
(\cref{eq:priors-window}), so the edges of the band see few windows, and the loud sources the flow
misses above about \qty{\ResultsFullSplitMhz}{\milli\hertz} are ones the prior rarely produces
(\cref{sec:results-full}). Moving the weights of \cref{eq:priors-window} towards log-uniform, or a
device per subband, which stratifies every batch over the spectrum, would test whether more
examples close the gap.

Two extensions reach beyond the \acp{gb}. \emph{Other source classes.} The architecture already
treats sources as one token stream among the data and frame streams
(\cref{sec:method-streams}); massive black-hole binaries, extreme mass-ratio inspirals and other
classes would each enter as a stream of their own, with their own parameter embeddings, attending
to the same data and to each other, so the posterior of one class accounts for the others as a
global fit does. \emph{Population.} The prior is now fixed, fitted once to a simulated catalogue
(\cref{eq:priors-window-prior}). Making its parameters, such as the shape of the Galaxy and the
distribution of $f_0$, $\fdot$ and $\Amp$, random variables drawn per training window, and inferring
them jointly with the sources, would turn the flow into a hierarchical model that learns the
population from the resolved binaries rather than assuming it.

\section*{Acknowledgements}
We thank Matteo Barbon, Iuliu Cuceu, Coen Rondeel, Natalia Korsakova, Alice Perego and Riccardo Ali for useful
discussions, and Nina Kunert for generating the catalogue from which we built the approximate
\ac{gb} prior.

\section*{Software and Data}
The code is at \url{https://github.com/davidesartor/canna}: the simulator, network, training
and evaluation, and the scripts that make every number, figure and table in this paper.
It is written in JAX \citep{bradbury2018jax} with Equinox \citep{kidger2021equinox}, Optax
\citep{deepmind2020jax} and Orbax \citep{gaffney2026orbax}. The \ac{gb} response is JaxGB's
\citep{jaxgb} on the orbits of LISA Orbits \citep{lisaorbits,martens2021trajectory}, and the
reference chains run on jexplore \citep{jexplore}.

\section*{Impact Statement}

This paper presents work whose goal is to advance the field of Machine Learning. There
are many potential societal consequences of our work, none which we feel must be
specifically highlighted here.

\bibliography{references}
\bibliographystyle{icml2026}

\newpage
\appendix
\counterwithin{equation}{section}
\renewcommand{\theequation}{\thesection\arabic{equation}}
\onecolumn

\section{Related Work}
\label{app:related}

This appendix compares our work with the prior work that bears on its claims: the
likelihood-based and simulation-based inference of \acp{gb} (\cref{tab:related-closest}), how crowded
our windows are against what earlier \acs{sbi} work tested (\cref{fig:related-crowding}), and the
methods from which each ingredient of ours is taken or departs (\cref{tab:related-ingredients}).
It supports \cref{sec:background}.

\begin{table}[!htbp]
\caption{Inference of galactic binaries, as each work reports it. \emph{Joint}: one posterior
over overlapping sources, not a product over them. \emph{Likelihood}: evaluated to draw a
posterior sample. \emph{Reference}: what the posteriors are checked against; P--P, a
 probability--probability plot of the credible levels of the true parameters.}
\label{tab:related-closest}
\centering
\small
\setlength{\tabcolsep}{4pt}
\renewcommand{\arraystretch}{1.15}
\newcolumntype{L}[1]{>{\raggedright\arraybackslash}p{#1}}
\newcolumntype{C}{>{\centering\arraybackslash}p{1.25cm}}
\newcommand{\rwyes}{$\checkmark$}
\newcommand{\rwno}{\textcolor{gray}{--}}
\newcommand{\rwgroup}[1]{\multicolumn{7}{@{}l}{\textsc{#1}}\\[1pt]}
\begin{tabular}{@{}L{2.9cm}L{2.8cm}CCCL{3.9cm}L{1.8cm}@{}}
\toprule
work & sources per posterior & joint & \acs{sbi} & likelihood & signal, data and prior & reference \\
\midrule
\rwgroup{likelihood-based}
\ac{rjmcmc} global fit \citep{littenberg2020global,katz2024efficient,strub2024global}
  & all resolvable ones, their number inferred
  & \rwyes & \rwno & \rwyes
  & full response; the whole band, segment by segment, with the noise and other source classes
  & \rwno \\
flows as priors and proposals \citep{korsakova2024neural,sasli2026learned}
  & 1 per flow, fitted to chains; proposed for the global fit
  & \rwno & \rwno & \rwyes
  & inside the global fit's sampler
  & its chains \\
\midrule
\rwgroup{SBI}
\citet{delmond2026neural}, \ac{nsf}
  & 1; 2 as a proof of concept, ordered by $f_0$
  & 2 only & \rwyes & \rwno
  & full response; $A$, $E$; $\Tobs=\qty{1}{\year}$; $f_0$ within $\pm\qty{2}{\micro\hertz}$ of a fixed centre; \acs*{snr} \numrange{10}{100}
  & \acs*{mcmc}; P--P \\
\citet{mao2025robust}, \ac{fmpe}
  & 1, with 3 free parameters
  & \rwno & \rwyes & \rwno
  & \acs*{gb}-like; \acs*{tdi} channel $A$ only; \qtyrange{30}{90}{\day} with gaps; narrow boxes around the truth
  & \acl*{maf} \\
SlotFlow \citep{houba2025slotflow}, \acs*{nsf} per slot
  & up to 10, their number inferred; slots matched to injections
  & factorised & \rwyes & \rwno
  & sinusoids in white noise, no response
  & \acs*{rjmcmc} \\
\midrule
this work, flow matching
  & 2 to 16, fixed per run, ordered in $f_0$
  & \rwyes & \rwyes & \rwno
  & full response; $A$, $E$, $T$; $\Tobs=\qty{\RelatedWorkTobs}{\year}$; $(f_0, \fdot, \Amp)$ from a
    Galactic catalogue over \qtyrange{\RelatedWorkFzeroLow}{\RelatedWorkFzeroHigh}{\milli\hertz}; most sources
    below the noise, visible ones at \acs*{snr} \numrange{\RelatedWorkSnrLow}{\RelatedWorkSnrHigh}
  & joint \acs*{mcmc} \\
\bottomrule
\end{tabular}
\end{table}

\subsection{Inference of Galactic Binaries}

\paragraph{Likelihood-based inference.}
The \ac{lisa} global fit samples all resolvable sources with reversible-jump
\citep{green1995reversible}, parallel-tempered \citep{earl2005parallel} \ac{mcmc}, segment by segment
for the \acp{gb} and jointly with the other source classes and the noise
\citep{littenberg2020global,littenberg2023prototype,katz2024efficient,strub2024global,deng2025modular}.
Neural density estimators enter it as fitted priors, proposals and compact catalogues, each flow
fitted to one source's chains \citep{korsakova2024neural}; \citet{sasli2026learned} show on
non-\ac{gb} benchmarks that learned birth proposals help once the fit has equilibrated rather than
while it is assembled. We use the sampler only as a reference (\cref{app:evaluation,app:mcmc}), a
joint \ac{mcmc} over all \rev{$8S$} parameters of each evaluation injection. A window
posterior from \acs{sbi} could in turn serve a global fit as a proposal or as the initial state of its
chains.

\paragraph{The global-fit problem.}
\label{app:related-globalfit}
The Milky Way holds tens of millions of double white dwarfs in the \ac{lisa} band
\citep{nelemans2001gravitational,korol2017prospects,lamberts2019predicting}; the simulated Galaxy
behind our prior holds \num{\RatesCatalogue} between \qty{\PriorsFullLowMhz}{\milli\hertz} and
\qty{\PriorsFullHighMhz}{\milli\hertz}. Over \qty{\ForwardModelTobsYears}{\year} that band spans
about $10^6$ Fourier bins, and below a few millihertz a window of \RatesNarrowWindow{} bins holds
\numrange{\RatesNarrowAllLo}{\RatesNarrowAllHi} of them (\cref{sec:gw-rates}). Only
\qty{\RelatedWorkVisibleShare}{\percent} reach an \ac{snr} of \RelatedWorkSnrLow{} alone in their window;
the rest form a confusion foreground \citep{timpano2006characterizing,karnesis2021characterization}.
Of order $10^4$ will be resolved, with eight parameters each: a posterior over some $10^5$
dimensions in which the number of sources is itself unknown. The same data hold other source
classes, a noise \ac{psd} to be estimated alongside them, gaps and glitches
\citep{colpi2024lisa,baghi2019gravitational,dey2021effect,baghi2022detection}, which the global fit
infers together.

The pipelines that have analysed the year-long Sangria data challenge \citep{baghi2022lisa} are
compared in \cref{tab:gw-globalfit}. Of the four, GLASS \citep{littenberg2023prototype} and the
pipeline of \citet{strub2024global} are \ac{cpu} samplers, Erebor runs on \acp{gpu}
\citep{katz2024efficient}, and Gee-Moo splits the band into \num{4663} narrow segments
\citep{deng2025modular}. Erebor and Gee-Moo now also fit the consortium's Mojito Light data
\citep{lisapriors}, with at most five \acp{gb} per segment for Gee-Moo where Erebor allows fifteen to
twenty.

\subsection{Simulation-Based Inference for Galactic Binaries}
\label{app:related-amortised}

\citet{delmond2026neural} train a neural spline flow on the frequency-domain $A$ and $E$ data
of a narrow band. For one source their posteriors match an \ac{mcmc} at \qty{1}{\milli\hertz}, come out slightly
wider at \qty{5}{\milli\hertz} and clearly wider in $f_0$ and sky at \qtyrange{10}{15}{\milli\hertz} and over a \qty{0.5}{\milli\hertz} band. For two sources they impose
$f_{0,1}\le f_{0,2}$, since an unordered flow put one mode on each labelling; their pair
\RelatedWorkDelmondEasiest{} bins apart is close to \ac{mcmc}, the pair \RelatedWorkDelmondHardest{}
bins apart broad and sometimes biased. \citet{mao2025robust} apply \ac{fmpe} to gapped data of one
source with three free parameters. SlotFlow \citep{houba2025slotflow} infers the number of
components with a classifier and gives each a slot of a shared flow trained with Hungarian
matching; its posterior factorises over the slots given a shared context, which cannot carry
correlations between close sources, and on sinusoid mixtures its frequency widths are two to
three times those of \ac{rjmcmc}. None of these has, to our knowledge, been extended to a joint
posterior over more than two sources under the \ac{lisa} response.

\begin{figure}[t]
\centering
\includegraphics[width=3.25in]{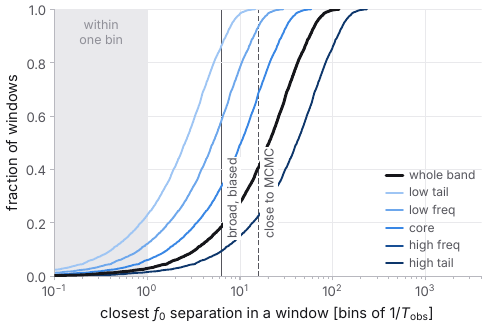}
\caption{How crowded the training windows are: the distribution of the closest $f_0$ separation
among the \RelatedWorkSourcesWord{} sources of a window, over \RelatedWorkCrowdingWindows{}
training windows (\cref{eq:priors-window}) of the whole band at $R=\NarrowBandHighFreqR$ (the full run) and
of each of its five subbands at the subband's own $R$ (\cref{tab:narrow-band-bands}), in each analysis's own
bins of $1/\Tobs$ (\qty{\RelatedWorkDelmondTobs}{\year} for \citet{delmond2026neural},
\qty{\RelatedWorkTobs}{\year} here). Vertical lines: the two source pairs of
\citet{delmond2026neural}, \RelatedWorkDelmondEasiest{} bins apart (close to \ac{mcmc}) and
\RelatedWorkDelmondHardest{} (broad, sometimes biased). Over the whole band
\qty{\RelatedWorkCloserFull}{\percent} of windows hold a pair closer than their hardest one; per
subband from low to high $f_0$, \qtylist{\RelatedWorkCloserLowTail;\RelatedWorkCloserLowFreq;\RelatedWorkCloserCore;\RelatedWorkCloserHighFreq;\RelatedWorkCloserHighTail}{\percent}.}
\label{fig:related-crowding}
\end{figure}

\paragraph{Crowding.}
Our sources are drawn from the prior, not placed by hand, and \cref{fig:related-crowding} shows what
that implies. Over the whole band the closest two sources of a window are a median
\RelatedWorkClosestFull{} bins apart, and \qty{\RelatedWorkCloserFull}{\percent} of windows hold a
pair closer than Delmond et al.'s hardest. A window is $4R$ bins wide, so the median doubles from one
subband to the next, from \RelatedWorkClosestLowTail{} bins in the low tail to
\RelatedWorkClosestHighTail{} in the high tail, and \qty{\RelatedWorkSubBinCore}{\percent} of core
windows hold a pair within one bin. Among the Galaxy's resolvable sources overlap is rarer: fewer
than \qty{1}{\percent} of pairs above \ac{snr} 7 have a match above one half after a year
\citep{littenberg2020global}.

Overlap costs the flow width. On the narrow band, sources at $\snr\ge7$ whose nearest injected
neighbour is more than \ResultsSepFarBins{} bins away have $f_0$ posteriors \ResultsNarrowFourSepFar{}
($S=4$) and \ResultsNarrowEightSepFar{} ($S=8$) times the joint \ac{mcmc}'s width; within
\ResultsSepCloseBins{} bins, \ResultsNarrowFourSepClose{} and \ResultsNarrowEightSepClose{} times
(\ResultsNarrowFourSepCloseN{} and \ResultsNarrowEightSepCloseN{} sources; models of
\cref{tab:results-widths}). The chains resolve such pairs, so the excess is the flow's. On the full band
the separation is mixed up with frequency, since the loud sources the flow misses sit at high $f_0$
(\cref{app:results-mcmc}).

\begin{table}[t]
\caption{The methods behind each ingredient of ours, and where we depart from them.}
\label{tab:related-ingredients}
\centering
\small
\renewcommand{\arraystretch}{1.15}
\newcolumntype{L}[1]{>{\raggedright\arraybackslash}p{#1}}
\begin{tabular}{@{}L{2.5cm}L{5.0cm}L{6.8cm}@{}}
\toprule
ingredient & prior work & ours \\
\midrule
phase and polarisation copies
  & phase left at the prior by the network, or drawn from the likelihood
    \citep{dax2023neural,du2024advancing,cole2026sequential,kofler2026flexible};
    linear amplitudes predicted directly \citep{delmond2026neural}
  & helicity amplitudes as flat complex coordinates; the four $(\psi,\phi_0)$ copies fold into one
    point and each phase is the angle of a coordinate, with no likelihood (\cref{app:chart}) \\
label switching
  & relabelling or an ordering constraint in \acs*{mcmc} \citep{stephens2000dealing,buscicchio2019label,littenberg2020global};
    ordered sources \citep{delmond2026neural}; slots matched to injections in the loss
    \citep{houba2025slotflow}; permutation-equivariant models with optimal-transport
    couplings \citep{klein2023equivariant,tong2024improving}
  & slots ordered in $f_0$ as in the global fit, so that each slot has a frame of its own
    (\cref{app:chart-slots}); the ordering folds the posterior at a boundary between close sources \\
objective and base
  & flow matching \citep{lipman2023flow,dax2023flow}, with fixed time weights
    \citep{karras2022elucidating,kingma2023understanding}; a point-estimate endpoint
    \citep{yim2023fast}
  & a Gaussian base placed and scaled per source by a frame read off the data, and an endpoint
    Gaussian trained on its likelihood, which weights each point's error by its own irreducible
    part \citep{kendall2018multi} (\cref{app:fm}) \\
network
  & Dingo-T1 reads the strain as tokens \citep{kofler2026flexible}; the Simformer is a transformer
    over parameters and data \citep{gloeckler2024allinone}; t-\acs*{fmpe} generates a variable number
    of components \citep{vandermeulen2026fast}
  & a multimodal diffusion transformer \citep{esser2024scaling} with sources and short-time Fourier
    tokens of \RelatedWorkPatch{} bins as two streams, one-way attention so that the data are encoded
    once per observation, and rotary embeddings on a shared frequency axis
    \citep{su2024roformer} (\cref{app:network}) \\
validation
  & P--P plots in most gravitational-wave work; calibration studies of flow-matching posteriors
    \citep{luo2026artifact,liu2026assessing}; distributional metrics \citep{baattrup2026pointwise}
  & ranks of data-space quantities, known-answer mocks and a joint \acs*{mcmc} reference for widths
    (\cref{app:evaluation}) \\
\bottomrule
\end{tabular}
\end{table}

\subsection{Methods We Build On}
\label{app:related-methods}
Beyond \acp{gb}, \cref{tab:related-ingredients} sets each ingredient of our recipe against its
closest precedents. Two points need more than a table row. \Acs{sbi} is established
for ground-based detectors \citep{chua2020learning,green2020gravitational,green2021complete,gabbard2022bayesian,dax2021real,dax2025real}
and for \ac{lisa}'s massive black-hole binaries \citep{du2024advancing,liang2024rapid,spadaro2026accurate}, but
their overlapping-signal analyses cover pairs
\citep{langendorff2023normalizing,alvey2023crowded,papalini2025can}, and \citet{papalini2025can}, the
closest in spirit, use a discrete flow, a fixed pair and neither a per-source frame nor ordered
slots. And the one precedent for a data-dependent frame, labrador \citep{roulet2026domain}, heterodynes the data against
a reference waveform and folds the parameter space to remove multimodality, but takes its reference
from a per-event optimisation where our frame comes from the network.

\subsection{Recent Work}
\label{app:related-recent}
Neural posterior estimation is now deployed in production for the follow-up of low-latency alerts
\citep{marx2026likelihood}, and a GPU nested sampler returns binary-neutron-star posteriors from an
uninformed prior in minutes \citep{yallup2026abinitio}. For one source, speed alone is a weakening
argument for \acs{sbi}; the joint posterior of many overlapping sources, whose chains take hours
(\cref{app:mcmc-cost}), is where it still pays. \citet{kolmus2024tuning} find that the accuracy of
neural posterior estimation follows how many similar samples the training prior supplies for each
region, which bears on our catalogue prior, where loud sources are rare (\cref{app:priors}). Around
the \ac{gb} problem, autoencoders trained on simulated confusion-limited segments learn a latent
representation of the foreground that helps flag resolvable sources
\citep{cain2026manifold,cain2026global}, a continuous-time trans-dimensional sampler is motivated by
the global fit \citep{astorino2026samsara}, and sequential inference is adequate where sources
couple weakly, as found for extreme mass-ratio inspirals and massive black-hole binaries in
simulated data \citep{agrawal2026untangling}, unlike overlapping \acp{gb}. Fast wavelet-domain
templates \citep{cornish2026time} with a differentiable wavelet transform
\citep{vajpeyi2026explicit} could replace our heterodyned simulator, and consistency models
\citep{schmitt2024consistency} would cut our \FlowMatchingOdeSteps{} sampler steps to a few; we
have tried neither.

\section{The Galactic-Binary Signal and the \acs{lisa} Response}
\label{app:forward-model}

This appendix defines the simulator of \cref{sec:gw}, so that it can be re-implemented: the waveform,
the response of one laser link and its heterodyned form, the \ac{tdi} channels, and the windows the
band is cut into. It ends with the accuracy of the truncated response, which backs the choice of $R$
in \cref{sec:narrow-band}.

\subsection{Waveform}
\label{app:forward-model-waveform}

The transverse-traceless strain of a \ac{gb} at the observer is
\begin{equation}
  h_{ij}(\xi) = h_+(\xi)\,e^+_{ij} + h_\times(\xi)\,e^\times_{ij},\qquad
  \xi = t - \frac{\hat{\mathbf k}\cdot\mathbf x(t)}{c},
  \label{eq:forward-model-xi}
\end{equation}
with $\hat{\mathbf k}$ the direction of propagation, $\mathbf x(t)$ the observer's position and
$e^{+,\times}_{ij}$ the polarisation tensors of the source frame. For a circular binary with a constant
frequency derivative,
\begin{equation}
\begin{aligned}
  h_+(\xi) &= -\Amp\,\bigl(1+\cos^2\iota\bigr)\cos\Psi(\xi),\\
  h_\times(\xi) &= -2\Amp\cos\iota\,\sin\Psi(\xi),\\
  \Psi(\xi) &= -\phi_0 + 2\pi f_0\,\xi + \pi\fdot\,\xi^2.
\end{aligned}
  \label{eq:forward-model-waveform}
\end{equation}
Radiation reaction ties the drift and the amplitude to the chirp mass $\Mc$ and the luminosity distance
$d_L$ \citep{maggiore2008gravitational},
\begin{equation}
  \fdot = \frac{96}{5}\,\pi^{8/3}\left(\frac{G\Mc}{c^3}\right)^{5/3} f_0^{11/3},\qquad
  \Amp = \frac{2\,(G\Mc)^{5/3}}{c^4\,d_L}\,(\pi f_0)^{2/3},
  \label{eq:forward-model-fdot}
\end{equation}
but the simulator keeps $\fdot$ and $\Amp$ free, with the prior of \cref{app:priors-sources}.

The sky position $(\alpha,\delta)$ defines the orthonormal triad
\begin{equation}
  \hat{\mathbf u} = (\sin\alpha,\,-\cos\alpha,\,0),\quad
  \hat{\mathbf v} = (-\sin\delta\cos\alpha,\,-\sin\delta\sin\alpha,\,\cos\delta),\quad
  \hat{\mathbf k} = (-\cos\delta\cos\alpha,\,-\cos\delta\sin\alpha,\,-\sin\delta)
  \label{eq:forward-model-triad}
\end{equation}
in the equatorial frame of the orbit files, with $\epsilon^+ = \hat{\mathbf u}\otimes\hat{\mathbf u} -
\hat{\mathbf v}\otimes\hat{\mathbf v}$ and $\epsilon^\times = \hat{\mathbf u}\otimes\hat{\mathbf v} +
\hat{\mathbf v}\otimes\hat{\mathbf u}$. The source frame is rotated by $\psi$ about $\hat{\mathbf k}$, so
$e^+ = \cos2\psi\,\epsilon^+ + \sin2\psi\,\epsilon^\times$ and
$e^\times = -\sin2\psi\,\epsilon^+ + \cos2\psi\,\epsilon^\times$, and
$h_{ij} = \mathrm{Re}\bigl\{(A_+\epsilon^+_{ij} + A_\times\epsilon^\times_{ij})e^{\ii\Psi}\bigr\}$ with
\begin{equation}
  A_+ = -\Amp\bigl[(1+\cos^2\iota)\cos2\psi + 2\ii\cos\iota\sin2\psi\bigr],\quad
  A_\times = -\Amp\bigl[(1+\cos^2\iota)\sin2\psi - 2\ii\cos\iota\cos2\psi\bigr].
  \label{eq:forward-model-complex-amp}
\end{equation}
In the circular basis $\epsilon^{\rm R,L} = (\epsilon^+ \pm \ii\epsilon^\times)/2$ this is
\begin{equation}
  A_+\epsilon^+ + A_\times\epsilon^\times
  = -\Amp\Bigl[(1+\cos\iota)^2 e^{2\ii\psi}\,\epsilon^{\rm L} + (1-\cos\iota)^2 e^{-2\ii\psi}\,\epsilon^{\rm R}\Bigr],
  \label{eq:forward-model-helicity}
\end{equation}
so the data depend on $(\Amp,\iota,\psi,\phi_0)$ only through the two complex numbers
$\Amp(1\pm\cos\iota)^2e^{-\ii(\phi_0\mp2\psi)}$\rev{, up to normalisation the complex conjugates of the flow's
helicity amplitudes $\zeta_\mp$ of \cref{eq:chart-helicity}, with the phase referred to the start rather than
mid-observation}.

\subsection{Link Response and Heterodyning}
\label{app:forward-model-link}\label{app:forward-model-heterodyne}

\ac{lisa} measures relative frequency fluctuations on six laser links. For link $ij$, received at
spacecraft $i$ at time $t_i$ and emitted at $j$ at $t_j = t_i - L_{ij}$, the response is
\citep{finn2009response,cornish2009alternative}
\begin{equation}
  y_{ij}(t_i) = -\frac12\,\frac{H_{ij}(\xi_i) - H_{ij}(\xi_j)}{1 - \hat{\mathbf k}\cdot\hat{\mathbf x}_{ij}},
  \qquad
  H_{ij}(\xi) = \hat x^k_{ij}\hat x^l_{ij}\,h_{kl}(\xi),
  \label{eq:forward-model-link}
\end{equation}
with $\hat{\mathbf x}_{ij}$ the unit vector from emitter to receiver. We take the emitter's position at
the time of reception and all six light travel times equal and constant,
$L = \qty{\ForwardModelArmGm}{\giga\metre}/c$, so the transfer frequency is
$f_\star = (2\pi L)^{-1} = \qty{\ForwardModelFStar}{\milli\hertz}$. With
$\mathcal A_{ij} = A_+F^+_{ij} + A_\times F^\times_{ij}$, where $F^+_{ij}$ and $F^\times_{ij}$ are the
antenna patterns, and $\alpha_{ij} = \pi L(1 - \hat{\mathbf k}\cdot\hat{\mathbf x}_{ij})(f_0 + \fdot\xi_i)$,
the response separates, for $q = \operatorname{rint}(f_0\Tobs)$, into a carrier and a slow envelope,
$y_{ij}(t) = \mathrm{Re}\{\mathcal G_{ij}(t)\,e^{2\pi\ii qt/\Tobs}\}$, with
\begin{equation}
  \mathcal G_{ij}(t) = -\frac{\ii}{2}\,\frac{f_0 + \fdot\,\xi_i}{f_{\star}}\,\mathcal A_{ij}\,
  \operatorname{sinc}\alpha_{ij}\;
  \exp\Bigl\{-\ii\Bigl[\alpha_{ij} + 2\pi f_0\frac{\hat{\mathbf k}\cdot\mathbf x_i(t)}{c}
  - \Bigl(-\phi_0 + 2\pi\bigl(f_0 - \tfrac{q}{\Tobs}\bigr)t + \pi\fdot\,\xi_i^2\Bigr)\Bigr]\Bigr\}.
  \label{eq:forward-model-slow}
\end{equation}
We sample the envelope at $t_m = m\Tobs/R$ and Fourier-transform it with a rectangular window, which
gives the positive-frequency spectrum on the $R$ bins starting at $q - R/2$,
\begin{equation}
  \tilde y_{ij}\Bigl(\frac{q+k}{\Tobs}\Bigr) = \frac{\Tobs}{2R}\sum_{m=0}^{R-1}\mathcal G_{ij}(t_m)\,
  e^{-2\pi\ii km/R},\qquad k = -\tfrac R2,\dots,\tfrac R2 - 1,
  \label{eq:forward-model-heterodyne}
\end{equation}
normalised to the continuous transform that the noise of \cref{eq:noise-data} refers to.

The envelope varies on the time scale of the orbit, so its spectrum is confined to the annual Doppler
sideband $\pm f_0v_\oplus/c$, the chirp drift $\fdot\Tobs$, and the antenna-pattern sidebands at
harmonics of $1/\unit{\year}$ \citep{cutler1998angular}. A source therefore needs
\begin{equation}
  n_{\rm span}(f_0,\fdot) = 2\left\lceil\Bigl(\frac{f_0 v_\oplus}{c}
  + \fdot\,\Tobs + \frac{\ForwardModelAntennaHarmonics{}}{\unit{\year}}\Bigr)\Tobs\right\rceil \le R
  \label{eq:forward-model-span}
\end{equation}
bins, with $v_\oplus = \qty{\ForwardModelVelocity}{\kilo\metre\per\second}$. Most responses fall
below the noise well inside this bound, so $R$ is set by the bins the data see
(\cref{app:forward-model-accuracy}).

\subsection{Time-Delay Interferometry}
\label{app:forward-model-tdi}

\Ac{tdi} cancels the laser frequency noise by combining delayed link responses
\citep{tinto2021time,babak2021lisa}. With the delay operator $D_{i_1i_2\cdots i_n}x(t) =
x(t - \sum_{h}L_{i_hi_{h+1}})$, the first Michelson combination is
\begin{equation}
  X = y_{13} + D_{13}y_{31} + D_{131}y_{12} + D_{1312}y_{21}
  - \bigl[y_{12} + D_{12}y_{21} + D_{121}y_{13} + D_{1213}y_{31}\bigr],
  \label{eq:forward-model-tdi-time}
\end{equation}
and $Y$, $Z$ follow by the cyclic permutation $1\to2\to3\to1$; for equal arms this is \acs{tdi}
generation 1.5. In the frequency domain each delay is a phase $e^{-2\pi\ii fL}$, applied to the
envelopes of \cref{eq:forward-model-slow} before the transform. The noise-orthogonal channels are
\citep{prince2002lisa}
\begin{equation}
  A = \frac{Z-X}{\sqrt2},\qquad E = \frac{X-2Y+Z}{\sqrt6},\qquad T = \frac{X+Y+Z}{\sqrt3}.
  \label{eq:forward-model-aet}
\end{equation}

\subsection{Windows}
\label{app:forward-model-windows}

A window $w$ is $K = \ForwardModelWindowResponses{}R$ consecutive bins with first bin and centre
\begin{equation}
  k_0(\fw) = \ForwardModelWindowStride{}\,\operatorname{rint}\Bigl(\frac{\fw\Tobs - K/2}{\ForwardModelWindowStride{}}\Bigr),
  \qquad \fw = \frac{k_0 + K/2}{\Tobs}.
  \label{eq:forward-model-window-start}
\end{equation}
A source's response fills bins $q-R/2,\dots,q+R/2-1$, which lie inside the window when
$k_0+R/2\le q\le k_0+K-R/2$. Since both bounds are integers this holds whenever $f_0\Tobs$ lies
between them, so a guard band of $R/2$ bins at each edge leaves an interior of
$\ForwardModelInteriorResponses{}R$ bins in which a source belongs to the window,
\begin{equation}
  \max\Bigl(\frac{k_0 + R/2}{\Tobs},\, f_{\min}\Bigr) \le f_0 \le
  \min\Bigl(\frac{k_0 + K - R/2}{\Tobs},\, f_{\max}\Bigr).
  \label{eq:forward-model-window-interior}
\end{equation}
The window index is clipped to the range whose interiors just cover the band.

\subsection{Accuracy of the Truncated Response}
\label{app:forward-model-accuracy}

\begin{figure}[t]
  \centering
  \includegraphics{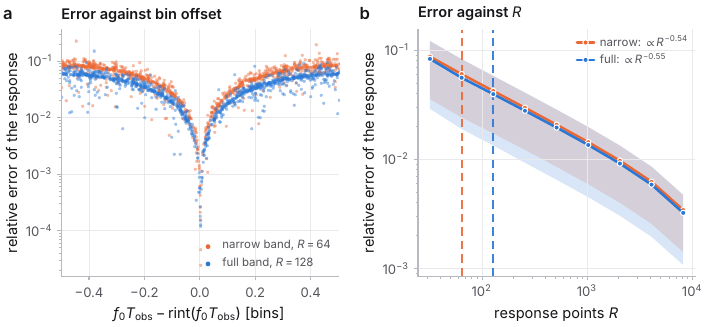}
  \caption{The error of the truncated response is leakage between bins. (a)~Each source's relative error against a
  \num{\ForwardModelReferencePoints}-point response, against its signed offset from the nearest bin
  centre: below \qty{1}{\percent} on a bin centre, several per cent halfway (narrow band orange,
  full band blue; 1024 of \num{4096} prior sources per band). (b)~The median and 16--84\% range of
  that error over a ladder of $R$, falling close to $R^{-1/2}$, with each band's own $R$ dashed.}
  \label{fig:forward-model}
\end{figure}

Against a \num{\ForwardModelReferencePoints}-point reference, the $R$-bin response keeps a source's
power: at most \qty{\ForwardModelNarrowTruncOutsideMax}{\percent} lies outside its $R$ bins and the
\ac{snr} is kept to \num{\ForwardModelFullTruncRhoDev}. It does not keep the exact shape of the
spectrum, because the rectangular window leaks a tone that falls between bins: the median error is
\qty{\ForwardModelNarrowTruncRelMedian}{\percent} (narrow band) and
\qty{\ForwardModelFullTruncRelMedian}{\percent} (full band), below \qty{1}{\percent} on a bin centre
and up to \qty{\ForwardModelNarrowTruncRelOffBin}{\percent} halfway between bins
(\cref{fig:forward-model}). Training, the test injections and the reference \ac{mcmc} share this
response, so the comparison between the flow and the chains is unaffected.

\section{Noise, Likelihood and Signal-to-Noise Ratio}
\label{app:noise}\label{app:noise-psd}\label{app:noise-likelihood}\label{app:noise-snr}\label{app:noise-validation}

The noise of \cref{eq:noise-data} is the instrument noise of the \ac{lisa} requirements
\citep{lisasst2018lisa}, in the single-link form of \citet[Issue~1.1]{babak2021lisa}: an optical
metrology noise and a test-mass acceleration noise,
\begin{align}
  P_{\rm oms}(f) &= \bigl(\qty{\NoiseOms}{\pico\metre}\bigr)^2
  \left[1+\Bigl(\frac{\qty{2}{\milli\hertz}}{f}\Bigr)^4\right]\unit{\per\hertz}, \nonumber\\
  P_{\rm acc}(f) &= \bigl(\qty{\NoiseAcc}{\femto\metre\per\second\squared}\bigr)^2
  \left[1+\Bigl(\frac{\qty{0.4}{\milli\hertz}}{f}\Bigr)^2\right]
  \left[1+\Bigl(\frac{f}{\qty{8}{\milli\hertz}}\Bigr)^4\right]\unit{\per\hertz},
  \label{eq:noise-single-link}
\end{align}
converted to fractional frequency by $S_{\rm oms} = (2\pi f/c)^2P_{\rm oms}$ and
$S_{\rm acc} = P_{\rm acc}/[(2\pi f)^2c^2]$. For equal arms, $x = 2\pi fL$, the \acs{tdi}-1.5 channels
have the one-sided \acp{psd} \citep[eqs.~19, 127--129]{babak2021lisa}
\begin{align}
  S_A = S_E &= 8\sin^2x\,\bigl[(2+\cos x)\,S_{\rm oms}
  + 4\,(1 + \cos x + \cos^2 x)\,S_{\rm acc}\bigr], \nonumber\\
  S_T &= 16\sin^2x\,(1-\cos x)\,\bigl[S_{\rm oms} + 2\,(1-\cos x)\,S_{\rm acc}\bigr].
  \label{eq:noise-psd}
\end{align}
The channel $T$ is quiet ($S_T/S_A = \num{\NoiseTOverALow}$ at \qty{\NoiseBandLow}{\milli\hertz}) but
carries almost no \ac{snr} in our band. The noise is instrumental only: no Galactic foreground
\citep{karnesis2021characterization}, gaps or glitches. Signals are compared through
\begin{equation}
  \inner{a}{b} = 4\,\mathrm{Re}\sum_{c}\sum_{k}\frac{a_{k,c}\,b^*_{k,c}}{S_c(f_k)\,\Tobs},
  \label{eq:inner}
\end{equation}
so that the log-likelihood of \cref{eq:loglik} is $-\tfrac12\inner{d-h}{d-h}$ up to a constant and
$\snr_s = \inner{h(\theta_s)}{h(\theta_s)}^{1/2}$. The simulator matches independent references: over
\num{\NoiseGeneratorWindows} pure-noise \rev{test windows of \num{\NoiseGeneratorBins} bins in $A$ and $E$,} $\inner{n}{n}$ follows $\chi^2_{\NoiseGeneratorDof{}}$
(Kolmogorov--Smirnov $p = \NoiseGeneratorKs{}$), and for the verification binary of
\citet[Table~VII]{strub2022bayesian} our \acs{snr} is \NoiseStrubSnr{} against their
\NoiseStrubSnrTheirs{}.

\section{Priors and the Training Distribution}
\label{app:priors}

This appendix states what the flow is trained on, which \cref{sec:method-training} uses: the per-source prior,
the prior over windows, the number of sources a window holds, and the stream of simulated windows.
\Cref{tab:priors-bands} collects what differs between the two bands; $S={}$\ResultsNarrowCounts{} on
the narrow band and \ResultsFullCounts{} on the full band in the runs of \cref{app:results}.

\subsection{The Sources of a Window}
\label{app:priors-sources}

\begin{figure*}[t]
  \centering
  \includegraphics{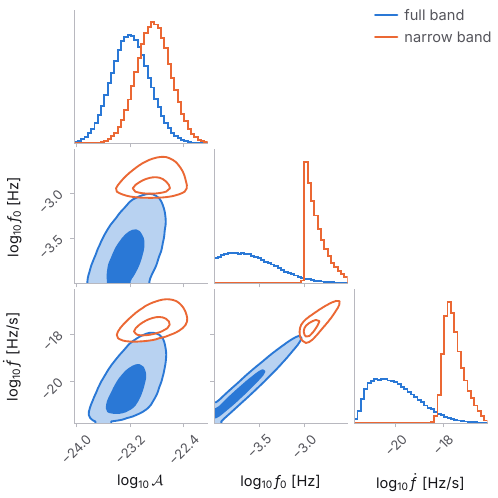}
  \caption{The per-source prior over each band (\cref{eq:priors-source}): amplitude, frequency and
  drift in $\log_{10}$, from \num{\PriorCornerDraws} draws per band with $f_0$ truncated to the
  band. Contours enclose \qty{\PriorCornerInner}{\percent} and \qty{\PriorCornerOuter}{\percent} of
  the draws (one and two standard deviations in two dimensions); diagonals are normalised
  marginals. The sky does not depend on the band (\cref{fig:priors}b).}
  \label{fig:prior-corner}
\end{figure*}

The $S$ sources of a window are independent draws from the per-source prior of
\cref{tab:priors-source}, so they are exchangeable. Only $f_0$ depends on the window directly. For
frequency, drift and amplitude we fit one Gaussian by maximum likelihood to the logarithms of the
binaries above \qty{\PriorsCatalogueCutMhz}{\milli\hertz} of a simulated catalogue of double white
dwarfs \citep{perego2025towards,breivik2020cosmic},
\begin{equation}
  \begin{pmatrix}\log_{10}f_0\\ \log_{10}\fdot\\ \log_{10}\Amp\end{pmatrix}
  \sim \Normal(\mu_\pi, \Sigma_\pi),
  \label{eq:priors-source}
\end{equation}
with $f_0$ truncated to the window (\cref{eq:priors-interior}), and, with $f_0$ in \unit{\hertz} and
$\fdot$ in \unit{\hertz\per\second},
\begin{equation}
  \mu_\pi = \begin{pmatrix}\PriorsMeanF\\ \PriorsMeanFdot\\ \PriorsMeanAmp\end{pmatrix},\qquad
  \Sigma_\pi = \begin{pmatrix}
    \PriorsCovFF & \PriorsCovFFdot & \PriorsCovFAmp\\
    \PriorsCovFFdot & \PriorsCovFdotFdot & \PriorsCovFdotAmp\\
    \PriorsCovFAmp & \PriorsCovFdotAmp & \PriorsCovAmpAmp
  \end{pmatrix}.
  \label{eq:priors-mean}
\end{equation}
Other population models of the Galaxy's double white dwarfs
\citep{toonen2012supernova,korol2020populations,georgousi2022gravitational,thiele2023applying}
differ in their binary physics \citep{toubiana2024interacting,boileau2025gravitational}; refitting
changes only $\mu_\pi$ and $\Sigma_\pi$. \Cref{fig:prior-corner} shows the prior over each band. Most sources barely drift: the full band's median drift over
the observation is \rev{\num{\PriorsFullDriftMedian}} of a bin, and \qty{\PriorsFullDriftOver}{\percent} of its
sources drift by more than one. The sky is a mixture of a bulge, a thick and a thin disc and an
isotropic floor in galactic longitude $l$ and $\sin b$,
\begin{equation}
  \pi(l, \sin b) = w_{\rm b}\,\Normal(l; 0, \sigma_l^2)\,\Normal(\sin b; 0, \sigma_b^2)
  + \sum_{d\in\{\rm thick, thin\}} w_d\,\mathcal{L}_{\pi}(l; \lambda_{l,d})\,\mathcal{L}_{1}(\sin b; \lambda_{b,d})
  + \frac{w_{\rm f}}{4\pi},
  \label{eq:priors-sky}
\end{equation}
with $\mathcal{L}_c(x;\lambda)\propto e^{-|x|/\lambda}$ a Laplace density truncated to $|x|\le c$,
rotated to the equatorial $(\alpha,\delta)$ of the response. It puts \qty{\PriorsBulgeWeight}{\percent}
of its weight in the bulge, \qty{\PriorsThickWeight}{\percent} and \qty{\PriorsThinWeight}{\percent} in
the thick and thin discs and \qty{\PriorsFloorWeight}{\percent} in the floor, and traces the plane:
\qty{\PriorsPlaneShare}{\percent} of the sources lie within \ang{\PriorsPlaneDeg} of the galactic
plane (\cref{fig:priors}b).

\subsection{The Window Prior}
\label{app:priors-window}

A window is $4R$ consecutive bins starting at an even bin $k_0 = \PriorsStride{}j$, with centre
$\fw = (k_0 + 2R)/\Tobs$ and interior
\begin{equation}
  f_0\Tobs \in \bigl[\max(k_0 + R/2,\ f_{\min}\Tobs),\ \min(k_0 + 4R - R/2,\ f_{\max}\Tobs)\bigr]
  \label{eq:priors-interior}
\end{equation}
(\cref{app:forward-model-windows}). Windows centred on the catalogue's $f_0$ marginal would sit low in
the band, where sources are faint: the upper half of the full band in $\log f_0$, which holds most of
the visible sources, would get \qty{\PriorsFullTopShare}{\percent} of the windows, and
\qty{\PriorsFullWithVisible}{\percent} of windows would hold a visible source. Training therefore
centres windows where sources can be seen. Of \PriorsCandidates{} candidates $\log_{10}f_c^{(m)}$
drawn uniformly over the band, each with one prior sky and orientation and alone in its window at the
conditional mean $\fdot$, one is picked with weight
\begin{equation}
\begin{gathered}
  w_m \propto \Normal\bigl(\log_{10}f_c^{(m)};\, \PriorsMeanF, \PriorsStdF^2\bigr)\,
  \Phi\!\Bigl(\frac{\PriorsInterceptAmp + \PriorsSlopeAmp\log_{10}f_c^{(m)} - \log_{10}\Amp_k^{(m)}}{\PriorsResidualAmp}\Bigr),\\
  j = \operatorname{clip}\!\Bigl(\operatorname{round}\frac{f_c\Tobs - 2R}{\PriorsStride{}},\ j_{\rm lo},\ j_{\rm hi}\Bigr),
\end{gathered}
  \label{eq:priors-window}
\end{equation}
the catalogue's $f_0$ marginal times the chance that $\log_{10}\Amp$ exceeds $\log_{10}\Amp_k^{(m)}$,
the amplitude at which the candidate reaches $\snr=k=\PriorsVisibleSnr$; $j_{\rm lo}$ and
$j_{\rm hi}$ are the first and last windows whose interior reaches the band. A share of
\qty{\PriorsUniformWindowsPercent}{\percent} of windows skips the weights and is centred
log-uniformly, so training also sees the empty windows that most of the band holds. This puts
\qty{\PriorsFullTrainingTopShare}{\percent} of the full band's windows in its upper half and
\qty{\PriorsFullTrainingWithVisible}{\percent} with a visible source; in the narrow band the share
rises from \qty{\PriorsNarrowWithVisible}{\percent} to \qty{\PriorsNarrowTrainingWithVisible}{\percent}.
Given the window, the prior is the catalogue's; the weights only choose which windows training sees,
so they do not change the posterior the flow learns.

\begin{table}[t]
\caption{The two bands. Every other prior is shared. Source and window shares are over
\num{\PriorsPriorWindows} windows per band centred on the catalogue's $f_0$ marginal, and
\num{\PriorsCutWindows} centred by \cref{eq:priors-window} (training); a source is visible at
$\snr\ge\PriorsVisibleSnr$ alone in its window. \rev{Steps, windows and sources seen are those of the reported runs of the
B network (\cref{tab:results-runs}), at the batch of \cref{app:priors-training}}; the simulation share is the time spent drawing a batch
inside one training step on an \PriorsNarrowStepChip{} (narrow band) and an \PriorsFullStepChip{} (full band).}
\label{tab:priors-bands}
\centering
\footnotesize
\setlength{\tabcolsep}{3pt}
\begin{tabular}{@{}>{\raggedright\arraybackslash}p{0.4\columnwidth}rr@{}}
\toprule
 & narrow & full \\
\midrule
band & \qtyrange{\PriorsNarrowLowMhz}{\PriorsNarrowHighMhz}{\milli\hertz} & \qtyrange{\PriorsFullLowMhz}{\PriorsFullHighMhz}{\milli\hertz} \\
response $R$ / window $4R$ [bins] & \num{\PriorsNarrowR} / \num{\PriorsNarrowW} & \num{\PriorsFullR} / \num{\PriorsFullW} \\
distinct windows & \num{\PriorsNarrowWindows} & \num{\PriorsFullWindows} \\
catalogue above \qty{\PriorsCatalogueCutMhz}{\milli\hertz} in the band & \qty{\PriorsNarrowCatalogueShare}{\percent} & \qty{\PriorsFullCatalogueShare}{\percent} \\
visible sources & \qty{\PriorsNarrowVisible}{\percent} & \qty{\PriorsFullVisible}{\percent} \\
windows with a visible source, catalogue / training & \qty{\PriorsNarrowWithVisible}{\percent} / \qty{\PriorsNarrowTrainingWithVisible}{\percent} & \qty{\PriorsFullWithVisible}{\percent} / \qty{\PriorsFullTrainingWithVisible}{\percent} \\
chirp drift above one bin & \qty{\PriorsNarrowDriftOver}{\percent} of sources & \qty{\PriorsFullDriftOver}{\percent} of sources \\
\midrule
training steps & \rev{\numrange{1.41e6}{1.79e6}} & \rev{\numrange{2.11e6}{3.20e6}} \\
\rev{windows seen} & \rev{\numrange{0.9e8}{1.2e8}} & \rev{\numrange{1.4e8}{2.0e8}} \\
\rev{sources seen} & \rev{\numrange{2.3e8}{7.2e8}} & \rev{\numrange{8.2e8}{2.2e9}} \\
simulation share of a step & \qty{\PriorsNarrowDrawShare}{\percent} (\rev{\num{\PriorsNarrowDrawMs}~of~\qty{\PriorsNarrowStepMs}{\milli\second}}) & \qty{\PriorsFullDrawShare}{\percent} (\rev{\num{\PriorsFullDrawMs}~of~\qty{\PriorsFullStepMs}{\milli\second}}) \\
\bottomrule
\end{tabular}
\end{table}

\subsection{Sources per Window}
\label{app:priors-counts}

A window holds each of the \num{\RatesCatalogue} sources of the simulated Galaxy with the prior's share
of $f_0$ in it, so its count is binomial (\cref{eq:gw-rates}); the visible sources follow the same law
thinned by $P(\snr\ge\RatesVisibleSnr\mid f_0)$, with the amplitude integrated out in closed form.
\Cref{tab:narrow-band-bands} gives the counts in each subband. A window of the full run's
\RatesFullWindow{} bins holds up to \num{\RatesFullAllHi} sources and \RatesFullVisibleMax{} visible
ones, near \qty{\RatesFullBusiestMhz}{\milli\hertz}, and one of the narrow run's \RatesNarrowWindow{}
bins \numrange{\RatesNarrowAllLo}{\RatesNarrowAllHi} sources and
\numrange{\RatesNarrowVisibleLo}{\RatesNarrowVisibleMax} visible ones. The $S\le16$ slots of our runs
are therefore far below the source count of a real window at a few millihertz: they test how a joint
learned posterior scales with the number of overlapping sources, not a complete model of the
Galactic foreground.

\subsection{The Training Distribution}
\label{app:priors-training}

There is no fixed dataset: every example is a freshly simulated window, with a centre $\fw$ by
\cref{eq:priors-window}, sources $\theta_{1:S}\sim\pi_w$, noisy data in the window's $4R$ bins whitened
to the conditioning input $y$ (\cref{app:noise}), and the target $X_1 = \varphi(\theta_{1:S};\fw)$ in
flow coordinates, $S\times\PriorsCoords{}$, sources in order of $f_0$ (\cref{eq:chart}). Each window
is paired with \PriorsDraws{} standard normal base points $Z_0$ with their own flow times $t$
(\cref{eq:loss}, \cref{app:fm}), drawn inside the compiled training step, so a batch of
\num{\PriorsBatch} pairs is $\PriorsObservationsPerStep{}$ windows. A run sees
\rev{$\num{0.9e8}$ to $\num{2.0e8}$} independent windows
(\cref{tab:priors-bands}\rev{, \cref{tab:results-runs}}), so overfitting is not a concern and comparisons between runs are
comparisons of optimisation. The simulation takes \qty{\PriorsNarrowDrawShare}{\percent} of a step in
the narrow band and \qty{\PriorsFullDrawShare}{\percent} in the full band.

\section{Flow Coordinates}
\label{app:chart}

This appendix defines the flow coordinates of \cref{sec:method-chart}, the map from a window's
physical sources to $\R^{\ChartCoords{}S}$, and derives the properties the main text relies on:
the four phase copies of a source map to one point, the helicity amplitudes make the likelihood
Gaussian, the coordinates are flat, and the slots in order of $f_0$ remove the labelling. It opens
with the symmetries the chart removes and ends with the slot ordering. The chart depends on the
window through its centre frequency $\fw$, so converting between physical and flow coordinates
always needs the $\fw$ of the observation. \Cref{fig:chart} shows what the chart does to the prior.

\subsection{Exact Symmetries}
\label{app:priors-symmetries}

The likelihood (\cref{eq:loglik}) has two groups of exact symmetries, which a posterior sampler
must either reproduce or remove.

\paragraph{Polarisation and phase.}
A quarter turn of $\psi$ flips the sign of both polarisation tensors, and a half turn of $\phi_0$
flips the sign of both polarisations (\cref{app:forward-model}). Together they leave the signal
unchanged:
\begin{equation}
  (\psi, \phi_0) \;\mapsto\; \bigl(\psi + k\tfrac{\pi}{2},\ \phi_0 + k\pi\bigr)
  \pmod{2\pi},\qquad k = 0,1,2,3,
  \label{eq:psi-phi-symmetry}
\end{equation}
so every source's posterior has four exact copies in $(\psi,\phi_0)\in[0,2\pi)^2$.

\paragraph{Exchangeable sources.}
The sources are independent and identically distributed and enter the likelihood through their
sum, so the likelihood is invariant under the \rev{$S!$ relabellings ($\PriorsPermutations{}$ for $S=\PriorsSources{}$)}: the
label-switching problem of mixture models \citep{stephens2000dealing,jasra2005markov}, which
\ac{mcmc} analyses of gravitational-wave sources meet as well \citep{buscicchio2019label}. With
\cref{eq:psi-phi-symmetry}, every mode of an unordered posterior has
\rev{$S!\,4^S$ equivalent copies, $\num{\PriorsCopies}$ for $S=\PriorsSources{}$}. The
chart folds the phase copies (\cref{app:chart-helicity}), and the slots in order of $f_0$ remove
the permutations (\cref{app:chart-slots}).

\subsection{The Map}
\label{app:chart-map}

For a source with frequency $f_0$, frequency derivative $\fdot$, sky position $(\alpha,\delta)$
and the helicity amplitudes $z_\pm$ of \cref{eq:chart-helicity}, the chart is
\begin{equation}
\begin{aligned}
  x_f &= \frac{2\ln f_0 - \ln f_{\rm lo} - \ln f_{\rm hi}}{\ln f_{\rm hi} - \ln f_{\rm lo}}, &
  x_{\fdot} &= \frac{\log_{10}\fdot - m(f_0)}{\sigma_{\fdot}}, &
  (x_l, x_b) &= \Bigl(\frac{l}{\sigma_l},\,\frac{b}{\sigma_b}\Bigr), \\
  x &= \bigl(x_f,\, x_{\fdot},\, x_l,\, x_b,\, \operatorname{Re}z_+,\, \operatorname{Im}z_+,\,
  \operatorname{Re}z_-,\, \operatorname{Im}z_-\bigr).
\end{aligned}
\label{eq:chart}
\end{equation}
The frequency, drift and sky coordinates are shaped so that their prior has a scale of order one;
the helicity amplitudes are in noise units, so the prior's mostly sub-threshold sources sit close
to their origin (\cref{app:chart-flat}). The chart has no singular or multiply covered point where
the prior has mass.

\paragraph{Frequency.}
$[f_{\rm lo}, f_{\rm hi}]$ is the window interior of \cref{eq:priors-interior}. A window is
narrow against the prior's spread in $\log f_0$ (\cref{app:priors-sources}), so $x_f$ is close to
$\Unif(-1,1)$: its standard deviation is \ChartXfStd{} in both bands, against $1/\sqrt3$. One unit
of $x_f$ spans $(4R-R)/2$ bins, \ChartBinsPerUnitNarrow{} in the narrow band and
\ChartBinsPerUnitFull{} in the full band. The flow always predicts an order-one position inside the
window; the absolute frequency reaches the network through its conditioning on $\fw$
(\cref{app:network}). In single precision, $x_f$ resolves \num{\ChartFThirtyTwoBinsFull} bins near
the window edges in the full band, far below any posterior width.

\paragraph{Drift.}
$\log_{10}\fdot$ is standardised by its conditional mean and standard deviation given $f_0$ under
the catalogue Gaussian of \cref{tab:priors-source}: $m(f_0)$ is linear in $\log_{10}f_0$ with
slope \ChartFdotSlope{}, and $\sigma_{\fdot} = \ChartFdotStd{}$~dex. Under the prior
$x_{\fdot}\sim\mathcal N(0,1)$ exactly, independent of $x_f$; the drift the data see grows with
$f_0$, but the coordinate the flow transports does not.

\paragraph{Sky.}
The chart rotates $(\alpha,\delta)$ to galactic longitude and latitude, $l\in(-\pi,\pi]$ and
$b\in[-\pi/2,\pi/2]$, and divides each by its standard deviation under the sky prior
(\cref{eq:priors-sky}), $\sigma_l = \qty{\ChartSigmaL}{\radian}$ (\ang{\ChartSigmaLDeg}) and
$\sigma_b = \qty{\ChartSigmaB}{\radian}$ (\ang{\ChartSigmaBDeg}). The bulge and \rev{discs} then sit at the
origin with unit scale. The chart's two defects lie where the prior is thinnest: the seam at the
anticentre, $l=\pm\pi$ ($x_l=\pm\ChartSeamX{}$), which holds \qty{\ChartSeamMass}{\percent} of the prior within
\ang{\ChartEdgeDeg}, and the galactic poles, $x_b=\pm\ChartPoleX{}$, which hold \qty{\ChartPoleMass}{\percent}
within \ang{\ChartEdgeDeg}. The inverse rotation maps any real $(l, b)$ to a direction, so the
flow's output never needs wrapping.
Galactic coordinates suit the prior, which is what the flow returns for the large majority of sources,
those below the noise. Ecliptic coordinates would suit the likelihood instead: \ac{lisa} orbits in the
ecliptic, and the second sky mode of a source at moderate \ac{snr}, reflected across the ecliptic, would
be a sign flip of one coordinate. We have not compared the two charts, and since the sky is where the
flow stays widest against the \ac{mcmc}, an ecliptic chart is worth a test.

\paragraph{Amplitude unit.}
\label{app:chart-unit}
The chart measures amplitudes in a unit that depends only on the window,
$u(\fw) = \sqrt{S_n(\fw)/\Tobs}$, with $S_n$ the sky-averaged sensitivity of
\citet[eq.~1]{robson2019construction} without the confusion term,
\begin{equation}
  S_n(f) = \frac{10}{3\,(cL)^2}\left[P_{\rm oms} + \frac{4P_{\rm acc}}{(2\pi f)^4}\right]
  \left[1+0.6\Bigl(\frac{f}{f_\star}\Bigr)^2\right],\qquad f_\star = \frac{1}{2\pi L}.
  \label{eq:noise-sky}
\end{equation}
Their eq.~13 carries $2\bigl(1+\cos^2(f/f_\star)\bigr)$ in place of the 4; over
\qtyrange{\NoiseBandLow}{\NoiseBandHigh}{\milli\hertz} the two differ by at most
\qty{\NoiseRobsonPower}{\percent} in $S_n$ (\qty{\NoiseRobsonAmplitude}{\percent} in amplitude),
because the acceleration term is negligible where they differ. $S_n$ enters nothing but this
unit: the likelihood, the \ac{snr} and the prior all use the full response and
\cref{eq:noise-psd}. For a monochromatic source averaged over sky, polarisation, inclination and
phase, $\E[\snr^2] = C\,\Amp^2\Tobs/S_n(f)$. Rather than take $C$ from a paper with other
normalisation conventions, we measure it with our own $\snr$ on \num{\NoiseCSources} sources:
$C = \num{\NoiseC \pm \NoiseCError}$ for an isotropic sky
(\num{\NoiseCGalactic} for the Galactic sky prior), constant to \qty{\NoiseCFlat}{\percent} below
\qty{\NoiseCFlatBelow}{\milli\hertz} and \qty{\NoiseCRise}{\percent} higher at
\qty{\NoiseCTop}{\milli\hertz}. The unit $u$ is thus the amplitude of a source at a sky-averaged
\ac{snr} of $\sqrt C\approx\NoiseSqrtC{}$; the chart needs a scale that tracks the noise across
the band, not an exact \acs{snr}.

\paragraph{Signal-to-noise per unit amplitude.}
In these units the \ac{snr} of a source is $\snr = \kappa\,\Amp/u(\fw)$, and $\kappa$ is set almost
entirely by the inclination (\cref{fig:chart}b): averaged over the sky,
$\snr^2\propto\lvert\zeta_+\rvert^2+\lvert\zeta_-\rvert^2\propto 1+6\cos^2\iota+\cos^4\iota$, so
\begin{equation}
  \kappa^2 \approx C\,\frac{1+6\cos^2\iota+\cos^4\iota}{16/5},
  \label{eq:chart-kappa}
\end{equation}
where $16/5$ is the average of the numerator over $\cos\iota$ and $C = \rev{\NoiseCGalactic{}}$ is the
mean of $\kappa^2$ over the \rev{Galactic} sky prior. The drawn sources follow \cref{eq:chart-kappa}
within \qty{\ChartKappaLawRms}{\percent} root-mean-square; $\kappa$ runs from \ChartKappaNarrowLow{}
(edge-on) to \ChartKappaNarrowHigh{} (face-on) in the narrow band and from \ChartKappaFullLow{} to
\ChartKappaFullHigh{} in the full band (1st and 99th percentiles), where the response departs from
its low-frequency limit. The $\operatorname{asinh}$ warp of the radius is linear for faint
amplitudes and logarithmic for loud ones, so a source's coordinates stay of order one from the
noise floor to the loudest sources.

\subsection{The Inverse Map}
\label{app:chart-inverse}

The frequency, drift and sky invert directly. For the helicity amplitudes, with
$r_\pm = \sqrt{\sinh\lvert z_\pm\rvert}$,
\begin{equation}
  \sqrt{\Amp/u(\fw)} = r_+ + r_-, \qquad
  \cos\iota = \frac{r_- - r_+}{r_+ + r_-}, \qquad
  \psi = \tfrac14\bigl(\arg z_+ - \arg z_-\bigr), \qquad
  \phi_0^{\rm mid} = \tfrac12\bigl(\arg z_+ + \arg z_-\bigr),
  \label{eq:chart-inverse}
\end{equation}
since $\sqrt{\lvert\zeta_+\rvert}+\sqrt{\lvert\zeta_-\rvert}=\sqrt{\Amp/u}$ for every $\iota$.
\Cref{eq:chart-inverse} returns one of the four copies of \cref{eq:psi-phi-symmetry}, all
equivalent. The inverse is defined everywhere except at $\Amp=0$, where both $z_\pm$ vanish and
the orientation is undefined, a set of measure zero under the prior.

\subsection{Why Helicity Amplitudes}
\label{app:chart-helicity}

\paragraph{Linearity.}
At fixed $(f_0, \fdot, \alpha, \delta)$ the waveform (\cref{app:forward-model}) is a linear
combination of two fixed templates, one per circular polarisation, with complex coefficients
$\zeta_\pm$. The likelihood \cref{eq:loglik} is therefore Gaussian in
$(\operatorname{Re}\zeta_\pm, \operatorname{Im}\zeta_\pm)$: the structure the
$\mathcal F$-statistic maximises over \citep{jaranowski1998data}, written in the
circular-polarisation coordinates of \citet{whelan2014new}. The four extrinsic parameters
$(\Amp, \iota, \psi, \phi_0)$ enter the data only through these four real numbers.

\paragraph{Exact symmetries.}
The map $(\psi,\phi_0)\mapsto(\phi_0^{\rm mid}+2\psi,\ \phi_0^{\rm mid}-2\psi) \bmod 2\pi$ has a kernel
generated by $\psi\mapsto\psi+\pi$ and $(\psi,\phi_0)\mapsto(\psi+\pi/2,\ \phi_0+\pi)$: exactly
the four copies of \cref{eq:psi-phi-symmetry}. All four map to one chart point, so where the
posterior in $(\psi,\phi_0)$ has four equivalent modes, the posterior in the chart has one. This
matters for flow matching beyond economy: with the copies as separate modes, the conditional
mean velocity the network regresses on (\cref{app:fm}) averages over the four copies early in
the path, and the phase information is washed out.

\begin{figure*}[t]
  \centering
  \includegraphics[width=\textwidth]{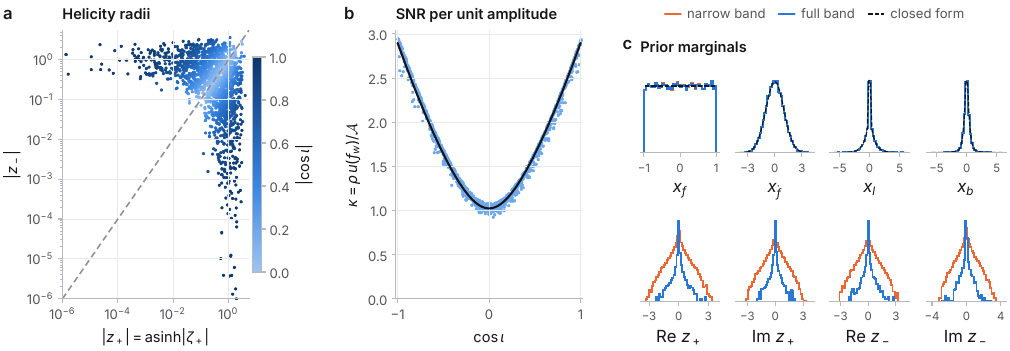}
  \caption{The helicity chart on prior draws. (a)~Narrow-band sources by the radii of their two
  helicity amplitudes, coloured by $\lvert\cos\iota\rvert$. Face-on sources lie far from the
  diagonal, their weak helicity orders of magnitude below the dominant one. (b)~\ac{snr} per unit
  amplitude of the same sources against inclination, with the sky-averaged law
  \cref{eq:chart-kappa} (line). (c)~The prior's marginal in each of the \ChartCoords{} flow
  coordinates, narrow band orange and full band blue, with the full band's closed-form marginal
  dashed where it exists. The $z_\pm$ densities are on a logarithmic axis; their spike at zero is
  the many sources far below one noise unit.}
  \label{fig:chart}
\end{figure*}

\paragraph{Face-on sources.}
At $\cos\iota=\pm1$ the wave is circularly polarised and only $\phi_0\pm2\psi$ is observable.
In the chart this is the point where one of $z_\pm$ vanishes, a regular point of the plane, not
a coordinate pole: a polar chart in $(\Amp, \cos\iota, \phi_0^{\rm mid}\pm2\psi)$ would have an
undefined phase there. Since $\operatorname{asinh}(r)/r$ is an even analytic function of $r$,
the warp $\zeta\mapsto\operatorname{asinh}\lvert\zeta\rvert\,\zeta/\lvert\zeta\rvert$ is a
diffeomorphism of $\mathbb C$, the origin included, with inverse
$z\mapsto\sinh\lvert z\rvert\,z/\lvert z\rvert$. The prior has much of its mass there
(\cref{fig:chart}a,c).

\paragraph{Amplitude and inclination.}
Away from edge-on one helicity dominates,
$\lvert\zeta_-\rvert/\lvert\zeta_+\rvert = \bigl((1+\cos\iota)/(1-\cos\iota)\bigr)^2$, and the
data measure the dominant amplitude much better than the weak one. In $(\Amp,\cos\iota)$ this
gives the familiar curved degeneracy, a distant face-on source against a nearer inclined one.
In the chart the same posterior is a compact blob in the dominant helicity's
plane and a roughly isotropic blob around the origin in the weak one's, whose radius is set by
the noise: the chart straightens the banana by construction.

\paragraph{Phase reference.}
The phase enters the chart referred to mid-observation,
\begin{equation}
  \phi_0^{\rm mid} = \phi_0 - \pi f_0\Tobs - \tfrac{\pi}{4}\fdot\Tobs^2,
  \label{eq:chart-phase}
\end{equation}
with $\fdot$ the source's frequency derivative (\cref{app:forward-model}). The signal phase is
$-\phi_0 + 2\pi f_0 t + \pi\fdot t^2$, so $-\phi_0^{\rm mid}$ is the carrier phase at
$t=\Tobs/2$, the time centroid of a monochromatic signal, which the data fix independently of
$f_0$. Referred to $t=0$ instead, an error of $\delta$ bins in $f_0$ moves the phase by
$\pi\delta$~\unit{\radian}, which couples phase and frequency across the whole posterior. For each
$(f_0, \fdot)$, \cref{eq:chart-phase} is a rotation of the circle, so $\phi_0^{\rm mid}$ is
uniform and independent of the other parameters under the prior, as $\phi_0$ is.

\subsection{Flat Coordinates and Straight Paths}
\label{app:chart-flat}

All \ChartCoords{} coordinates are flat: the flow acts on $\R^{\ChartCoords{}}$ per source, with
no manifold, no angle with a branch cut and no exponential or logarithm map. The flow-matching
path between a base point and a target is the straight line of constant velocity
(\cref{app:fm}), and the sampler's steps stay in the chart whatever the network predicts. The
frame of \cref{app:fm-frame} is an affine map per coordinate, so a straight path in the frame is a
straight path in the chart. Neither training nor sampling evaluates a density of the prior, and
the chart's Jacobian never enters.

The pushed-forward prior is not simple, but it is well scaled (\cref{fig:chart}c): $x_f$ is close
to uniform, the drift and sky coordinates have standard deviations between \ChartUnitStdLow{} and
\ChartUnitStdHigh{}, and no pair of the \ChartCoords{} coordinates is correlated by more than
$\lvert r\rvert = \ChartMaxCorr{}$ over \ChartPriorDraws{} prior sources per band. The helicity
components are the exception: in noise units the prior's sources are faint, and their standard
deviations lie between \ChartZStdLow{} and \ChartZStdHigh{}. The scale that matters for them is the
posterior's, about one noise unit for every source the data resolve. The frame starts from these
prior moments and is trained to the posterior's (\cref{app:fm-frame}), so the flow sees every
coordinate on the posterior's scale, whatever its scale under the prior.

A flat chart of the sphere distorts lengths away from $b=0$, but the prior sits near the galactic
plane, where the distortion is small, and little of its mass lies at the poles or across the cut in
longitude. A Riemannian flow on the product of the sphere and the phase torus would remove both at
the cost of geodesic paths and exponential maps.

\subsection{Sources in Frequency Order}
\label{app:chart-slots}

The $S$ sources of a window are exchangeable under the prior and the likelihood
(\cref{app:priors-symmetries}), so every mode of the posterior has $S!$ copies, one per labelling.
We break the symmetry by an ordering: the slots hold the sources in order of $f_0$, in training
targets and in the frame (\cref{app:fm}), and the flow learns the posterior restricted to
$f_{0,1}\le\dots\le f_{0,S}$, which has one copy of each mode. This is the label-switching remedy
of mixture models \citep{stephens2000dealing,jasra2005markov} and the ordering of
\citet{delmond2026neural}; slots matched to the targets in the loss by the Hungarian algorithm
\citep{houba2025slotflow} would leave the network to learn the labelling as well. Each slot has a
frame of its own, centred on the $k$-th lowest source.

Every ordering has a boundary where two labellings meet. In frequency it falls between sources
close in $f_0$, the overlap that is hardest to resolve: where two sources' $f_0$ posteriors overlap,
the ordered posterior is folded at $f_{0,k} = f_{0,k+1}$, and its marginals per slot are cut there.
The frame takes the moments of the folded marginals, and the flow places the edge. Its draws are
not forced into order, and close sources may swap in a draw's tail; the order matters only for
which slot learns which source.

Scores against the injections still need a labelling. They pair drawn sources with injected ones
by the cheapest assignment over frequency, amplitude and sky, and the calibration tests relabel
the truth and every draw by the same truth-independent rule, each set's own \ac{snr}, loudest
first (\cref{app:evaluation}).

\section{Flow Matching: Objective and Sampler}
\label{app:fm}

This appendix derives the objective and the sampler of \cref{sec:method-fm}: the frame and the
straight paths, the endpoint likelihood and what it optimises, the Gaussian model of one coordinate
that shows why that loss and a frame are needed for narrow posteriors
(\cref{prop:fm-sensitivity}), and the sampler with its cost. \Cref{tab:fm-recipe} lists the choices.
The network is described in \cref{app:network}, the optimiser in \cref{app:training}.

\begin{table*}[t]
  \centering
  \caption{The objective and the sampler at a glance.}
  \label{tab:fm-recipe}
  \small
  \begin{tabular}{@{}lll@{}}
    \toprule
    component & setting & reason \\
    \midrule
    sources & slots in order of $f_0$ & one copy of each mode, no labels to learn \\
    frame & $(\mu,\log\sigma)$ per slot from the data, around the prior's moments & posterior widths of order one in $Z$ \\
    base & $Z_0\sim\Normal(0,I)$ in the frame & matched to the frame \\
    path & straight, $Z_s = Z_0 + s(Z_1 - Z_0)$ & flat coordinates (\cref{app:chart}) \\
    output & endpoint mean $\hat Z_1$ and $\log\tau$ & bounded output, a width per coordinate \\
    loss & Gaussian likelihood of $X_1$ in the frame and of $Z_1$ in the flow & rewards narrowing relative to each width \\
    time & uniform, stratified over the batch & every stratum in every batch \\
    sampler & \FlowMatchingOdeSteps{} midpoint steps, \FlowMatchingEvaluations{} evaluations & data encoded once per window \\
    \bottomrule
  \end{tabular}
\end{table*}

\subsection{Straight Paths in the Frame}
\label{app:fm-paths}

A window with centre frequency $\fw$ holds $S$ sources $\theta_{1:S}$, and its data $d$ enter the
network as the input $y = y(d,\fw)$ (\cref{app:network}). Write $X = \varphi(\theta_{1:S};\fw)\in
\R^{8S}$ for the sources in the flow coordinates of \cref{eq:chart}, listed in order of $f_0$
(\cref{app:chart-slots}). The target is the posterior $p(X\mid y,\fw)\propto p(d\mid
X)\,\pi(X\mid\fw)$ over ordered sources, with the likelihood of \cref{eq:loglik} and the prior of
\cref{app:priors}. Given the frame $(\mu,\sigma)$ of the window (\cref{app:fm-frame}), a function
of $(y,\fw)$ alone, the flow transports standard normal draws $Z(0)$ along
\begin{equation}
  \frac{\dd Z}{\dd s} = \vnet\bigl(Z, s; y, \fw\bigr),\qquad s\in[0,1],
  \label{eq:fm-ode}
\end{equation}
so that $X = \mu + \sigma Z(1)$ is a posterior draw, elementwise. The position $s$ along the path is
the network's time input, written $t$ in \cref{app:network}.

Training uses conditional flow matching \citep{lipman2023flow,liu2023flow,albergo2023building}.
One training observation draws $\fw$ and $\theta_{1:S}$ from the prior, simulates the data
(\cref{app:forward-model,app:noise}) and sets $X_1 = \varphi(\theta_{1:S};\fw)$ and
$Z_1 = (X_1-\mu)/\sigma$ in the frame the network reads off the data. It then draws
\FlowMatchingDraws{} independent base points $Z_0\sim\Normal(0,I)$, each with its own time; a batch
holds \FlowMatchingWindows{} observations, \FlowMatchingBatch{} points in all. On the path of
\cref{eq:fm-path} the target velocity is $u = Z_1 - Z_0$. Regressing a field on $u$ has the marginal
field $v^\star(Z,s;y,\fw) = \E[u\mid Z_s = Z, y,\fw]$ as its minimiser, which carries $\Normal(0,I)$
to the posterior of $Z_1$ \citep{lipman2023flow}. The frame is an affine map per coordinate, so
straight lines in $Z$ are straight lines in $X$, and all eight coordinates are flat
(\cref{app:chart-flat}): the path needs no geodesics. \Ac{fmpe} with a Gaussian base is studied by
\citet{dax2023flow}, and has been applied to \ac{lisa} massive black-hole binaries
\citep{liang2024rapid} and to a single \ac{gb}-like source in gapped data \citep{mao2025robust};
here the Gaussian is placed and scaled by the data.

\subsection{The Frame}
\label{app:fm-frame}

The frame is the network's first output (\cref{app:network-frame}): for each of the $S$ slots and
each coordinate, a shift and a log scale $(\delta_\mu,\delta_s)$ around the prior's moments,
\begin{equation}
  \mu = \mu_\pi + \sigma_\pi\,\delta_\mu,\qquad \log\sigma = \log\sigma_\pi + \delta_s,
  \label{eq:fm-frame}
\end{equation}
with $(\mu_\pi,\sigma_\pi)$ the mean and standard deviation of the slot's coordinate under the prior,
sources in order of $f_0$. For $x_f$ these are the moments of the $k$-th of $S$ uniform order
statistics on $[-1,1]$, mean $2k/(S+1)-1$ and standard deviation
$2\sqrt{k(S+1-k)/((S+1)^2(S+2))}$; $x_{\fdot}$ and the sky are standardised by the chart and have
$(0,1)$; for $z_\pm$ the mean is zero and the variance $\E\lvert z_\pm\rvert^2/2$, by quadrature over
the amplitude prior at the window's centre and the inclination. The frame head starts at zero, so an
untrained network's frame is the prior's.

\paragraph{Loss.}
The frame is trained on the Gaussian negative log-likelihood of $X_1$, up to a constant and a
factor of two, with a weight and a ridge,
\begin{equation}
  \mathcal L_{\rm frame} = \E\;\frac{1}{8S}\sum_{k=1}^{S}\sum_{i=1}^{8}
  \Bigl[\operatorname{sg}\bigl(e^{\delta_{s,ki}}\bigr)\bigl(\kappa_{ki}^2 + 2\delta_{s,ki}\bigr)
  + \lambda\bigl(\delta_{\mu,ki}^2 + \delta_{s,ki}^2\bigr)\Bigr],\qquad
  \kappa = \frac{X_1-\mu}{\sigma},
  \label{eq:fm-frame-loss}
\end{equation}
with $\operatorname{sg}$ the stop-gradient and $\lambda = \FlowMatchingFrameDecay{}$. The frame
depends on $(y,\fw)$ alone, so the weight is the same for every $X_1$ the posterior holds and does
not move the minimiser of each observation's term: up to the ridge, $\mu = \E[X_1\mid y,\fw]$ and
$\sigma^2 = \Var(X_1\mid y,\fw)$, the posterior's marginal moments. What the weight changes is the
balance across observations. The gradient of $\kappa^2$ in $\delta_\mu$ is $-2\kappa\,e^{-\delta_s}$,
which grows as the frame narrows: a loud source, whose frame is many decades narrower than the
prior, would dominate the batch. With the weight it is $-2\kappa$, the same for every source in
standardised units. The curvature in $\delta_\mu$ is still $2e^{-\delta_s}$, so a loud source's
term is far stiffer than a quiet one's, and the optimiser's steps, the same size for every source,
can throw it many widths off. The ridge sits outside the weight and bounds that: as
$e^{\delta_s}\to 0$ its pull on $\delta_s$ stays $2\lambda\delta_s$ while the likelihood's fades, so
the frame's width settles no lower than $\sigma \approx \lambda\ln(1/\lambda)$ prior widths. A loud
source's frame is then wider than its posterior and the flow narrows the rest in $Z_1$; a quiet
source's is unchanged, the ridge being small against its likelihood.

\paragraph{Student-t variant.}
The continuation of the full band with $S=4$ (\cref{app:results-runs}) replaces $\kappa^2$ in
\cref{eq:fm-frame-loss} by the Student-t term $(\nu+1)\ln(1+\kappa^2/\nu)$ with $\nu=4$, weight and
ridge unchanged. Under the weight, the Gaussian term's gradients are $-2\kappa$ in $\delta_\mu$ and
$2e^{\delta_s}(1-\kappa^2)$ in $\delta_s$, unbounded in $\kappa$: one entry whose frame has
collapsed far below its error fills a clipped step on its own. The Student-t's are
$-2(\nu+1)\kappa/(\nu+\kappa^2)$, at most $(\nu+1)/\sqrt\nu$ in size, and
$2e^{\delta_s}\bigl[1-(\nu+1)\kappa^2/(\nu+\kappa^2)\bigr]$, at most $2\nu e^{\delta_s}$. The
Student-t is a scale mixture of Gaussians, and its expected precision weight given the residual,
$(\nu+1)/(\nu+\kappa^2)$, discounts an outlier instead of following it. The frame is then no longer
the moment match: its scale lies below the posterior's standard deviation, and the flow, which
reads it through the stop-gradient, takes up the difference in $Z_1$. The logged frame loss stays
the Gaussian one, so it does not compare across the switch.

\subsection{The Endpoint Distribution and the Loss}
\label{app:fm-loss}

For each source and coordinate the flow outputs the mean $\hat Z_1$ and the log standard deviation
$\log\tau$ of a Gaussian over the endpoint, given the point on the path, and its velocity is
\cref{eq:fm-velocity}. The output is a position in the frame, whose scale does not change along the
path, while the velocity grows as $1/(1-s)$ near the end (\cref{app:fm-sensitivity}); endpoint
prediction converted to a velocity is also used on structured data \citep{stark2024harmonic}. The
width $\tau$ serves the loss only; the sampler follows the mean.

The second term of \cref{eq:loss} is minimised by the conditional mean and variance of each
coordinate, $\hat Z_1 = \E[Z_1\mid Z_s,y,\fw]$ and $\tau^2 = \Var(Z_1\mid Z_s,y,\fw)$, so the
velocity of \cref{eq:fm-velocity} is the marginal field $v^\star$. On the straight path
$Z_1 - Z_s = (1-s)\,u$, so $\hat Z_1 - Z_1 = (1-s)(\vnet - u)$ and
$\Var(Z_1\mid Z_s) = (1-s)^2\Var(u\mid Z_s)$. At the optimal $\tau$ each term is therefore, up to
a constant, the ratio of \cref{eq:fm-relative}: the factors of $1-s$ cancel, so a source the data pin
down weighs as much as one whose posterior is its prior, with no weights over time or coordinates.
The learned $\tau$ acts as the learned task uncertainties of \citet{kendall2018multi}, one per point
rather than one per task. The frame's loss, \cref{eq:fm-frame-loss}, is the same Gaussian likelihood
at $s=0$.

\subsection{Stratified Time}
\label{app:fm-clock}

Each point's time is $s\sim\Unif(0,1)$, stratified over the $B=\FlowMatchingBatch{}$ points of a
batch: with $\pi$ a random permutation, point $i$ gets $s_i = (\pi_i + U_i)/B$,
$U_i\sim\Unif(0,1)$. Every stratum of $s$ is then covered in every batch, which lowers the
variance of the batch loss at no cost, as the low-discrepancy sampler of variational diffusion
models does \citep{kingma2021variational}. The network sees $s$ and $\log_{10}(1-s)$, the latter
clipped to $[\FlowMatchingTimeDecades{},0]$ (\cref{app:network}), so it can tell apart times that
differ only in how close they are to the end.

Non-uniform time sampling is standard for diffusion models
\citep{karras2022elucidating,kingma2023understanding,esser2024scaling}, and \ac{fmpe} up-weights late
times with a power law in $t$ \citep[Sec.~3.3]{dax2023flow}. Here time stays uniform: under
\cref{eq:loss} the reward for narrowing a posterior does not fade with its width and is collected
along the whole path (\cref{app:fm-sensitivity}), so the end of the path needs no extra share of
the batch.

\subsection{What the Loss and the Sampler See of a Narrow Posterior}
\label{app:fm-sensitivity}

The posteriors of loud sources are orders of magnitude narrower than the prior in $x_f$ and
$x_{\fdot}$. A Gaussian model of one coordinate shows how much a loss rewards getting such a width
right and how well a sampler resolves it. The velocity loss of standard flow matching hardly
rewards it, the endpoint likelihood of \cref{eq:loss} does at every width, and a uniform sampler
grid resolves it only once the frame has divided it out.

\paragraph{Model.}
Take one coordinate with base $x_0\sim\Normal(0,s_0^2)$, posterior $x_1\sim\Normal(0,w^2)$, the
straight path $x_s = (1-s)x_0 + s x_1$ and the target $u = x_1 - x_0$. With $\epsilon=1-s$ the
remaining path, the marginal is $x_s\sim\Normal(0,m^2)$ with $m^2 = \epsilon^2s_0^2 + s^2w^2$, and
the optimal field is linear,
\begin{equation}
  v^\star(x,s) = g(s)\,x,\qquad
  g(s) = \frac{s\,w^2 - \epsilon\,s_0^2}{m^2},\qquad
  V \equiv \Var(u\mid x_s) = \frac{s_0^2w^2}{m^2}.
\end{equation}
Without a frame, $s_0$ is the prior's width; in the frame, $s_0=1$ and $w$ is the posterior's width
over the frame's. A model that takes the posterior to be the wider $\Normal(0,w_m^2)$ has the field
$g_m(s)\,x$ and the conditional variance $V_m$, with $w\to w_m$ in $g$, $m$ and $V$. Under uniform
time the velocity loss pays, by the orthogonality of the regression residual, an excess over its
irreducible part $V$ of
\begin{equation}
  \Delta L_{\rm vel}(w_m;w) = \int_0^1 m^2\,\bigl[g_m - g\bigr]^2\,\dd\epsilon ,
  \label{eq:fm-excess}
\end{equation}
and the flow's term of \cref{eq:loss}, whose minimiser is $(g, V)$ by \cref{eq:fm-relative}, pays
\begin{equation}
  \Delta L_{\rm nll}(w_m;w) = \int_0^1 \Bigl[\frac{m^2(g_m-g)^2}{V_m}
  + \frac{V}{V_m} - 1 - \ln\frac{V}{V_m}\Bigr]\dd\epsilon .
  \label{eq:fm-excess-nll}
\end{equation}

\begin{proposition}
\label{prop:fm-sensitivity}
For $w\ll w_m\ll s_0$,
\begin{equation}
  \Delta L_{\rm vel} \simeq \frac{\pi}{4}\,s_0\,w_m ,
  \label{eq:fm-sensitivity-uniform}
\end{equation}
to leading order independent of the true width $w$, and for $w,w_m\ll s_0$
\begin{equation}
  \Delta L_{\rm nll} \simeq \ln\frac{w_m^2}{w^2} + \frac{w^2}{w_m^2} - 1
  = 2\,\mathrm{KL}\bigl(\Normal(0,w^2)\,\big\|\,\Normal(0,w_m^2)\bigr),
  \label{eq:fm-sensitivity-nll}
\end{equation}
which depends on the widths only through $w_m/w$.
\end{proposition}

\begin{proof}
Set $s_0=1$. Where $\epsilon\gg w$, $m^2\simeq\epsilon^2$ and $g\simeq-1/\epsilon$, so up to a
relative $O(\epsilon)$
\begin{equation*}
  m^2(g_m-g)^2 \simeq \frac{w_m^4}{(\epsilon^2+w_m^2)^2},
\end{equation*}
and the region $\epsilon\lesssim w$ contributes a relative $O(w/w_m)$. With $\epsilon=w_m z$,
\cref{eq:fm-excess} is $w_m\int_0^\infty(1+z^2)^{-2}\dd z = \frac{\pi}{4}w_m$. In
\cref{eq:fm-excess-nll}, $V_m\simeq w_m^2/(\epsilon^2+w_m^2)$, so the first term is $\simeq
w_m^2/(\epsilon^2+w_m^2)$ and integrates to $\frac{\pi}{2}w_m$. Where $\epsilon\gg w_m$,
$V/V_m\simeq w^2/w_m^2$ up to a relative $O(w_m^2/\epsilon^2)$, and the rest of the integrand is the
constant \cref{eq:fm-sensitivity-nll}; the region $\epsilon\lesssim w_m$ changes the integral by
$O(w_m\ln(w_m/w))$. The factors of $s_0$ follow from dimensions: $\Delta L_{\rm vel}$ scales as
$s_0^2$ and depends on $w_m/s_0$ alone, $\Delta L_{\rm nll}$ is dimensionless.
\end{proof}

\begin{figure*}[t]
  \centering
  \includegraphics[width=\textwidth]{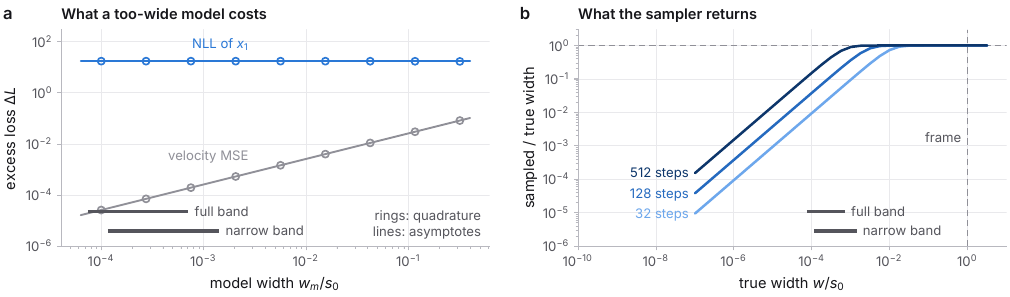}
  \caption{The Gaussian model of one coordinate. The bars mark the ideal $f_0$ widths $w/s_0$ of
  each band over its \ac{snr} range, with $s_0$ the prior's width. (a)~Excess loss of a model $w_m$
  wide when the true posterior is $10^4$ times narrower, under the velocity loss (grey) and the
  endpoint likelihood of \cref{eq:loss} (blue): quadrature of \cref{eq:fm-excess,eq:fm-excess-nll}
  (rings) against \cref{eq:fm-sensitivity-uniform,eq:fm-sensitivity-nll} (lines). The velocity
  loss's reward for narrowing vanishes linearly with the width; the likelihood's stays the same at
  every width. (b)~Width the sampler of \cref{app:fm-sampling} returns, run on the exact mean
  $\E[x_1\mid x_s]$, over the true width. With \FlowMatchingOdeSteps{} steps from the prior's scale
  it collapses well inside the narrow band's range and over all of the full band's, and more steps
  only move the collapse to narrower widths (darker with more steps); in a calibrated frame the width is $s_0$ (dashed line), where
  the sampler is exact.}
  \label{fig:fm-sensitivity}
\end{figure*}

Quadrature of \cref{eq:fm-excess,eq:fm-excess-nll} confirms both: at $w_m = 10^{-3}s_0$ and a true
posterior $10^4$ times narrower, the velocity loss's excess is \FlowMatchingCheckUniform{} of
\cref{eq:fm-sensitivity-uniform} and the likelihood's \FlowMatchingCheckNll{} of
\cref{eq:fm-sensitivity-nll} (\cref{fig:fm-sensitivity}a).

\paragraph{Consequence for the loss.}
Under the velocity loss the excess is linear in the model's width: halving a width of
$10^{-3}s_0$ is worth a hundredth of halving one of $10^{-1}s_0$. Nothing in the objective grows as
a narrow posterior approaches its true width, and the reward for the last decades of narrowing is
far below the scatter of a logged loss. Reweighting time towards the end of the path raises it, but
only down to the latest time the weighting reaches. The likelihood instead pays $2$ for every e-fold
by which the model is too wide, whatever the true width, until $w_m$ nears $w$; a model too narrow
pays more still, through $w^2/w_m^2$. The reward is collected along the whole path rather than at
its end: wherever $\epsilon\gg w_m/s_0$, $V$ is the posterior variance itself, and the loss compares
the model's width with it directly. Because $\tau$ is predicted per point, the comparison is made
for each source separately, not against a mean pooled over the batch.

\paragraph{Where the bands sit.}
The $x_f$ coordinate is uniform on $[-1,1]$ under the prior, so without a frame $s_0 = 1/\sqrt3$,
and one unit of $x_f$ spans \FlowMatchingNarrowBinsPerUnit{} frequency bins in the narrow band and
\FlowMatchingFullBinsPerUnit{} in the full band (\cref{app:chart}). For a sinusoid of unknown
amplitude and phase the Fisher bound on the frequency is $\sqrt3/(\pi\snr)$ bins. Over the visible
sources, from \ac{snr} \FlowMatchingSnrLow{} to the prior's \FlowMatchingSnrQuantile{}\% quantile
(\FlowMatchingNarrowSnrHigh{} in the narrow band, \FlowMatchingFullSnrHigh{} in the full band),
this gives an ideal $w/s_0$ from \num{\FlowMatchingNarrowLoud} to \num{\FlowMatchingNarrowWeak} in
the narrow band and from \num{\FlowMatchingFullLoud} to \num{\FlowMatchingFullWeak} in the full band
(bars in \cref{fig:fm-sensitivity}). A frame that matches the posterior's width brings every one of
these to $w/s_0\approx1$, and the flow's term then sees order-one widths whatever the source's
\ac{snr}; the decades the frame does not remove are left to the likelihood's relative reward.

\paragraph{Consequence for the sampler.}
The time grid is uniform, so the last step starts at $1-s = 1/K_t$ and evaluates its midpoint at
$1-s = \FlowMatchingLastMid{}$, where the velocity carries the draw to $2\hat Z_1' - \hat Z_1$, with
$\hat Z_1$ and $\hat Z_1'$ the means at the start and the midpoint. Where a posterior is much
narrower than $s_0/2K_t$, the exact means at both points hardly vary across draws: they are close to
the posterior mean, and the last step puts every draw near it. Run from the prior's scale on the
exact Gaussian mean, the sampler returns \FlowMatchingWidthWide{} of the true width at $w = 0.1\,s_0$,
but \FlowMatchingWidthNarrowWeak{} of it at the narrow band's ideal width for \ac{snr}
\FlowMatchingSnrLow{} (\FlowMatchingWidthNarrowWeakMore{} with \FlowMatchingOdeStepsMore{} steps,
\FlowMatchingWidthNarrowWeakMost{} with \FlowMatchingOdeStepsMost{}) and
\num{\FlowMatchingWidthFullWeak} at the full band's (\FlowMatchingWidthFullWeakMost{} with
\FlowMatchingOdeStepsMost{} steps; \cref{fig:fm-sensitivity}b). In a frame of the posterior's width,
$w = s_0$, it returns \FlowMatchingWidthFrame{} of it with \FlowMatchingOdeSteps{} steps: the frame
removes the collapse, as long as it is not decades wider than the posterior. A trained frame and
flow are not exact, so the widths of trained models are measured instead; they no longer change
with the number of steps at \FlowMatchingOdeSteps{} (\cref{app:evaluation-sampler}).

\subsection{Sampling}
\label{app:fm-sampling}

To sample a window, the data are encoded once and the frame $(\mu,\sigma)$ read off; $N$ base draws
$Z\sim\Normal(0,I)$ are carried over $K_t = \FlowMatchingOdeSteps{}$ uniform midpoint steps of
\cref{eq:fm-ode} with the velocity of \cref{eq:fm-velocity}, and mapped back with $X = \mu + \sigma
Z$ and $\varphi^{-1}$ (\cref{alg:fm-sample}). A step from $s$ to $s+\Delta s$ evaluates the velocity
at its start, takes a half step to the midpoint and takes the full step with the velocity there, so
a draw costs $2K_t=\FlowMatchingEvaluations{}$ network evaluations. The data stream of the network
depends only on $(y,\fw)$, so it serves every draw and every step (\cref{app:network}). The draws
keep the order of the slots but are not forced into it: two close sources may swap, as they do in
the posterior's tail across the boundary of the ordered region (\cref{app:chart-slots}).

\begin{figure}[t]
  \centering
  \includegraphics{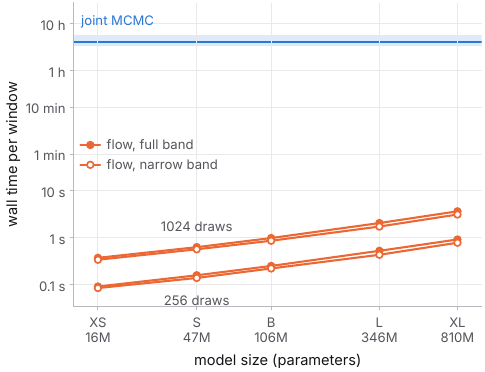}
  \caption{Wall time to draw posterior samples of one window, against model size: the flow on
  one A100 (orange; full band filled markers, narrow band hollow; best of three windows), and the joint
  \ac{mcmc} reference over all sources of a window on \McmcJointCores{} CPU cores (blue; median
  and range over the chains, one per injection). \rev{Measured on an earlier revision of the network, whose sizes had the parameter counts on the
  axis; \cref{tab:network-sizes} gives those of the reported network.}}
  \label{fig:fm-cost}
\end{figure}

\paragraph{Cost.}
On one A100, \FlowMatchingCostDraws{} draws of one window take \qty{\FlowMatchingCostNarrow}{\second} with the
narrow-band B model and \qty{\FlowMatchingCostFull}{\second} with the full-band one, and
\FlowMatchingCostDrawsMore{} draws take \qty{\FlowMatchingCostFullMore}{\second}: the cost is linear in the
draws and almost independent of the band, since the network evaluations dominate and the data are
encoded once (\cref{fig:fm-cost}). Across sizes, \FlowMatchingCostDraws{} draws take
\qty{\FlowMatchingCostSmallest}{\second} (XS) to \qty{\FlowMatchingCostLargest}{\second} (full-band XL). The \ac{mcmc}
reference (\cref{app:mcmc}) runs one chain per injection over all the sources of its window, and a
chain takes \rev{\qtyrange{\McmcCostJointLo}{\McmcCostJointHi}{\hour} on two CPU cores in the narrow band and about
\qtyrange{\McmcCostFullSixteenLo}{\McmcCostFullSixteenHi}{\hour} in the full band with sixteen sources
(\cref{app:mcmc-cost}). The flow is about $10^{4.8}$ to $10^{5.8}$ times faster per window.} The hardware
differs, so the ratio is a wall-time comparison, not one of compute; and the flow's cost is paid
once more in training (\cref{app:training}).

\begin{algorithm}[t]
  \caption{One training step, written for one observation; the batch holds
  \FlowMatchingWindows{} observations of \FlowMatchingDraws{} points each.}
  \label{alg:fm-train}
  \begin{algorithmic}
    \STATE {\bfseries Input:} network weights $\eta$
    \STATE draw $\fw$ and $\theta_{1:S}\sim\pi(\cdot\mid\fw)$ (\cref{app:priors})
    \STATE simulate the data $d$ and form $y = y(d,\fw)$ (\cref{app:forward-model,app:noise})
    \STATE $X_1\leftarrow\varphi(\theta_{1:S};\fw)$, sources in order of $f_0$
    \STATE encode $(y,\fw)$ once: frame $(\mu,\sigma)$ \hfill (\cref{eq:fm-frame})
    \STATE $Z_1\leftarrow\operatorname{sg}\bigl((X_1-\mu)/\sigma\bigr)$
    \FOR{each of the \FlowMatchingDraws{} points}
      \STATE $Z_0\sim\Normal(0,I)$;\ \ $s$ from its stratum
      \STATE $Z_s\leftarrow Z_0 + s(Z_1 - Z_0)$ \hfill (\cref{eq:fm-path})
      \STATE $(\hat Z_1,\log\tau)\leftarrow\mathrm{net}_\eta\bigl(Z_s, s; \operatorname{sg}(\mu,\sigma), y, \fw\bigr)$
    \ENDFOR
    \STATE take an optimiser step on \cref{eq:loss} in $\eta$ (\cref{app:training})
  \end{algorithmic}
\end{algorithm}

\begin{algorithm}[t]
  \caption{Sampling the posterior of one window.}
  \label{alg:fm-sample}
  \begin{algorithmic}
    \STATE {\bfseries Input:} the input $y$ of a window $\fw$, number of draws $N$, steps $K_t=\FlowMatchingOdeSteps{}$
    \STATE encode $(y,\fw)$ once: frame $(\mu,\sigma)$;\ \ $\Delta s\leftarrow1/K_t$
    \STATE $\vnet(Z,s) = \bigl(\hat Z_{1,\eta}(Z, s; y, \fw) - Z\bigr)/(1-s)$ \hfill (\cref{eq:fm-velocity})
    \FOR{$n=1$ {\bfseries to} $N$ (in parallel)}
      \STATE $Z\sim\Normal(0,I)$
      \FOR{$j=0$ {\bfseries to} $K_t-1$}
        \STATE $s\leftarrow j\Delta s$;\ \ $Z'\leftarrow Z + \tfrac{\Delta s}{2}\,\vnet(Z, s)$
        \STATE $Z\leftarrow Z + \Delta s\,\vnet\bigl(Z', s+\tfrac{\Delta s}{2}\bigr)$ \hfill (\cref{eq:fm-ode})
      \ENDFOR
      \STATE {\bfseries output} $\varphi^{-1}(\mu + \sigma Z;\fw)$
    \ENDFOR
  \end{algorithmic}
\end{algorithm}

\section{Network}
\label{app:network}

This appendix specifies the network of \cref{sec:method-network} well enough to rebuild it: the
forward pass in the order of the data flow, then the sizes. The network is a one-way multi-stream
transformer, the \ac{mmdit} of \citet{esser2024scaling} with the adaptive-normalisation
conditioning of \citet{peebles2023scalable}. An encoder runs the data tokens and one frame register
per slot, conditioned on the window alone, and returns the frame of \cref{app:fm}, a mean and a log
standard deviation per coordinate of each source slot. A decoder runs the source tokens and
\NetworkSinksWord{} sink tokens, conditioned on the flow time and the window, and returns the mean
and log standard deviation of the endpoint $Z_1$ in that frame. \Cref{fig:network-architecture}
shows the layout, \cref{tab:network-embeddings} what each token is built from and
\cref{tab:network-sizes} the sizes.

\paragraph{Notation.}
All feed-forward maps are SwiGLU \citep{shazeer2020glu}, $\mathrm{FF}(h) =
W_2\bigl(a\odot\mathrm{SiLU}(g)\bigr)$ with $[a;g] = W_1h$. A scalar $\xi$ enters as sinusoids
$E(\xi) = \bigl[\cos(2\pi\xi/P_k);\ \sin(2\pi\xi/P_k)\bigr]_k$ over a set of periods $P_k$,
followed by an $\mathrm{FF}$. The absolute frequencies of the data tokens, the sources and the
window share one set of periods in hertz, octaves $P_k = 2^kP_0$ from $P_0 =
\NetworkFreqFinest{}$ of a bin ($1/\Tobs$) up to past twice the top bin of the band. The fine
periods give the sub-bin resolution that posteriors this narrow require. The turns $\xi/P_k$ are
reduced modulo one in float64 before the cast, so that even the finest period keeps its phase.
Every embedding runs in float32, whatever the precision of the backbone
(\cref{app:network-sizes}).

\begin{table*}[t]
  \centering
  \caption{What each token is built from. Frequencies use sinusoids with periods octave-spaced in
  hertz from \NetworkFreqFinest{} of a bin up to past twice the top bin of the band (period counts
  for the narrow and the full band).}
  \label{tab:network-embeddings}
  \small
  \begin{tabular}{@{}lll@{}}
    \toprule
    input & embedding & numbers or periods \\
    \midrule
    \multicolumn{3}{@{}l}{\emph{data token}} \\
    amplitudes and phase parts & linear & \NetworkTokenScalars{} numbers \\
    lowest and highest frequency & sinusoidal, $\mathrm{FF}_y$ & \NetworkFreqPeriodsNarrow{} / \NetworkFreqPeriodsFull{} periods each \\
    \midrule
    \multicolumn{3}{@{}l}{\emph{frame register}} \\
    slot & learned & one vector per slot, plus $\nu_f$ \\
    \midrule
    \multicolumn{3}{@{}l}{\emph{source token}} \\
    $Z$, $X$, harmonics, $\mu$, $\log\sigma$ & linear & \NetworkSourceScalars{} numbers \\
    $f_0$ & sinusoidal, $\mathrm{FF}_{f_0}$ & \NetworkFreqPeriodsNarrow{} / \NetworkFreqPeriodsFull{} periods \\
    \midrule
    \multicolumn{3}{@{}l}{\emph{sink token}} \\
    sink index & learned & one vector per sink, plus $\nu$ \\
    \midrule
    \multicolumn{3}{@{}l}{\emph{conditioning}} \\
    $t$, $\log_{10}(1-t)$ & sinusoidal, $\mathrm{FF}_\nu$ & \NetworkSinusoidFreqs{} over \NetworkSinusoidDecades{} decades each \\
    lowest and highest $f_0$ of the window & sinusoidal, $\mathrm{FF}_f$ & \NetworkFreqPeriodsNarrow{} / \NetworkFreqPeriodsFull{} periods each \\
    \bottomrule
  \end{tabular}
\end{table*}

\subsection{Data Tokens}
\label{app:network-input}

\paragraph{Whitened input.}
The input $y$ is the window whitened as in \cref{sec:gw-windows}: the real and imaginary parts of
$A$, $E$ and $T$ in each of the $K$ bins from the window start $k_0$ (\cref{app:forward-model}).
Below, $y_{k,c}$ is the complex whitened value of bin $k = 0,\dots,K-1$ and channel $c\in\{A, E,
T\}$. Under noise alone its real and imaginary parts have unit variance, and a source's whitened
signal $\hat h$ has $\sum_{k,c}|\hat h_{k,c}|^2 = \snr^2$ (\cref{eq:inner}).

\paragraph{Short-time spectra.}
The window is cut into bands of $Q=\NetworkPatch{}$ consecutive bins, one token each:
\NetworkTokensNarrow{} tokens on the narrow band ($K=\NetworkBinsNarrow{}$) and
\NetworkTokensFull{} on the full band ($K=\NetworkBinsFull{}$). An orthonormal inverse discrete
Fourier transform over each band turns its spectrum into $Q$ time samples of the band's signal, per
channel,
\begin{equation}
  \check y_{n,j,c} = \frac{1}{\sqrt Q}\sum_{l=0}^{Q-1} y_{nQ+l,\,c}\,e^{2\pi\ii jl/Q},
  \qquad j = 0,\dots,Q-1,
  \label{eq:network-stft}
\end{equation}
an \ac{stft} with the tokens as its bands. It is critically sampled and untapered, so it is
invertible, keeps the noise white, and neither leaks nor duplicates power between tokens, as an
overlapping tapered transform would. Within a token, a source's sub-bin frequency becomes a phase
slope over the samples, its initial phase the intercept, and the annual Doppler and the amplitude
modulation that carry its sky position a slow evolution in time.

\paragraph{Compression.}
The transform is unitary, so a source's samples carry the same $\snr^2$ as its bins, and a single
sample of its signal has $|\check h| \le \snr$, which reaches \NetworkSnrTop{} at the prior's
\qty{\NetworkSnrQuantile}{\percent} quantile in the full band (\cref{app:priors}), against
unit-variance noise. Each complex sample therefore enters as its compressed amplitude
$\operatorname{asinh}|\check y|$, linear in the noise and logarithmic for loud signals, and its
gated phase $\check y/\sqrt{1+|\check y|^2}$, which is $\check y$ in the noise and a unit phasor
for a loud signal. A token is $3Q\times3 = \NetworkTokenScalars{}$ numbers:
\NetworkAmplitudeScalars{} amplitudes and \NetworkPhaseScalars{} phase parts, mapped linearly to
$\R^D$.

\paragraph{Absolute frequency.}
Whitening and windowing remove the absolute frequency, which still matters: the Doppler width
scales with $f$ and the prior's typical drift $\fdot$ with $f_0^{\ChartFdotSlope}$
(\cref{app:chart}). Each data token therefore receives the absolute frequencies of its lowest and
highest bin, $(k_0 + nQ)/\Tobs$ and $(k_0 + nQ + Q - 1)/\Tobs$, through $\mathrm{FF}_y$.

\subsection{Frame Registers and Frame Head}
\label{app:network-frame}

The $S$ frame registers are tokens that run beside the data tokens in every block, one per source
slot, each a learned slot embedding plus $\nu_f$. They attend to the data and to each other, and
the data attend to them; they are registers in the sense of \citet{darcet2024vision}, tokens that
hold global state, here one per source. The slots are the sources in order of $f_0$
(\cref{app:chart-slots}), so register $k$ learns to frame the $k$-th lowest source.

After the last block, a final layer normalisation with a shift, scale and gate from $\nu_f$ and a
linear map $\R^D\to\R^{\NetworkHeadOutputs{}}$ without bias give each slot the shift and log scale
$(\delta_\mu,\delta_s)$ of its frame, per coordinate (\cref{eq:fm-frame}). The modulation and the
gate are zero-initialised, so a fresh network's frame is the prior's moments. The frame costs almost
nothing: the registers add $S$ tokens to the data tokens, and the head is a small part of the
encoder.

\subsection{Source Tokens and Sinks}
\label{app:network-sources}

Each source token carries \NetworkSourceScalars{} scalars, mapped linearly to $\R^D$:
\begin{itemize}
  \item its \NetworkCoordsWord{} coordinates in the frame, $Z$, and in the chart, $X = \mu +
  \sigma Z$ (\cref{app:chart});
  \item the higher harmonics of its two helicity amplitudes in $X$, $r\cos k\phi$ and $r\sin
  k\phi$ for $k=2,\dots,\NetworkHarmonics{}$, with $(r,\phi)$ the polar form of each helicity pair
  (\NetworkHarmonicScalars{} scalars), which give the network functions of the phases that vanish
  with the amplitude;
  \item its slot's frame, $\mu$ and $\log\sigma$, through a stop-gradient, so that the flow cannot
  train the frame (\cref{app:fm-frame}).
\end{itemize}
To these the token adds a sinusoidal embedding of its absolute frequency $f_0$, obtained from $x_f$
through the inverse chart in float64 before any cast. A source token carries no label of its slot:
it reads its frame through $\mu$ and $\log\sigma$ and its slot's keys and values in the cache, and
every map applied to a source token is shared across sources, as in set transformers
\citep{zaheer2017deep,lee2019set}.

The \NetworkSinksWord{} sink tokens, each a learned embedding plus $\nu$, join the source stream
with their own maps. They give the sources' attention somewhere to go other than the data, and since
they start from $\nu$ they can carry per-draw state \citep{xiao2024efficient,darcet2024vision}.
They are neither cached nor read out.

\subsection{One-Way Blocks}
\label{app:network-blocks}

Each of the $N$ blocks updates four streams: the data tokens $h^y$ and the frame registers $h^r$,
conditioned on $\nu_f$, and the source tokens $h^x$ and the sink tokens $h^s$, conditioned on $\nu$
(\cref{app:network-conditioning}). For each stream $\ell$, a zero-initialised linear map of
$\mathrm{SiLU}(\nu_f)$ or $\mathrm{SiLU}(\nu)$ gives a shift, a scale and a gate
$(\beta^\ell,\gamma^\ell,\varpi^\ell)$, so that every residual branch starts as the identity
\citep{peebles2023scalable}:
\begin{align}
  u^\ell &= \mathrm{LN}(h^\ell)\odot(1+\gamma^\ell) + \beta^\ell,\qquad
  [q^\ell; k^\ell; v^\ell] = \bigl[W^\ell_{q}u^\ell; W^\ell_{k}u^\ell; W^\ell_{v}u^\ell\bigr], \nonumber\\
  [o^y; o^r] &= \mathrm{Attn}\bigl([q^y; q^r];\ [k^y; k^r], [v^y; v^r]\bigr),\nonumber\\
  [o^x; o^s] &= \mathrm{Attn}\bigl([q^x; q^s];\ [k^x; k^s; k^y; k^r], [v^x; v^s; v^y; v^r]\bigr), \nonumber\\
  h^\ell &\leftarrow h^\ell + \varpi^\ell\odot W^\ell_{o}\,o^\ell, \nonumber\\
  h^\ell &\leftarrow h^\ell + \varpi'^\ell\odot\mathrm{FF}^\ell\bigl(\mathrm{LN}(h^\ell)\odot(1+\gamma'^\ell) + \beta'^\ell\bigr),
  \label{eq:network-block}
\end{align}
with a second shift, scale and gate for the feed-forward branch, $\mathrm{LN}$ a layer
normalisation without learned affine parameters, and feed-forward hidden width $\NetworkExpand{}D$.
Each stream has its own maps and there are no biases inside the blocks. The query, key and value
maps are separate matrices, so that the optimiser orthogonalises each on its own
(\cref{app:training-optimiser}). Activations are rematerialised in the backward pass
\citep{chen2016training}.

\paragraph{Cache.}
The data and the registers never read the sources, the sinks or $t$, so their keys and values in
every block and the registers' outputs depend on $(y,\fw)$ alone. One encoding of an observation
therefore gives its frame and serves all \NetworkDraws{} base draws that share it in training, and
in sampling all \rev{$2K_t=\FlowMatchingEvaluations{}$} evaluations along every draw's path (\cref{app:fm-sampling}).
In size B, encoding a full-band window takes \NetworkEncodeGflopFull{} GFLOP of matrix products,
\NetworkEncodeOverDecodeFull{} times one evaluation on the cache (\NetworkDecodeGflopFull{} GFLOP),
and the cache holds \qty{\NetworkCacheMibFull}{\mebi\byte} in bfloat16; a narrow window takes
\NetworkEncodeGflopNarrow{} and \NetworkDecodeGflopNarrow{} GFLOP, with
\qty{\NetworkCacheMibNarrow}{\mebi\byte} of cache. The last block's data and sink outputs reach
nothing, so it omits their output projections, feed-forward modulations and feed-forward maps; its
register outputs feed the frame head.

\paragraph{Attention.}
Attention has $H$ heads of dimension \NetworkHeadDim{}. Queries and keys are normalised by their
\ac{rms} per head, and the queries carry a learned per-head gain initialised to one
\citep{henry2020query,dehghani2023scaling}. The source and sink queries take one softmax over their
own keys and the cached keys, computed in parts that are never concatenated in memory, since
concatenation would copy the shared cache once per draw.

\subsection{Conditioning}
\label{app:network-conditioning}

The data stream, the registers and the frame head are conditioned on the window alone,
\begin{equation}
  \nu_f = \mathrm{FF}_f\bigl(\bigl[E(f_{\rm lo});\ E(f_{\rm hi})\bigr]\bigr),
\end{equation}
with $f_{\rm lo}$ and $f_{\rm hi}$ in hertz the lowest and highest $f_0$ a source of the window can
have, the prior's range cut to the window less its guard bands (\cref{app:forward-model}). The
source stream, the sinks and the endpoint head are conditioned on
\begin{equation}
  \nu = \mathrm{FF}_\nu\Bigl(\bigl[E(t);\ E\bigl(\max(\log_{10}(1-t), -\NetworkTimeDecades{}) +
  \NetworkTimeDecades{}\bigr)\bigr]\Bigr) + \nu_f,
  \label{eq:network-c}
\end{equation}
where $t$ is the path position $s$ of \cref{app:fm} and each time scalar's periods are geometric
down from twice its span. The second input resolves the last decades of the path, where the fine
structure of a posterior forms (\cref{app:fm-sensitivity}).

\subsection{Positions}
\label{app:network-positions}

The keys, queries and values of the data and source tokens are rotated by frequency
\citep{su2024roformer,miyato2024gta}: each of the \NetworkRopePairs{} pairs of a head's dimensions
turns by $2\pi f/P_j$, with periods $P_j = 2K^{j/(\NetworkRopePairs{}-1)}/T$ geometric from 2 bins
to twice the window. A score then depends on how far apart its two tokens are in frequency, and the
weighted sum of rotated values is turned back by the query's own angle, so that a value also
arrives relative to its query. A data token's position $f$ is its centre frequency and a source's
its $f_0$, absolute in hertz, with the turns $f/P_j$ reduced modulo one in float64 before the cast.
A source thus finds the data tokens near its own $f_0$ by relative distance, instead of learning
where each absolute position sits, and attends to the other sources by their distance in bins.
Many of the periods are shorter than a token, so the rotation also resolves frequency inside a
patch. The registers and the sinks carry no position: their keys and values are not rotated, and
their outputs are not turned back.

\subsection{Endpoint Head}
\label{app:network-head}

The source tokens pass through a final layer normalisation with a shift and scale from $\nu$ and an
$\mathrm{FF}:\R^D\to\R^{\NetworkHeadOutputs{}}$ with hidden width $D$, shared over sources. It gives
each of the \NetworkCoords{} coordinates the mean $\hat Z_1$ and the log standard deviation
$\log\tau$ of the endpoint in the frame (\cref{app:fm-loss}). With the frame head, these are the
network's only outputs.

\subsection{Sizes and Precision}
\label{app:network-sizes}

\Cref{tab:network-sizes} gives the five sizes, all with heads of dimension \NetworkHeadDim{}. The
two bands differ only in the number of data tokens, which adds no parameters, and in the number of
periods of the two absolute-frequency embeddings, which makes the narrow-band totals
\numrange{\NetworkNarrowSavedMin}{\NetworkNarrowSavedMax} parameters lower. Each block has $64D^2 +
H$ parameters: each of the four streams has $6D^2$ in its modulations, $4D^2$ in its attention maps
and $6D^2$ in its feed-forward maps, and the queries carry $H$ gains. The last block lacks the
$20D^2$ of the data and sink streams' unused output projections, modulations and feed-forward maps,
so the backbone has $N(64D^2 + H) - 20D^2$ parameters.

\begin{table}[t]
\caption{Network sizes. Backbone parameters are $N(64D^2 + H) - 20D^2$; the embeddings, the
conditioning maps and the two heads add the rest. The source stream has \NetworkSourcesWord{}
source and \NetworkSinksWord{} sink tokens and the encoder \NetworkSourcesWord{} frame registers on
both bands; the data stream has \NetworkTokensNarrow{} tokens on the narrow band and
\NetworkTokensFull{} on the full band.}
\label{tab:network-sizes}
\centering
\small
\begin{tabular}{@{}lrrrrrr@{}}
\toprule
size & width $D$ & heads $H$ & blocks $N$ & backbone & total (narrow) & total (full) \\
\midrule
XS & \NetworkXsWidth{} & \NetworkXsHeads{} & \NetworkXsBlocks{} & \num{\NetworkXsBackbone} & \num{\NetworkXsTotalNarrow} & \num{\NetworkXsTotalFull} \\
S  & \NetworkSWidth{}  & \NetworkSHeads{}  & \NetworkSBlocks{}  & \num{\NetworkSBackbone}  & \num{\NetworkSTotalNarrow}  & \num{\NetworkSTotalFull} \\
B  & \NetworkBWidth{}  & \NetworkBHeads{}  & \NetworkBBlocks{}  & \num{\NetworkBBackbone}  & \num{\NetworkBTotalNarrow}  & \num{\NetworkBTotalFull} \\
L  & \NetworkLWidth{}  & \NetworkLHeads{}  & \NetworkLBlocks{}  & \num{\NetworkLBackbone}  & \num{\NetworkLTotalNarrow}  & \num{\NetworkLTotalFull} \\
XL & \NetworkXlWidth{} & \NetworkXlHeads{} & \NetworkXlBlocks{} & \num{\NetworkXlBackbone} & \num{\NetworkXlTotalNarrow} & \num{\NetworkXlTotalFull} \\
\bottomrule
\end{tabular}
\end{table}

\paragraph{Precision.}
Parameters are kept in float32. One setting chooses what runs in bfloat16
\citep{micikevicius2018mixed,kalamkar2019study}: the whole backbone, forward and backward (the
default), only the attention scores and weighted sums, or nothing. In every mode the embeddings, the
conditioning maps and the two heads run in float32, and every softmax is taken in float32. The
embeddings stay in float32 because bfloat16 keeps \NetworkBfBits{} significant bits: rounding a
whitened value above \NetworkBfUnity{} can move it by a unit of noise, and the loudest samples reach
\NetworkSnrTop{}. In bfloat16 on Ampere or newer GPUs the data stream's self-attention runs as a
fused flash-attention kernel \citep{dao2022flashattention}.

Evaluating the trained models with a float32 backbone \rev{changes them little against their gap to the
chains. In the separate re-evaluation of \cref{app:results-mcmc}, whose bfloat16 values differ from those of
\cref{tab:results-widths},} the median $f_0$ width
relative to the joint chains' at $\snr=15$--40 goes from \ResultsNarrowFourFpBfMid{} to
\ResultsNarrowFourFpThirtyTwoMid{} (narrow band, $S=4$) and from \ResultsFullEightFpBfMid{} to
\ResultsFullEightFpThirtyTwoMid{} (full band, $S=8$), and on loud full-band sources below
\qty{10}{\milli\hertz} from \ResultsFullEightFpBfLoud{} to \ResultsFullEightFpThirtyTwoLoud. The
bfloat16 backbone is therefore not what keeps the loud sources wide\rev{, though at $\snr=15$--40 on the full band it
accounts for about a sixth of the width}.

\section{Training and Compute}
\label{app:training}

This appendix gives the recipe that trains the flow of \cref{app:fm,app:network}
(\cref{sec:method-training}) and the compute it needs. \Cref{tab:training-recipe} collects the
hyperparameters, which every band and size shares; only the step counts depend on the band. The
subsections explain the batch, the optimiser, the schedule, the training loop and the cost.

\begin{table}[htb]
\caption{Training recipe, shared by every band and size. $\gamma_t$ is the Muon learning rate of
\cref{app:training-schedules}.}
\label{tab:training-recipe}
\centering
\small
\begin{tabular}{@{}lll@{}}
\toprule
group & setting & value \\
\midrule
Muon & peak learning rate & \num{\TrainingMuonRate} \\
\quad (hidden matrices) & momentum $\upsilon$ & \TrainingMomentum{}, Nesterov, stored in bfloat16 \\
& orthogonalisation & \TrainingNewtonSchulzSteps{} quintic Newton--Schulz iterations, bfloat16 \\
& update RMS & $\TrainingMuonRms{}\,\gamma_t$ for every matrix shape \\
& weight decay & $\varkappa = \TrainingWeightDecay{}$, decoupled, step $\gamma_t\varkappa\vartheta$ \\
\addlinespace
AdamW (everything else) & peak learning rate & \num{\TrainingAdamRate} \\
& moments & $\beta_1 = \TrainingBetaOne{}$ (float32), $\beta_2 = \TrainingBetaTwo{}$, $\epsilon = \num{\TrainingAdamEps}$ \\
& weight decay & none \\
\addlinespace
Schedule & gradient clipping & global norm \TrainingClipNorm{} \\
& warmup & linear from zero, \rev{\qty{\TrainingWarmupPercent}{\percent} of the run} \\
& after warmup & constant, then $1-\sqrt{\cdot}$ to zero over the last \qty{\TrainingCooldownPercent}{\percent} \\
& total steps & per run \rev{(\cref{tab:results-runs}), \numrange{1.4e6}{3.2e6}} \\
& epoch & \rev{one 64th of the run,} one checkpoint each \\
\addlinespace
Batch & samples per step & \TrainingBatch{} $=$ \TrainingObservations{} observations $\times$ \TrainingDraws{} draws of $(Z_0, t)$ \\
& data & simulated on the device each step, never repeated \\
\addlinespace
Precision & backbone & bfloat16 compute, float32 weights \\
& everything else & float32 \\
\bottomrule
\end{tabular}
\end{table}

\subsection{Batch}
\label{app:training-batch}

A batch holds \TrainingBatch{} $(Z_0, t)$ samples: \TrainingObservations{} simulated observations,
each with its own centre frequency $\fw$ and sources (\cref{app:priors}), and \TrainingDraws{} draws
of the base point and time per observation. The data stream of an observation is encoded once and
shared by its draws (\cref{app:network}), so the draws share the cost of both the simulation and the
encoding. The times are stratified over the whole batch, one per stratum of \TrainingBatch{} equal
strata (\cref{app:fm-clock}).

Every batch is simulated on the device inside the step, so no observation is seen twice. The
simulation takes \rev{\qty{\PriorsNarrowDrawShare}{\percent}} of a narrow-B step and
\rev{\qty{\PriorsFullDrawShare}{\percent} of a full-B step on an A100 (\cref{tab:priors-bands})}, whereas on
\TrainingDrawCpuCores{} CPU cores the full-band batch alone takes \qty{\TrainingDrawCpuFullSeconds}{\second},
\rev{about 22} times the whole GPU step. Simulating on the host is therefore not an
option. The gradient can be accumulated over equal micro-batches of whole observations, which
leaves the mean gradient unchanged; by default the batch is not split.

\subsection{Optimiser}
\label{app:training-optimiser}

The gradient of the loss (\cref{eq:loss}) is clipped to a global norm of \TrainingClipNorm{}. The
parameters are then split between Muon \citep{jordan2024muon} and AdamW, with the settings with
which \citet{liu2025muon} scale Muon to large language models.

\paragraph{Muon.}
For a weight matrix $W \in \R^{m \times n}$ (output by input) with clipped gradient $G$, Muon keeps
a momentum $M \leftarrow \upsilon M + (1-\upsilon)\, G$, takes the Nesterov look-ahead $N =
\upsilon M + (1-\upsilon)\, G$, and replaces it by an approximation $O$ of the nearest
semi-orthogonal matrix $UV\T$ (from $N = U\Sigma V\T$), computed with \TrainingNewtonSchulzWord{}
quintic Newton--Schulz iterations \citep{jordan2024muon}. The step is
\begin{equation}
  W \leftarrow W - \gamma_t \left(\TrainingMuonRms{}\sqrt{\max(m, n)}\; O + \varkappa W\right).
  \label{eq:training-muon}
\end{equation}
Its orthogonal part has an RMS of $\TrainingMuonRms{}\gamma_t$ for every shape, the scale at which
\citet{liu2025muon} match AdamW's update size, so one learning rate serves every matrix. Weight
matrices stacked over the backbone's blocks are orthogonalised one block at a time. The decoupled
decay $\varkappa$ applies only to the matrices Muon updates, and through $\gamma_t$ it ends with the
cooldown; Muon's updates have a fixed size whatever the gradient's, so without decay the weights
grow and its lead over AdamW fades on long runs \citep{liu2025muon}.

\paragraph{AdamW.}
Every other parameter is updated by AdamW, without decay: decaying the AdamW parameters of an
earlier version of the network shrank the head's output rows for one coordinate to a small fraction
of the others. The first moment is float32, because a bfloat16 one loses the small increments of a
smooth gradient.

\paragraph{Assignment.}
Muon takes the weight matrix of every linear layer in \TrainingMuonFieldCount{} modules of the
network (\cref{app:network}): the backbone, the two conditioning \acp{mlp}, the two Fourier-feature
\acp{mlp}, the output modulations of the flow and of the frame, and the hidden layer of the
endpoint head. AdamW takes the rest: the linear embeddings of the sources and of the data tokens,
the register and sink embeddings, the frame's output layer, the query gains, every bias and the
endpoint head's output layer. Muon holds \qty{\TrainingMuonShareXS}{\percent} of the parameters of
full-XS, \qty{\TrainingMuonShareB}{\percent} of full-B and \qty{\TrainingMuonShareXL}{\percent} of
full-XL.

\paragraph{Learning rates and stability.}
The peak rates were not tuned by a sweep on this network, and neither was the batch. Muon's matches
AdamW's because its update RMS does. At these rates five of the six runs of \cref{app:results}
trained to their end with every loss finite. Full-band $S=4$ did not: from a few epochs before,
isolated single steps sent the frame's loss and the gradient norm up by many decades while the
epoch medians stayed flat, and at epoch \ResultsFullFourDivergedEpoch, just before its cooldown, a
loss turned non-finite (\cref{app:results-runs}). Its continuation from an earlier snapshot with the
Student-t frame loss of \cref{app:fm-frame}, at the same rates and schedule, trained to its end with
every loss finite, and its largest single-step gradient norm was \num{\ResultsTFourMaxGrad},
against up to \num{\ResultsTFourGaussMaxGrad} in the Gaussian run's last epochs. Whether lower
rates avoid such spikes, or higher ones train faster, is open.

Muon \citep{jordan2024muon}, its decay and update scale \citep{liu2025muon} and the
warmup--stable--decay schedule \citep{hu2024minicpm,hagele2024scaling} are established on language
models; none was tested on flow matching, and none was ablated here.

\subsection{Schedule}
\label{app:training-schedules}

Both learning rates rise linearly from zero over the first \qty{\TrainingWarmupPercent}{\percent} of
the run, stay constant, and fall to zero over the last \qty{\TrainingCooldownPercent}{\percent} as
$\gamma_t = \gamma\,\bigl(1 - \sqrt{(t - T_c)/(T - T_c)}\bigr)$, with $T$ the run's total steps and
$T_c$ the cooldown's start. This is the warmup--stable--decay schedule \citep{hu2024minicpm} with
the $1-\sqrt{\cdot}$ cooldown that \citet{hagele2024scaling} find best; with a cooldown of a fifth
of the run, they report it matching a cosine schedule. Only the cooldown depends on $T$, so raising
$T$ before the cooldown starts extends the run as if it had been planned that long.

The step counts are a wall-time budget, not a convergence criterion. Each run of \cref{app:results}
was sized from the epoch time of a short smoke test of its band and source count to
\ResultsEpochs{} epochs, about \ResultsNarrowHours\,h on the narrow band and \ResultsFullHours\,h on
the full band of A100 time. A later cooldown generally ends lower, and some configurations are not
converged at the planned epoch (\cref{tab:results-forks}).

\subsection{Training Loop}
\label{app:training-loop}

\paragraph{Epochs.}
An epoch is \rev{one 64th of a run, \numrange{22000}{50000} steps (\cref{tab:results-runs}),} compiled as one program: the simulator, the loss, the
gradient and the optimiser update of all its steps run in one loop on the device.

\paragraph{Eval-catalogue loss.}
After every epoch, the \EvaluationInjections{} eval injections with their noise
(\cref{app:evaluation-injections}) give the frame's and the flow's negative log-likelihoods, each
at the midpoints of \FlowMatchingLossBins{} equal steps in $s$. At the midpoints of
\FlowMatchingLossBins{} bins of $\log_{10}(1-s)$ down to $10^{\FlowMatchingLossDecades}$ they also
give, per bin and coordinate group, the predicted $\log_{10}\tau$ and the squared standardised
error $((Z_1-\hat Z_1)/\tau)^2$ (\cref{app:fm-loss}), and per bin of the source's \ac{snr} the
frame's $\log_{10}\sigma/\sigma_\pi$ and $\kappa^2$ (\cref{app:fm-frame}). The catalogue and its
draws of $Z_0$ come from a random stream that training never uses and are fixed for the run, so the
curves move only with the weights. Training stops before the checkpoint if any loss of the epoch is
not finite, so the last checkpoint is always finite.

\paragraph{Checkpoints.}
Every epoch saves, in the background while the next epoch trains, the weights, the optimisers'
moments, the random key and the loss histories, keeping only the latest. A second checkpoint keeps
the epoch with the lowest sum of the eval-catalogue frame and flow negative log-likelihoods, which
is the checkpoint used for evaluation (\cref{app:evaluation-weights}). Because the key is
checkpointed, a resumed run draws the same batches as an uninterrupted one.

\paragraph{Snapshots and cooldown forks.}
Every \ResultsSnapshotEvery{} epochs a run also keeps a snapshot of its whole state\rev{, kept apart from the
rolling checkpoint, which each epoch overwrites}. A fork restarts from the snapshot at epoch $k$ with the total cut to
$k/(1-c)$ epochs, $c$ the cooldown fraction, so it cools down at once, over $ck/(1-c)$ epochs, as a
run planned that long from the start would have; the schedule before the cooldown depends only on
the warmup, which every fork shares with its parent. Forks from epochs \ResultsForkEpochs{} of each
run give, at the cost of the cooldowns alone, the loss and the posteriors of four shorter runs
(\cref{tab:results-forks}).

\subsection{Precision}
\label{app:training-precision}

Parameters and AdamW's moments are float32. The backbone runs in bfloat16, forward and backward,
from bfloat16 copies of its float32 weights, while the embeddings, the conditioning \acp{mlp} and
the heads stay in float32 (\cref{app:network-sizes}); Muon's momentum and orthogonalisation run in
bfloat16. The bfloat16 backbone makes a step \rev{\numrange{\TrainingBackboneSpeedupAMin}{\TrainingBackboneSpeedupAMax}}
times faster than float32 on an A100, more for the larger sizes, whereas bfloat16 attention alone
gains at most \qty{\TrainingAttentionGainPercent}{\percent}: the dense layers carry the cost.

A float32 backbone is available, but it needs the GPU's matrix-multiplication autotuning switched
off and full float32 matrix products (no TF32). On an earlier version of the network, float32
programs compiled with autotuning gave silently wrong, over-confident posteriors on two GPU models,
while the same program without autotuning reproduced the CPU. This was not re-tested on the current
network; the bfloat16 backbone was not affected.

\subsection{Compute}
\label{app:training-compute}

\begin{figure}[t]
\centering
\includegraphics[width=\linewidth]{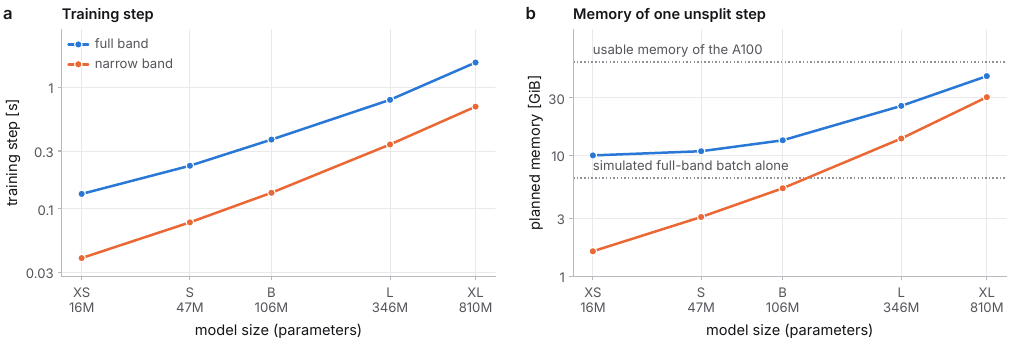}
\caption{Cost of every size on one A100 with the bfloat16 backbone, at the default batch of
\TrainingBatch{} samples; full band blue, narrow band orange. (a)~Training step, simulation
included. (b)~Planned device memory of the unsplit step, against the card's usable memory and the
full band's simulated batch alone (dotted). \rev{Measured on an earlier revision of the network, whose sizes had the parameter counts on the
  axis; \cref{tab:network-sizes} gives those of the reported network.}}
\label{fig:training-compute}
\end{figure}

\Cref{fig:training-compute} gives the measured cost of every band and size at the default batch on
an A100. The narrow band's step is the network's. The full band also simulates its observations at
many more frequency points (\cref{app:forward-model}), a fixed cost per batch that weighs most on
the small sizes and sets a memory floor, since the whole batch is simulated before it is split into
micro-batches. \rev{At the default length of \num{\TrainingStepsNarrow} and \num{\TrainingStepsFull} steps} a size-B run takes
\qty{\TrainingComputeNarrowBHours}{\hour} of steps on one A100 on the narrow band and \rev{\qty{\TrainingComputeFullBHours}{\hour}} on the full band\rev{; the reported
runs, set to 64 epochs, take \qty{32}{\hour} and \qtyrange{63}{64}{\hour} (\cref{tab:results-runs})}.
\rev{Every size's step fits on the A100 without splitting the batch; full-XL no longer fits the \qty{80}{\giga\byte} card
once the trainer's eval pass and checkpointing run beside it.}

\section{Evaluation Protocol and Metrics}
\label{app:evaluation}

This appendix defines how the flows of \cref{app:results} are judged, in support of \cref{sec:validation}. The
\emph{reference pass} compares the flow with joint \ac{mcmc} chains (\cref{app:mcmc}) on fixed windows and uses the
truth to pair draws with sources, so it is descriptive. The \emph{validation pass} never sees the truth or a chain: it
ranks the truth among the draws on windows simulated from the prior, and is the only test of calibration
(\cref{tab:evaluation-passes}). We show on
posteriors with known errors that its tests detect what they should (\cref{app:evaluation-sensitivity}), then give the
sampler settings and out-of-distribution checks.

\begin{table}[h]
  \centering
  \caption{The two evaluation passes.}
  \label{tab:evaluation-passes}
  \small
  \begin{tabular}{@{}l>{\raggedright\arraybackslash}p{0.27\textwidth}r>{\raggedright\arraybackslash}p{0.36\textwidth}@{}}
    \toprule
    Pass & Windows & Draws & Shows \\
    \midrule
    Reference, eval windows & \EvaluationInjections{} fixed windows, log-uniform over the band
         & \EvaluationPosterior & corner plots; width ratios and shapes against the chains \\
    Reference, scorecard & the eval windows plus \EvaluationRandom{} prior windows (\EvaluationWindows{} in all)
         & \EvaluationDraws & widths, detection and coverage per \acs{snr} band, with the truth \\
    Validation & \EvaluationValidation{} prior windows from a separate key
         & \EvaluationValidationDraws & rank tests; never sees the truth \\
    \bottomrule
  \end{tabular}
\end{table}

\subsection{Reference Pass}
\label{app:evaluation-reference}

\paragraph{Eval windows.}
\label{app:evaluation-injections}
The eval windows sit at \EvaluationInjections{} centre frequencies $\fw$, one log-uniform draw in each of
\EvaluationInjections{} equal bins of $\log f_0$, empty windows included. Each holds one prior draw of the run's $S$
sources (\cref{app:priors}), ordered by $f_0$, with window \acp{snr}
\rev{from 1.2 to 98 in the narrow band and from 0 to 380 in the full band, rising with $S$ (median 8.6 to 20 and
6.8 to 12)}. Windows and noise are fixed by the seed and band, so every checkpoint, run and chain sees the same data; as a
conditional sample they are never used for calibration.

\paragraph{Relabelling draws to the truth.}
\label{app:evaluation-matching}
The flow's draws list their sources in order of $f_0$ (\cref{app:chart-slots}), which breaks label switching
\citep{stephens2000dealing,jasra2005markov,buscicchio2019label} only where the frequencies are resolved. For display
and for the reference pass each draw $\theta_{1:S}$ is relabelled to the truth $\theta^\star_{1:S}$ by
\begin{gather}
  \hat\sigma = \operatorname*{arg\,min}_{\sigma\in\mathfrak S_S}
  \sum_{s=1}^{S}\bigl\|\mathbf m(\theta_{\sigma(s)}) - \mathbf m(\theta^\star_s)\bigr\|^2,
  \label{eq:evaluation-matching}\\
  \mathbf m(\theta) = \Bigl(\frac{f_0}{f_+ - f_-},\ \log_{10}\Amp,\ \hat n(\alpha,\delta)\Bigr),
  \nonumber
\end{gather}
found over the \rev{$S!$ permutations of the sources ($\EvaluationPermutations$ for $S=4$)}, with $(f_-, f_+)$ the window's range of $f_0$
(\cref{eq:forward-model-window-interior}) and $\hat n$ the unit sky vector. The orientation angles are left out, since
their symmetries (\cref{eq:psi-phi-symmetry}) make the cost multimodal. Because the truth is used, a posterior too wide
to separate two sources is sorted back onto them and looks better than it is.

\paragraph{Scorecard.}
\label{app:evaluation-scorecard}
Each source of the scorecard windows (\EvaluationWindowSources{} sources) gets the quantities of
\cref{tab:evaluation-scorecard} after relabelling, with $f_0$ in bins of $1/\Tobs$, grouped by true \ac{snr} in bands
$<\EvaluationBandOne$, \EvaluationBandOne--\EvaluationBandTwo, \EvaluationBandTwo--\EvaluationBandThree{} and
$\ge\EvaluationBandThree$, since the ideal width (below) runs from \EvaluationIdealLo{} to \num{\EvaluationIdealHi}
bins over the visible sources.

\begin{table}[h]
  \centering
  \caption{Reference-pass quantities per source, on the relabelled draws.}
  \label{tab:evaluation-scorecard}
  \small
  \begin{tabular}{@{}l>{\raggedright\arraybackslash}p{0.5\textwidth}l@{}}
    \toprule
    Quantity & Definition & Per band \\
    \midrule
    Width $w$ & $1.4826\,\MAD(f_0)$, with $\MAD$ the median absolute deviation: the standard deviation for a Gaussian, robust to stray draws
      & median over found \\
    Width ratio & $w/w^\star$ (scorecard), $w$ over the chains' width (eval windows), with $w^\star$ of \cref{eq:evaluation-ideal-width}
      & median over found \\
    Found & $w <{}$ \rev{one bin}: the frequency is pinned to the bin spacing
      & fraction \\
    Off-peak & an $f_0$ draw more than five chain widths and half a bin from the chains' median
      & fraction of draws \\
    Offset & median of $f_0$ minus the truth & median over found \\
    Coverage & truth inside the central \qty{68}{\percent} and \qty{95}{\percent} intervals & mean \\
    \bottomrule
  \end{tabular}
\end{table}

Coverage and offset are not a calibration test, since they use the truth, and coverage is optimistic: a flow that returns the prior still puts the truth in its central \qty{68}{\percent}
interval for a share \EvaluationMockPriorCover{} of the sources in the narrow band \rev{(\cref{tab:evaluation-mocks})} and \EvaluationMockPriorCoverFull{}
in the full band \rev{(the same mock, run on the full band)}, and only its width shows the failure.

\paragraph{The ideal width.}
For a sinusoid of \ac{snr} $\snr$ observed for $\Tobs$ in stationary noise, with the time origin at the centre, the
Fisher matrix in angular frequency and phase is diagonal, $\Gamma_{\omega\omega} = \snr^2\Tobs^2/12$ and
$\Gamma_{\phi\phi} = \snr^2$, and the frequency width is
\begin{equation}
  w^\star = \frac{\sqrt{12}}{2\pi\,\snr\,\Tobs}\,\Tobs = \frac{\sqrt3}{\pi\,\snr}\bins.
  \label{eq:evaluation-ideal-width}
\end{equation}
This is a scale, not a bound: other correlations widen the real width. Since $f_0$ is the frequency at the start of the observation (\cref{app:forward-model}), a measured chirp
widens it by a factor of \EvaluationStartFactor{} over the centred one, so a full-band ratio above one is not by itself a
failure. Ratios against the chains are the comparison that counts.

\paragraph{Widths and shapes against the chains.}
\label{app:evaluation-mcmc}
Chain and flow draws are compared with $(\psi,\phi_0)$ folded onto the copy nearest the truth, and orientation through
the phases $\phi_0 \pm 2\psi$, since $\phi_0$ alone inherits the spread of $\psi$. Widths are $1.4826\,\MAD$ per parameter, flow over chain, against
the joint chains of every eval window. The chains are an evaluation reference only.

The shapes of the marginals, offsets and tails included, are compared with the Jensen--Shannon divergence
\citep{lin1991divergence}, as \citet{delmond2026neural} do. For the densities $p$ and $q$ of one coordinate, with the
Kullback--Leibler divergence \citep{kullback1951information}
\begin{equation}
  D_{\mathrm{KL}}(p\,\|\,q) = \int p(x)\,\log_2\frac{p(x)}{q(x)}\,\dd x ,
  \label{eq:evaluation-kl}
\end{equation}
it is
\begin{equation}
  D_{\mathrm{JS}}(p\,\|\,q) = \tfrac12\,D_{\mathrm{KL}}(p\,\|\,m) + \tfrac12\,D_{\mathrm{KL}}(q\,\|\,m),
  \qquad m = \tfrac12\,(p+q).
  \label{eq:evaluation-js}
\end{equation}
It is symmetric, finite when one density vanishes where the other does not, invariant under one-to-one maps of the
coordinate, and with base-2 logarithms lies in $[0,1]$. We estimate it per source and coordinate from the flow's and the
chain's \ResultsJsDraws{} draws each, relabelled and folded as above, on $B=\ResultsJsBins$ bins whose edges are
quantiles of the pooled draws:
\begin{equation}
  \hat D_{\mathrm{JS}} = \frac12\sum_{b=1}^{B}\hat p_b\log_2\frac{\hat p_b}{\hat m_b}
  + \frac12\sum_{b=1}^{B}\hat q_b\log_2\frac{\hat q_b}{\hat m_b},
  \qquad \hat m_b = \tfrac12\,(\hat p_b+\hat q_b),
  \label{eq:evaluation-js-hat}
\end{equation}
with $\hat p_b$ and $\hat q_b$ the shares of the flow's and the chain's draws in bin $b$. For two samples of one continuous distribution its median is
\ResultsJsFloorMedian{} and its 95th percentile \ResultsJsFloorHigh{}, the floor of the comparison
(\num{2000} simulated pairs). Between two Gaussians it is \ResultsJsGaussWidthTwo{} for a width ratio of 2 and
\ResultsJsGaussOffsetOne{} for an offset of one width.
\Cref{app:results-js} gives the results.

\subsection{Validation Pass}
\label{app:evaluation-validation}

The validation pass draws \EvaluationValidation{} windows, each with its own $\fw$, sources and noise from the prior,
and $M = \EvaluationValidationDraws$ flow draws on each. If the flow is the exact posterior, the rank of any scalar test
quantity $g$ of the truth among the draws, the number of draws with $g(\theta_m) < g(\theta^\star)$ (ties broken at
random), is uniform on $\{0,\dots,M\}$ \citep{talts2018validating}. A failure is a non-uniform histogram: ranks mostly
high mean a bias, ranks piled in the middle a posterior too wide, ranks piled at the ends one too narrow.

\paragraph{Slots and strata.}
The same $g$ is applied to the truth and to every draw without using the truth. Slots are therefore defined by each
parameter set's own \ac{snr}: the truth and each draw are sorted independently by the \acp{snr} of their own sources, a
fixed function of the parameters, so the ranks stay uniform under an exact posterior, unlike those after
\cref{eq:evaluation-matching}. Strata follow the same rule: a source is binned in the bands of
\cref{app:evaluation-scorecard} by $\hat\snr$, the median own \ac{snr} of the draws in its slot, and window quantities by
the draws' median total \ac{snr}.

\paragraph{Tests.}
\Cref{tab:evaluation-tests} lists them. The coordinate and \ac{tarp} ranks test the parameters alone, which the prior passes; data-dependent
quantities catch it \citep{modrak2023simulation}. The likelihood is only a test quantity on draws already made.

\begin{table}[t]
  \centering
  \caption{Rank tests of the validation pass. ``Slot'' is per source slot, ``window'' over all
  \rev{$8S$} coordinates of a window. None needs the truth's labels or a reference posterior.}
  \label{tab:evaluation-tests}
  \small
  \begin{tabular}{@{}>{\raggedright\arraybackslash}p{0.30\textwidth}l>{\raggedright\arraybackslash}p{0.44\textwidth}@{}}
    \toprule
    Test quantity & Level & Failure shows \\
    \midrule
    The eight flow coordinates (\cref{eq:chart}) & slot & bias or wrong width in one coordinate \\
    \acs{tarp} distance to a prior reference point \citep{lemos2023sampling} & slot, window & width errors \\
    Overlap $\inner{d}{h_s}/\inner{h_s}{h_s}^{1/2}$ (\cref{eq:inner}) & slot & draws that do not fit the data: a wrong phase, sky or dependence between coordinates, a prior-like posterior \\
    Own \acs{snr} of the source & slot & a bias toward loud or faint sources; broken dependence \\
    Log-likelihood (\cref{eq:loglik}) & window & a window that does not fit the data as a whole \\
    \bottomrule
  \end{tabular}
\end{table}

\paragraph{Uniformity test.}
Ranks are tested with the graphical \ac{ecdf} test of \citet{sailynoja2022graphical}. With $N$ ranks on
$k = \EvaluationRankValues$ values, the statistic $\gamma$ is twice the smallest binomial tail probability of the rank
\ac{ecdf} at $z_i = i/k$, $i = 1,\dots,k-1$. Its $p$-value comes from \EvaluationGammaSims{} simulated sets of $N$
uniform ranks, which also give a simultaneous \qty{\EvaluationBandLevel}{\percent} band for the \ac{ecdf} minus $z$.
Each cell reports the $p$-value, the mean rank over $M$ minus one half (positive when the truth sits above the draws),
and the share of ranks in the middle half (above one half: too wide).

\subsection{Sensitivity of the Tests}
\label{app:evaluation-sensitivity}

Passing means something only if the tests fail on wrong posteriors, which we check on a toy and on mocks.

\paragraph{Size and power on an exact toy.}
\label{app:evaluation-power}
Each slot of the toy is a linear-Gaussian model in its eight coordinates, drawn exactly, with all test quantities
computed on its data. A test's rate (\cref{tab:evaluation-power}) is the share of chunks of \EvaluationPowerChunk{} of
\EvaluationPowerInjections{} injections (\EvaluationPowerDraws{} draws each) in which it rejects at
\qty{\EvaluationPowerAlpha}{\percent}. The rank tests hold their size. A phase or sky taken from the prior, or
coordinates shuffled across draws, keep every one-dimensional rank uniform and pass the coordinate ranks, as does most of a
\qty{\EvaluationPowerMixShare}{\percent} prior admixture; the overlap and the log-likelihood catch all four.
\Ac{tarp}\rev{, not shown in \cref{tab:evaluation-power}, rejects width errors at rates of} \EvaluationPowerTarpWidth{} but
\rev{the other errors at} most \EvaluationPowerTarpBlind{}.

\begin{table}[t]
  \centering
  \caption{Rejection rates of the validation tests on the exact-posterior toy, per error injected
  into the draws. Coordinate ranks give the range over the eight flow coordinates; overlap and own
  \acs{snr} are per slot, the log-likelihood per window.}
  \label{tab:evaluation-power}
  \small
  \begin{tabular}{@{}lcccc@{}}
    \toprule
    & & \multicolumn{3}{c}{Data-space ranks} \\
    \cmidrule(lr){3-5}
    Injected error & Coordinate ranks & Overlap & Own \acs{snr} & Log-likelihood \\
    \midrule
    None & \EvaluationPowerNullCoord & \EvaluationPowerNullOverlap & \EvaluationPowerNullSnr & \EvaluationPowerNullLike \\
    $z_+$ phase from the prior & \EvaluationPowerPhaseCoord & \EvaluationPowerPhaseOverlap & \EvaluationPowerPhaseSnr & \EvaluationPowerPhaseLike \\
    Sky from the prior & \EvaluationPowerSkyCoord & \EvaluationPowerSkyOverlap & \EvaluationPowerSkySnr & \EvaluationPowerSkyLike \\
    Coordinates shuffled across draws & \EvaluationPowerShuffleCoord & \EvaluationPowerShuffleOverlap & \EvaluationPowerShuffleSnr & \EvaluationPowerShuffleLike \\
    \qty{\EvaluationPowerMixShare}{\percent} prior draws & \EvaluationPowerMixedCoord & \EvaluationPowerMixedOverlap & \EvaluationPowerMixedSnr & \EvaluationPowerMixedLike \\
    Width $\times0.8$ & \EvaluationPowerNarrowCoord & \EvaluationPowerNarrowOverlap & \EvaluationPowerNarrowSnr & \EvaluationPowerNarrowLike \\
    Width $\times1.25$ & \EvaluationPowerWideCoord & \EvaluationPowerWideOverlap & \EvaluationPowerWideSnr & \EvaluationPowerWideLike \\
    $f_0$ shifted by $0.5\sigma$ & \EvaluationPowerShiftCoord & \EvaluationPowerShiftOverlap & \EvaluationPowerShiftSnr & \EvaluationPowerShiftLike \\
    \bottomrule
  \end{tabular}
\end{table}

\paragraph{Mock posteriors of the real problem.}
\label{app:evaluation-mocks}
Each mock replaces the flow's sampler by a posterior with a known error and goes through the whole pipeline. The base mock is a
linearised Gaussian in flow coordinates, with the expected Fisher matrix at the truth plus the prior precision, centred
on the linearised maximum a posteriori of the noisy data. The ladder perturbs it: \emph{wide} and \emph{narrow} (every
width times \EvaluationMockWide{} and \EvaluationMockNarrow), \emph{biased} ($f_0$ shifted by one width), \emph{prior}
(prior draws, a flow that ignores the data), \emph{phases-prior} (phases of $z_\pm$ redrawn uniform) and
\emph{loud-floor} (widths that stop shrinking above \ac{snr} \EvaluationMockFloorSnr{}, quiet sources exact).
\Cref{tab:evaluation-mocks} gives the readings on the narrow band with four sources, on catalogue-prior windows
(\EvaluationMockScorecardSources{} scorecard sources, \EvaluationMockInjections{} validation windows,
\EvaluationMockReferenceSources{} eval sources at $\snr\ge\EvaluationMockReferenceSnr$ with joint chains,
\cref{app:results-mcmc}); the full band, with \EvaluationMockInjectionsFull{} validation windows, agrees. The Gaussian
mock is also the \rev{reference} of \cref{tab:results-validation}.

\begin{table*}[t]
  \centering
  \caption{The mock ladder on the narrow band. Width over the ideal width
  (\cref{eq:evaluation-ideal-width}): median over the \EvaluationMockLoudSources{} scorecard sources with
  $\snr\ge\EvaluationMockLoudSnr$.
  Coverage: share of sources inside the central \qty{68}{\percent} interval after relabelling.
  Validation $p$-values: the smallest over the eight coordinate ranks, \acs{tarp}, and the overlap
  and log-likelihood ranks. Flow over \acs{mcmc}: range over
  the eight physical parameters of the median width ratio against the joint reference chains, over the
  eval sources with $\snr\ge\EvaluationMockReferenceSnr$ (\cref{app:results-mcmc}).}
  \label{tab:evaluation-mocks}
  \small
  \begin{tabular}{@{}lccccccc@{}}
    \toprule
    & \multicolumn{2}{c}{Scorecard} & \multicolumn{4}{c}{Validation $p$-value} & \\
    \cmidrule(lr){2-3}\cmidrule(lr){4-7}
    Mock & Width$/w^\star$ & Coverage & Coordinates & \acs{tarp} & Overlap & Log-lik. & Flow$/$\acs{mcmc} \\
    \midrule
    Gaussian & \EvaluationMockFisherIdeal & \EvaluationMockFisherCover & \EvaluationMockFisherCoord & \EvaluationMockFisherTarp & \EvaluationMockFisherOverlap & \EvaluationMockFisherLike & \EvaluationMockFisherMcmc \\
    Wide & \EvaluationMockWideIdeal & \EvaluationMockWideCover & \EvaluationMockWideCoord & \EvaluationMockWideTarp & \EvaluationMockWideOverlap & \EvaluationMockWideLike & \EvaluationMockWideMcmc \\
    Narrow & \EvaluationMockNarrowIdeal & \EvaluationMockNarrowCover & \EvaluationMockNarrowCoord & \EvaluationMockNarrowTarp & \EvaluationMockNarrowOverlap & \EvaluationMockNarrowLike & \EvaluationMockNarrowMcmc \\
    Biased & \EvaluationMockBiasedIdeal & \EvaluationMockBiasedCover & \EvaluationMockBiasedCoord & \EvaluationMockBiasedTarp & \EvaluationMockBiasedOverlap & \EvaluationMockBiasedLike & \EvaluationMockBiasedMcmc \\
    Prior & \EvaluationMockPriorIdeal & \EvaluationMockPriorCover & \EvaluationMockPriorCoord & \EvaluationMockPriorTarp & \EvaluationMockPriorOverlap & \EvaluationMockPriorLike & \EvaluationMockPriorMcmc \\
    Phases-prior & \EvaluationMockPhasesIdeal & \EvaluationMockPhasesCover & \EvaluationMockPhasesCoord & \EvaluationMockPhasesTarp & \EvaluationMockPhasesOverlap & \EvaluationMockPhasesLike & \EvaluationMockPhasesMcmc \\
    Loud-floor & \EvaluationMockFloorIdeal & \EvaluationMockFloorCover & \EvaluationMockFloorCoord & \EvaluationMockFloorTarp & \EvaluationMockFloorOverlap & \EvaluationMockFloorLike & \EvaluationMockFloorMcmc \\
    \bottomrule
  \end{tabular}
\end{table*}

Three readings shape how the results are read.
\begin{itemize}
  \item \emph{Marginal calibration proves nothing alone.} The prior mock passes \ac{tarp} and the coordinate ranks (its
    smallest $p$ of eight, \EvaluationMockPriorCoord, is what chance gives); the overlap and log-likelihood ranks reject
    it, and coverage reads it as conservative.
  \item \emph{The Gaussian mock is not a null.} Where sources are faint the linearisation fails: only $f_0$ and $\fdot$
    pass (or nearly pass), and the sky, the $z_\pm$ components and radii, \ac{tarp}, the overlap and the log-likelihood
    reject it. A flow's $p$-value or rank offset is therefore compared with the Gaussian mock's on the same test and
    windows, not with \qty{\EvaluationPowerAlpha}{\percent}.
    A second realisation of the same mock on the same windows, on other hardware, gave $p=0.14$ for $f_0$ and $0.42$ for
    $\fdot$ where this one gives $0.04$ and $0.26$, while the prior mock reproduces to the digit: the Gaussian draws differ
    between machines (an eigendecomposition of an ill-conditioned covariance), so $p$-values near the threshold move by that
    much and these readings do not change.
  \item \emph{Coverage hides a width floor.} The loud-floor mock covers \EvaluationMockFloorCoverQuiet{} of sources below
    \ac{snr} \EvaluationBandTwo{} and \EvaluationMockFloorCoverLoud{} above, so it reads conservative; only the width
    ratio sees it (\EvaluationMockFloorIdeal{} times the ideal, against \EvaluationMockFisherIdeal{} for the Gaussian mock).
    On loud full-band sources the Gaussian mock is itself \EvaluationMockFisherIdealFull{} times wider than the ideal width.
\end{itemize}

\subsection{Checkpoint and Sampler Settings}
\label{app:evaluation-weights}

\rev{Every pass evaluates the run's checkpoint with the lowest eval-catalogue loss (\cref{app:training-loop}),
which guards against a run that degrades late; it is chosen on the eval windows, so its results on them are
optimistic. The runs of \cref{app:results} were all evaluated at that checkpoint (\cref{tab:results-runs})}, except the full band with $S=4$, whose run diverged: its Student-t continuation is evaluated at
its last checkpoint, since its logged loss does not compare across the switch (\cref{app:results-runs}).
The flow predicts the endpoint $Z_1$ in the frame (\cref{app:fm}), and the sampler takes \EvaluationOdeSteps{} midpoint
steps toward the predicted mean from standard normal base draws in the frame (\cref{app:fm-frame}), in the precision the
run was trained in.

\subsection{Sampler Steps and Robustness}
\label{app:evaluation-sampler}

\paragraph{Step count.}
A coarse grid can collapse narrow posteriors late in the path (\cref{fig:fm-sensitivity}b), and a collapse only narrows.
On the source-count runs (narrow band, $S=4$, and full band, $S=8$; 128 draws per window) the median $f_0$ width of the
sources at $\snr\ge15$, over its width at \ResultsSamplerNarrowMax{} uniform midpoint steps, is
\ResultsSamplerNarrowEight, \ResultsSamplerNarrowSixteen, \ResultsSamplerNarrowThirtyTwo{} and
\ResultsSamplerNarrowSixtyFour{} at 8, 16, 32 and 64 steps on the narrow band, and \ResultsSamplerFullEight{} already at 8
steps on the full band. \EvaluationOdeSteps{} steps are within \qty{2}{\percent} of converged. The converged width does not
depend on the grid: the power-2 and log grids give \ResultsSamplerGridLo--\ResultsSamplerGridHi{} times the uniform
grid's from \ResultsSamplerGridFrom{} steps up, on both bands. Euler converges from below
(\ResultsSamplerEulerNarrowThirtyTwo{} and \ResultsSamplerEulerFullThirtyTwo{} of the converged width at
\EvaluationOdeSteps{} steps), so every width here is read with the midpoint rule.

\paragraph{Out of distribution.}
The flow is trained on one noise \ac{psd}, a fixed prior and exactly $S$ sources per window. Each case changes one of these
on the eval windows, keeping window, truth and noise: the \ac{psd} halved and doubled; the loudest source's $f_0$, $\fdot$
or amplitude past the prior's bulk; the quietest source's amplitude below it; one source removed or added; an empty window.
On the narrow band with $S=4$ (\ResultsRobustInjections{} windows, \ResultsRobustDraws{} draws) the flow is robust to its
prior's edges and fragile to its noise model. With the loudest source past the prior, or the noise halved, the share of
sources at $\snr\ge7$ whose truth lies within one flow width of its median is \ResultsRobustOtherWithinLo{} to
\qty{\ResultsRobustOtherWithinHi}{\percent}, against \qty{\ResultsRobustBaseLoudWithin}{\percent} in distribution\rev{ for the same sources}. With the
noise doubled the flow, which whitens by its own \ac{psd}, reads noise as signal: its slots claim a median \ac{snr} of
\ResultsRobustNoiseSlotSnr{} instead of \ResultsRobustBaseSlotSnr, the faint sources' $f_0$ widths shrink to \ResultsRobustNoiseFaintWidth{} of their in-distribution widths while their offsets
double, and only \qty{\ResultsRobustNoiseFaintWithin}{\percent} cover the truth within one width
(\qty{\ResultsRobustBaseFaintWithin}{\percent} \rev{of the faint sources} in distribution): confident and wrong. The full band with $S=8$ shows the same on its loud sources: with the noise doubled
\qty{\ResultsRobustFullNoiseLoudWithin}{\percent} cover the truth within one width
(\qty{\ResultsRobustFullBaseLoudWithin}{\percent} in distribution), and the other cases keep
\ResultsRobustFullOtherWithinLo{} to \qty{\ResultsRobustFullOtherWithinHi}{\percent}.

\section{Reference \acs{mcmc}}
\label{app:mcmc}

The flow's posteriors are judged against Markov chains that sample the window likelihood of
\cref{eq:loglik} directly; training never sees them. This appendix says what the reference is, how far it
can be trusted and what it costs, in support of the width comparisons of \cref{app:results-mcmc}. Unlike a
global fit \citep{littenberg2020global}, the number of sources and the noise are known, so no move changes
the model.

\subsection{Sampler}
\label{app:mcmc-sampler}

Each eval window (\cref{app:evaluation-injections}) is sampled jointly, one ensemble over all $8S$
parameters of its sources, with responses of \PriorsNarrowR{} bins in the narrow band and \PriorsFullR{} in
the full band, in double precision. Each source moves in
$(\ln f_0, \ln\fdot, \ln\Amp, \alpha, \delta, \psi, \cos\iota, \phi_0)$ under the training prior
(\cref{app:priors}), with one extra cut, $\log_{10}\fdot$ at most \McmcFdotSigmas{} standard
deviations above its mean given $f_0$ (prior mass \num{\McmcFdotCutMass}). The sampler is a
parallel-tempered ensemble sampler with differential-evolution \citep{terbraak2006markov} and
stretch moves \citep{goodman2010ensemble,foreman2013emcee}, \McmcTemps{} temperatures up to
\McmcTempMax{} \citep{vousden2016dynamic} (\cref{tab:mcmc-settings}). The cold draws are relabelled to the
injected sources (\cref{eq:evaluation-matching}) and folded onto the window's branch of the
$(\psi,\phi_0)$ symmetries (\cref{eq:psi-phi-symmetry}) before any width is taken.

\begin{table}[t]
  \centering
  \caption{Settings of the reference sampler, per problem. It stores every \McmcThin{}th cold iteration
  after the burn-in.}
  \label{tab:mcmc-settings}
  \small
  \begin{tabular}{@{}ll@{}}
    \toprule
    Likelihood & \cref{eq:loglik}, all sources of the window \\
    Dimension & \rev{$8S$: 16 to 128 ($S=2$ to 16)} \\
    Windows & \McmcInjections \\
    Walkers per temperature & \McmcWalkers \\
    Temperatures & \McmcTemps, up to \McmcTempMax \\
    Burn-in / kept iterations & \McmcBurn{} / \McmcKeep \\
    Thinning & \McmcThin \\
    Seeds / chains & \McmcSeeds{} / \McmcChains\rev{, \McmcChainsAll{} over the six problems} \\
    Draws per window & \McmcKeepDraws \\
    Start & Fisher draws \\
    \bottomrule
  \end{tabular}
\end{table}

\subsection{Initialisation from a Fisher forecast}
\label{app:fisher}

The chains start from draws of a Gaussian centred at the truth $\theta^\star$, with precision
\begin{equation}
  \Gamma = \frac1{N_{\rm F}}\sum_{r=1}^{N_{\rm F}}\nabla^2_\theta\bigl[-\log L(\theta; d_r)\bigr]_{\theta^\star}
  + \Lambda,
  \qquad N_{\rm F} = \EvaluationFisherDraws,
  \label{eq:evaluation-fisher}
\end{equation}
the Hessian of \cref{eq:loglik} averaged over noise realisations $d_r$, which is the Fisher matrix
\citep{cutler1994gravitational,vallisneri2008use}, plus $\Lambda$, a Gaussian stand-in for the prior that keeps
poorly measured directions at the prior's width. It is drawn with $\log\fdot$ in place of $\fdot$, and each
eigenvalue is floored at the prior's precision along its eigenvector. Being unimodal and Gaussian, it misses
the exact symmetries of \cref{eq:psi-phi-symmetry} and the curved amplitude--inclination degeneracy, and where
the data barely constrain a parameter it returns the prior's width. It is not a stand-in for the posterior:
from \ac{snr} \McmcFisherMidSnr{} up its median width per parameter is within
\McmcFisherHighMedianLo--\McmcFisherHighMedianHi{} times the chains', but below \ac{snr}
\McmcFisherFaintSnr{} the chains spread a faint source's $f_0$ over much of the window, as the prior does, and
the forecast does not.

\subsection{Convergence}
\label{app:mcmc-convergence}

Over the six references the largest split $\hat R$ \citep{gelman1992inference} is
\ResultsMcmcRhatMax{} (\ResultsMcmcRhatNarrowMax{} on the narrow band) and the smallest \ac{ess}
\ResultsMcmcEssMin{}; the full-band chains sit above the usual $\hat R<1.01$ for their worst
parameter. Longer chains of the narrow band with four sources, \McmcConvSeeds{} seeds on
windows \McmcConvRuns, show that the loud sources are settled once read without labels, while
two things drift below $\snr=7$: sources under $\snr=3$ widen towards the prior, and sources at
$\snr=3$--7 are slowly lost. The reference is therefore sound at $\snr\ge7$, somewhat narrow for
invisible sources, and an upper bound on how often a source at $\snr=3$--7 is localised. For the
problems with the most parameters the faint sources stay near their Fisher start, which
\cref{app:results-mcmc} takes into account when reading the width ratios. On the verification binary of
\citet{strub2022bayesian} the sampler reproduces the published \rev{\ac{snr} (\cref{app:noise-validation}) and posterior}.

\subsection{Cost}
\label{app:mcmc-cost}

On \McmcJointCores{} \acs{cpu} cores a joint chain of \num{400000} iterations takes
\qtyrange{\McmcCostNarrowFourLo}{\McmcCostNarrowFourHi}{\hour} for a narrow-band window with four
sources, \qtyrange{\McmcCostNarrowEightLo}{\McmcCostNarrowEightHi}{\hour} with eight, and about
\qtyrange{\McmcCostFullSixteenLo}{\McmcCostFullSixteenHi}{\hour} for a full-band window with sixteen
(doubled from \num{200000}-iteration chains), against a fraction of a second for the flow
(\cref{app:fm-sampling}).

\section{Results}
\label{app:results}\label{app:results-narrow}\label{app:results-full}

This appendix holds the evidence behind \cref{sec:experiments}: the six training runs and which
model of each problem is reported, the width and JS-divergence tables against the joint chains, the
validation pass with its data-space ranks, one corner plot per model, and the longer-run forks
that qualify the comparison. Every run uses the B network (\cref{app:network}), the recipe of
\cref{app:training} and the priors of \cref{app:priors}, and is evaluated on its own eval
catalogue of \EvaluationInjections{} windows log-uniform over its band
(\cref{app:evaluation-injections}) against the joint chains of the same injections and noise
(\cref{app:mcmc}). Widths are $1.4826\,\MAD$ of the draws, after relabelling to the truth and folding
$(\psi,\phi_0)$ (\cref{app:evaluation-matching}), as the flow over the chains, median over the
sources of a band of source \ac{snr}.

\subsection{The Runs}
\label{app:results-runs}

\begin{table}[h]
\centering
\caption{The six runs. Steps and warmup are set per run from the epoch time of a
\qty{30}{\minute} smoke test, so that each run has \ResultsEpochs{} epochs; the hours are those epochs
at the smoke test's A100 step time, without restarts. A cooldown over the last
\qty{\ResultsCooldownPercent}{\percent} closes each run. The last column is the lowest eval-catalogue
negative log-likelihood (NLL), frame plus flow (\cref{app:training-loop}), and its epoch.}
\label{tab:results-runs}
\small
\begin{tabular}{@{}lrrrrrll@{}}
\toprule
band & $S$ & steps [M] & epochs & warmup & hours & outcome & best NLL (epoch) \\
\midrule
\ResultsRunsRows
\bottomrule
\end{tabular}
\end{table}

Five runs trained to their end with every loss finite. The full-band run with $S=4$ diverged at
epoch \ResultsFullFourDivergedEpoch, just before its cooldown, after isolated single-step spikes of
the frame's loss (\cref{app:training-optimiser}). We report its continuation from the snapshot at
epoch 44 with the Student-t frame loss of \cref{app:fm-frame}, which trained the remaining epochs and
the cooldown with every loss finite and is evaluated at its last checkpoint, epoch \ResultsEpochs;
the best-checkpoint rule reads the logged, Gaussian frame loss, which does not compare across the
switch.

Two runs of each other problem trained to near their end: the 64-epoch main run and its fork cooled
from epoch 32 (\cref{app:results-forks}). They agree within the differences one seed cannot resolve,
and for each problem we report the one with the lower eval-catalogue loss at its best checkpoint:
the fork for the narrow band with $S=2$ and 8 and the full band with $S=16$, the main run for the
narrow band with $S=4$ and the full band with $S=8$. The tables name each model and the epoch
evaluated. Only the narrow band with $S=4$ is read before its last epoch (62 of 64), where the
narrow runs' eval-catalogue loss moves only within its epoch-to-epoch fluctuations of a few
hundredths; the full-band runs with $S=8$ and 16 were still improving at their last epoch.

\subsection{Widths Against the Joint \acs{mcmc}}
\label{app:results-mcmc}

\begin{table*}[t]
\centering
\caption{Flow over joint-\ac{mcmc} width per parameter, median over the sources of each band of
source \ac{snr} ($n$ sources), for the model reported for each band and source count, named under it (main: the 64-epoch run; e32: its fork
cooled from epoch 32; t44: full $S=4$'s continuation from epoch 44) with the epoch evaluated in brackets
(\cref{app:results-runs}). Found: share of the
sources with an $f_0$ width below one bin. Coverage: share of true flow coordinates inside the
central \qty{68}{\percent} and \qty{95}{\percent} intervals, over the eight coordinates and the sources
of the band, after relabelling to the truth (optimistic, \cref{app:evaluation-matching}). $\alpha$ and
$\delta$ are equatorial. Below $\snr=7$ (not shown) the ratios are
\ResultsFaintWidthLo--\ResultsFaintWidthHi{} for every parameter of the other four models. For narrow
$S=8$ ($f_0$ ratio \ResultsNarrowEightFaintFzero) and full $S=16$ ($f_0$ \ResultsFullSixteenFaintFzero)
the chains are stuck at their Fisher start there and the ratios measure the reference, not the flow
(\cref{app:results-mcmc}).}
\label{tab:results-widths}
\small
\begin{tabular}{@{}llrrrrrrrrrr@{}}
\toprule
model & \acs{snr} & $n$ & $f_0$ & $\log\fdot$ & $\log\Amp$ & $\alpha$ & $\delta$ & $\sin\iota$ & found & \multicolumn{2}{c}{coverage 68 / 95} \\
\midrule
\ResultsWidthRows
\bottomrule
\end{tabular}
\end{table*}

\begin{figure}[t]
\centering
\includegraphics[width=\textwidth]{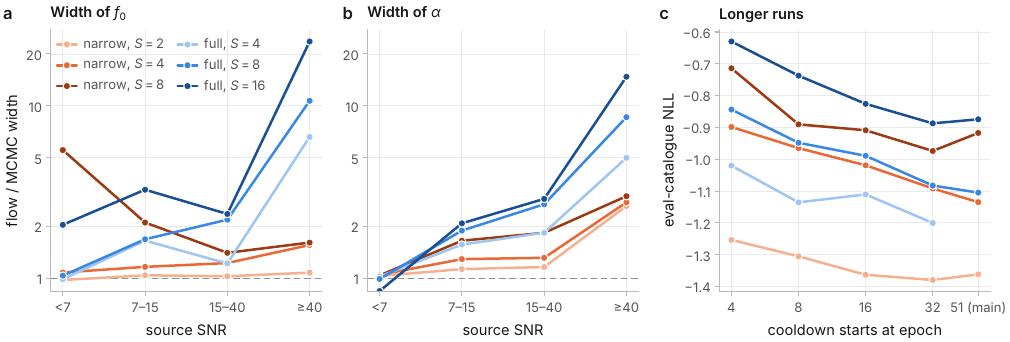}
\caption{(a), (b)~Flow over joint-\ac{mcmc} width of $f_0$ and of the right ascension against the
source \ac{snr}, for the models of \cref{tab:results-widths}. (c)~The
eval-catalogue loss of the cooldown forks against the epoch at which their cooldown starts, and of the
main runs at theirs (\cref{app:results-forks}); the full band with $S=4$ has no cooled main run (its Student-t continuation's
logged loss does not compare, \cref{app:fm-frame}).}
\label{fig:results}
\end{figure}

\Cref{tab:results-widths} is the evidence for the narrow- and full-band subsections of
\cref{sec:experiments}. Two further readings of it do not appear there. The loud draws of the full
band are off-peak in $f_0$ for \qty{\ResultsFullFourOff}{\percent}, \qty{\ResultsFullEightOff}{\percent}
and \qty{\ResultsFullSixteenOff}{\percent} of the $S=4$, 8 and 16 models, against
\qty{\ResultsFullFourMcmcOff}{\percent}, \qty{\ResultsFullEightMcmcOff}{\percent} and
\qty{\ResultsFullSixteenMcmcOff}{\percent} of the chains' own. And $\log\fdot$ is as poorly resolved
as $f_0$ on loud sources, at \ResultsFullFourFdotLoud, \ResultsFullEightFdotLoud{} and
\ResultsFullSixteenFdotLoud{} times the chains' width.

\paragraph{Where the reference is reliable.}
Below $\snr=7$ every width is within a few per cent of the chains' except for the narrow band with
$S=8$, whose faint sources come out \ResultsNarrowEightFaintFzero{} times wider in $f_0$ than the
chains', and the full band with $S=16$, whose $f_0$ is \ResultsFullSixteenFaintFzero{} times the
chains' width and whose sky is \ResultsFullSixteenFaintSkyLo--\ResultsFullSixteenFaintSkyHi{} times
it. In $f_0$ the chains are wrong: below $\snr=\ResultsInvisibleSnr$ the posterior is the prior, and
against the prior's own $f_0$ width the flow's is \ResultsNarrowEightFaintFlowPrior{} ($S=8$) and
\ResultsFullSixteenFaintFlowPrior{} ($S=16$) times it, the chains' only
\ResultsNarrowEightFaintMcmcPrior{} and \ResultsFullSixteenFaintMcmcPrior; for the other four
problems both lie within \ResultsFaintFlowPriorLo--\ResultsFaintFlowPriorHi. The chains start from
Fisher draws (\cref{app:mcmc}), which give even an $\snr=2$ source an $f_0$ width of a fraction of a
bin, and with 64 or 128 parameters the ensemble never leaves that start for the faint sources.

To check the rest of the reference we reran four injections of narrow $S=8$ with every source below
$\snr=7$ started from the prior, and read the chains without labels (\cref{app:mcmc-convergence}).
At $\snr\ge7$ the reference stands: the prior-started chains find every loud source, their $f_0$
widths are \McmcPriorStartLoudChainRef{} times the reference's, and the flow's
\McmcPriorStartLoudFlowChain{} times theirs. Below $\snr=3$ the prior-started chains reach
\McmcPriorStartFaintChainPrior{} of the prior's $f_0$ width, the flow \McmcPriorStartFaintFlowPrior{}
and the reference \McmcPriorStartFaintRefPrior. Between $\snr=3$ and 7 the two starts bracket the
answer from opposite sides: the share of draws with a source within \McmcLabelFreeFoundBins{} bins of
the true $f_0$ is \McmcPriorStartMidFoundRefLo--\McmcPriorStartMidFoundRefHi{} per injection in the
Fisher-started reference and \McmcPriorStartMidFoundLastLo--\McmcPriorStartMidFoundLastHi{} in the
prior-started chains, still rising (chance: \McmcPriorStartChance), and the flow's
\McmcPriorStartMidFoundFlowLo--\McmcPriorStartMidFoundFlowHi{} lies inside that bracket on three
injections and below it on one. Neither start reaches the balance, so a prior-started reference
would be no better, and below $\snr=7$ we score against the prior ($\snr<3$) and report the bracket
($\snr=3$--7) instead.

Full $S=16$'s prior-started chains (four injections) agree below $\snr=7$. At $\snr\ge7$ they find
every loud source, but their $f_0$ widths are
\McmcPriorStartSixteenLoudChainRefLo--\McmcPriorStartSixteenLoudChainRefHi{} times the reference's,
for \McmcPriorStartSixteenLoudGrowing{} of the \McmcPriorStartSixteenLoudN{} sources still growing at
the chains' end. The full band's reference is therefore likely too narrow from $\snr=7$ to about 20,
the full $S=16$ ratios there are upper bounds on the flow's excess width, and the flow's own widths
are \McmcPriorStartSixteenLoudFlowRefLo--\McmcPriorStartSixteenLoudFlowRefHi{} times the
reference's on these sources.

\subsection{Jensen--Shannon Divergence From the Joint \acs{mcmc}}
\label{app:results-js}

\begin{table*}[t]
\centering
\caption{Jensen--Shannon divergence in bits (\cref{eq:evaluation-js,eq:evaluation-js-hat}) of the
flow's one-dimensional marginals from the joint chains', median over the sources of each band of source
\ac{snr} ($n$ sources), for the models of \cref{tab:results-widths}. Mean: median over the sources of
the mean over the eight coordinates. Draws relabelled to the truth, $(\psi,\phi_0)$ folded, $\alpha$
and $\delta$ equatorial\rev{, $\fdot$ in units of $\qty{1}{\hertz\per\second}$}. Two samples of one distribution give \ResultsJsFloorMedian{} (median) and
\ResultsJsFloorHigh{} (95th percentile); two Gaussians give \ResultsJsGaussWidthTwo{} at a width ratio
of 2 and \ResultsJsGaussWidthTen{} at 10.}
\label{tab:results-js}
\small
\begin{tabular}{@{}llrrrrrrrrrr@{}}
\toprule
model & \acs{snr} & $n$ & $f_0$ & \rev{$\log\fdot$} & \rev{$\log\Amp$} & $\alpha$ & $\delta$ & $\psi$ & $\sin\iota$ & $\phi_0$ & mean \\
\midrule
\ResultsJsRows
\bottomrule
\end{tabular}
\end{table*}

\Cref{tab:results-js} compares the shapes of the marginals through \cref{eq:evaluation-js}, as
\citet{delmond2026neural} do. Below $\snr=3$, where both should return the prior, the seven
coordinates other than $f_0$ sit at \ResultsJsFaintLo--\ResultsJsFaintHi{} on the four models other
than narrow $S=8$ and full $S=16$; $f_0$ is at \ResultsJsFaintFullFzero{} on the full band with $S=4$
and 8, at \ResultsJsFaintNarrowTwoFzero{} and \ResultsJsFaintNarrowFourFzero{} on the narrow band
with $S=2$ and 4, where the Fisher-started chains drift slowly, and at
\ResultsJsFaintNarrowEightFzero{} and \ResultsJsFaintFullSixteenFzero{} for the two problems whose
reference is stuck. From $\snr=7$ the narrow band with two sources has a mean of
\ResultsJsNarrowTwoMean, about that of two Gaussians offset by half a width
(\ResultsJsGaussOffsetHalf) or with widths 1.5 apart (\ResultsJsGaussWidthOneHalf), and the mean
rises with $S$, mostly through $f_0$. That is about the \num{0.05} that \citet{delmond2026neural}
report for one source in their low-frequency bands, though their estimator, and so its floor, is not
stated. On the full band above $\snr=40$ the divergences \rev{for $S=4$, 8 and 16 reach
\ResultsJsFullFourLoudFzero{}, \ResultsJsFullEightLoudFzero{} and \ResultsJsFullSixteenLoudFzero{} in $f_0$,
\ResultsJsFullFourLoudFdot{}, \ResultsJsFullEightLoudFdot{} and \ResultsJsFullSixteenLoudFdot{} in $\log\fdot$, and
\ResultsJsFullFourLoudSky{}, \ResultsJsFullEightLoudSky{} and \ResultsJsFullSixteenLoudSky{} in the right ascension},
in line with the excess widths of
\cref{tab:results-widths}, for a Gaussian ten times too wide gives \ResultsJsGaussWidthTen. As for
the widths, full $S=16$'s divergences from $\snr=7$ to about 20 are upper bounds
(\cref{app:results-mcmc}).

\subsection{Validation Without the Truth}
\label{app:results-validation}

\begin{table*}[t]
\centering
\caption{The validation pass of the models of \cref{tab:results-widths}: \ResultsValidationInjections{}
prior injections per model, \ResultsValidationDraws{} draws each, ranks of the truth among the draws
(\cref{app:evaluation}), $p$-values of the \acs{ecdf} $\gamma$ test. Strata are by the draws' median own
\ac{snr}, a function of the data alone; $n$ is the number of sources in the stratum. Coordinates: the
smallest $p$ of the eight flow coordinates. \acs{tarp} per source and joint over the window. Overlap and
own \ac{snr} per source, log-likelihood per window. Offset: the mean rank over the number of draws minus
one half ($+0.5$: the truth's overlap or log-likelihood beats every draw's); middle: the share of
log-likelihood ranks in the middle half (one half for a calibrated posterior). Under each model's first
row, the Gaussian mock of \cref{app:evaluation-mocks} on the same problem and injections.}
\label{tab:results-validation}
\small
\setlength{\tabcolsep}{4.5pt}
\begin{tabular}{@{}llrccccccccc@{}}
\toprule
& & & \multicolumn{6}{c}{$p$-value} & overlap & \multicolumn{2}{c}{log-lik.\ rank} \\
\cmidrule(lr){4-9}\cmidrule(lr){10-10}\cmidrule(l){11-12}
model & $\widehat{\snr}$ & $n$ & coord. & \acs{tarp} & \acs{tarp} joint & overlap & own \acs{snr} & log-lik. & offset & offset & middle \\
\midrule
\ResultsValidationRows
\bottomrule
\end{tabular}
\end{table*}

\begin{figure*}[t]
  \centering
  \includegraphics[width=\textwidth]{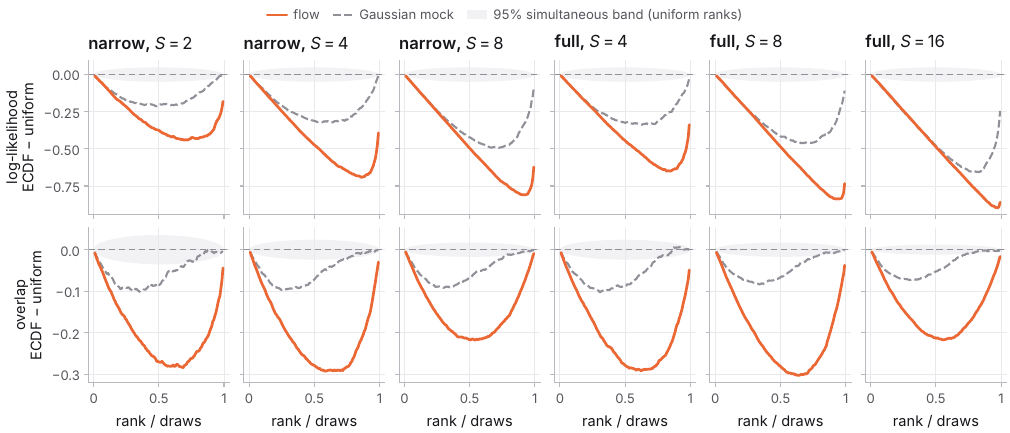}
  \caption{\acs{ecdf} of the truth's rank among the \ResultsValidationDraws{} draws, minus the uniform
  \acs{ecdf}, for the joint log-likelihood (top, one rank per window) and the overlap (bottom, one rank per
  source) of the validation pass of \cref{tab:results-validation}: each band and source count's model
  (solid) and its Gaussian mock on the same injections (dashed), with the simultaneous
  \qty{95}{\percent} band of the $\gamma$ test for uniform ranks (shaded). Below zero, the ranks pile up
  high: the true signal fits the data better than the drawn ones. Every curve leaves the band, the mock's
  too; the flows' log-likelihood curves run below their mocks', and their overlap curves well below.}
  \label{fig:results-rank-ecdf}
\end{figure*}

The validation pass draws every model on prior injections and ranks the truth among the draws,
so it needs neither chains nor relabelling (\cref{app:evaluation}). \Cref{tab:results-validation}
sets each model above the Gaussian mock of the same problem (\cref{app:evaluation-mocks}), which is
close to calibrated in its flow coordinates. With \ResultsValidationInjections{} injections every
test rejects the mock, so the $p$-values no longer separate a good posterior from a bad one and the
offsets carry the reading. The log-likelihood offset grows with the number of parameters for the
mock as well, and every flow sits above its mock (\cref{fig:results-rank-ecdf}, top). The overlap
rank is the clearer signal (\cref{fig:results-rank-ecdf}, bottom): the mock's offset is the same on
every problem, the flows' larger, the narrow band with $S=2$ included.

\paragraph{Which parameters carry the offset.}
To find the parameter, every draw of the validation pass had one group of its flow coordinates set
to the truth's (the frequencies $x_f, x_{\fdot}$; the sky $l, b$; the phases $\arg z_\pm$; the
moduli $|z_\pm|$) before its overlap was ranked again, for each model and for its mock. \rev{No single group
carries the offset.} Setting the frequencies or the sky to the truth leaves the mock's offset where it was
(\ResultsOverlapFixMockFreqSkyLo{} to \ResultsOverlapFixMockFreqSkyHi), but raises the flow's, to
\ResultsOverlapFixFlowLo--\ResultsOverlapFixFlowHi{} with the frequencies, the sky or the phases;
the phases raise the mock's too, by \ResultsOverlapFixMockPhasesLo{} to
\ResultsOverlapFixMockPhasesHi. Setting the frequencies and the phases together raises it less than
either alone (\ResultsOverlapFreqPhasesFlowLo--\ResultsOverlapFreqPhasesFlowHi): a drawn source
with the wrong $f_0$ carries a phase that partly makes up for it, so each draw is a coherent fit
displaced from the truth along a direction that mixes the groups. The amplitudes are right: the
ranks of $|z_\pm|$ are off by \ResultsOverlapAbsZFlowLo{} to \ResultsOverlapAbsZFlowHi{} and the own
\ac{snr} by \ResultsOverlapOwnSnrFlowLo{} to \ResultsOverlapOwnSnrFlowHi. Only below $\hat\snr=7$
does one group help: the moduli, whose ratio is the inclination, take
\ResultsOverlapModuliFaintFlowLo{} to \ResultsOverlapModuliFaintFlowHi{} off the flow's offset where
the mock's moves by \ResultsOverlapModuliFaintMockLo{} to \ResultsOverlapModuliFaintMockHi.

\subsection{Joint Posteriors on Crowded Windows}
\label{app:results-corners}

\begin{figure*}[p]
\centering
\includegraphics[width=\textwidth]{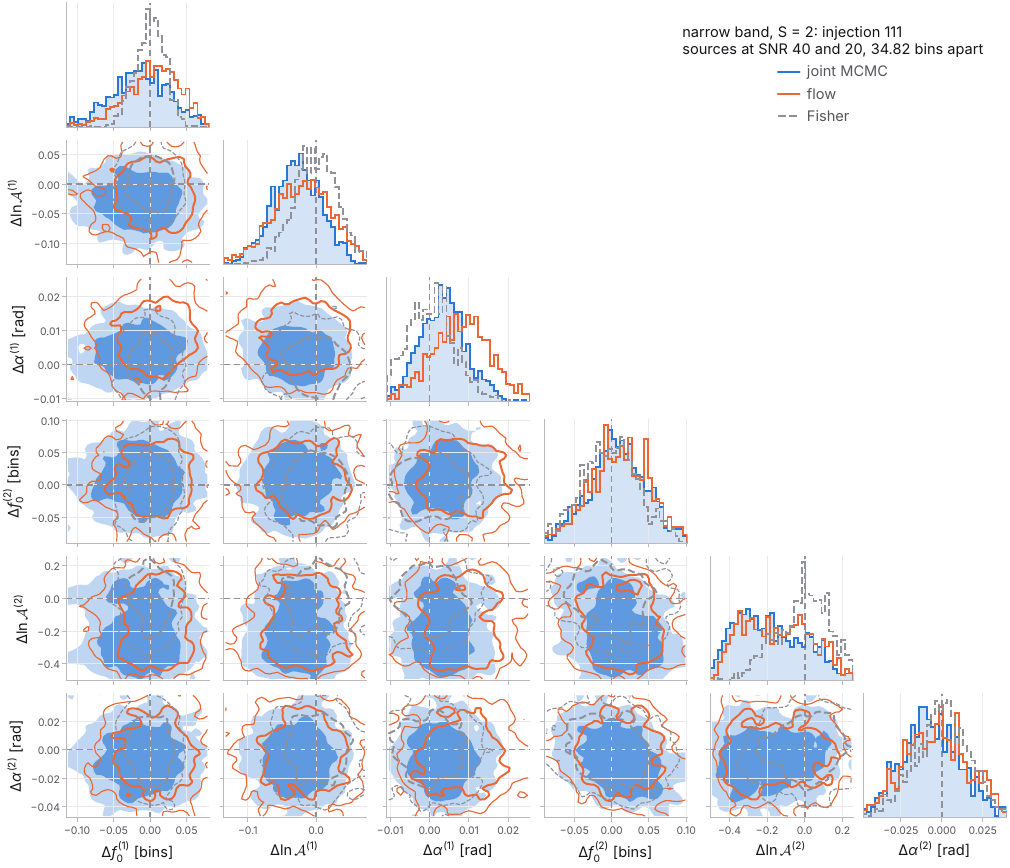}
\caption{Narrow band, $S=2$: the flow overlaps the chains, with Fisher close by, with about their
widths in every parameter. The pair is far apart, since that problem has no close loud pair.
Flow, joint \ac{mcmc} (filled) and Fisher (dashed), \qty{68}{\percent} and
\qty{95}{\percent} contours.}
\label{fig:corner-narrow-2}
\end{figure*}

\begin{figure*}[p]
\centering
\includegraphics[width=\textwidth]{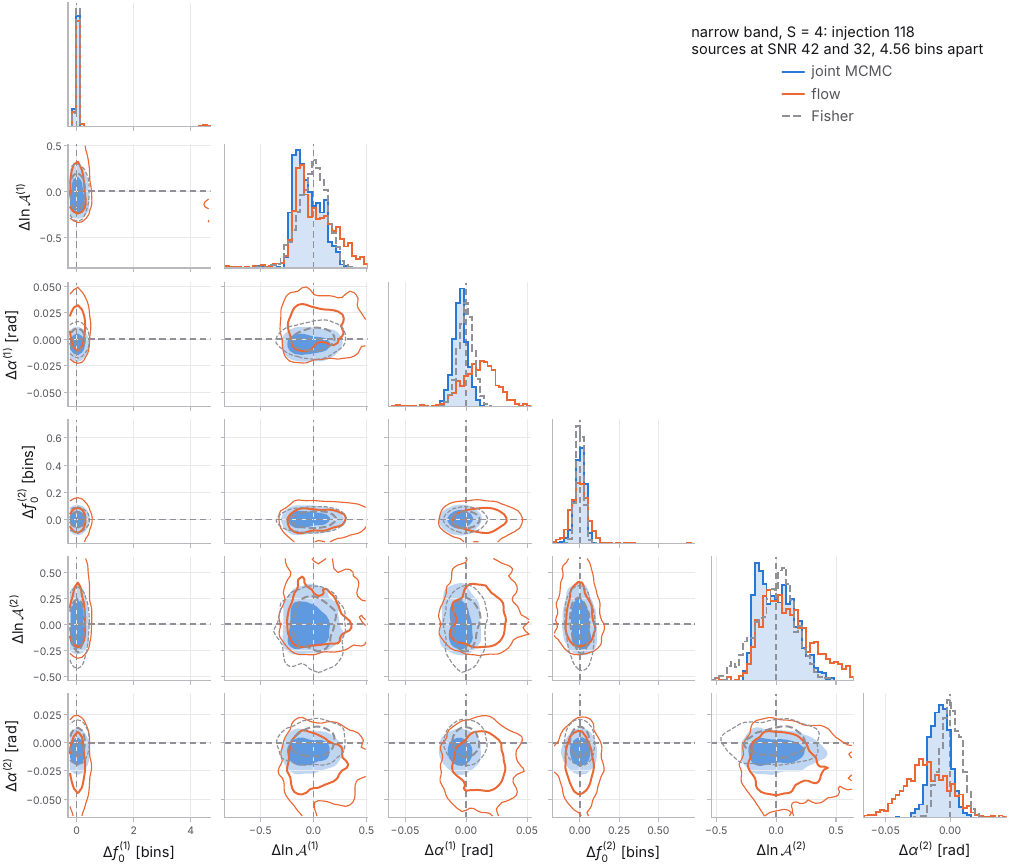}
\caption{Narrow band, $S=4$: the flow keeps the core of each source where the chains have it, and a few
per cent of the lower source's $f_0$ draws sit on its neighbour. Contours as in
\cref{fig:corner-narrow-2}.}
\label{fig:corner-narrow-4-pair}
\end{figure*}

\begin{figure*}[p]
\centering
\includegraphics[width=\textwidth]{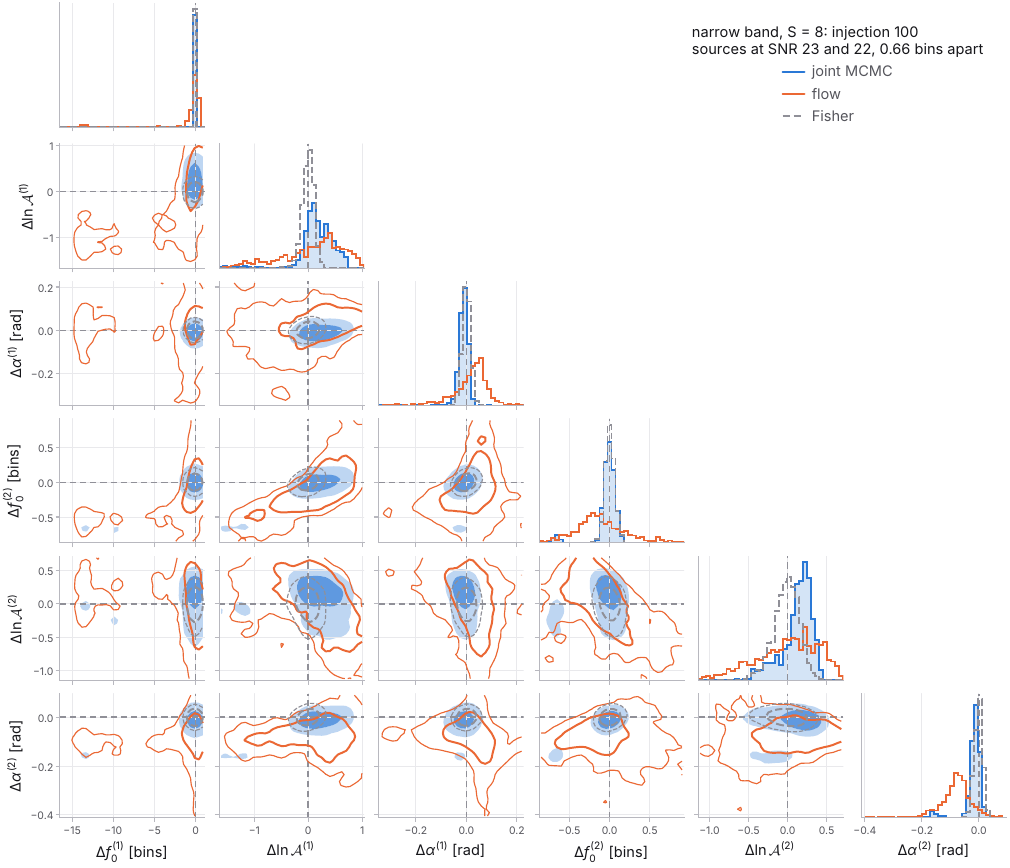}
\caption{Narrow band, $S=8$: for two sources less than a bin apart the flow is wider than the chains
in every parameter and its amplitudes have long tails. Contours as in
\cref{fig:corner-narrow-2}.}
\label{fig:corner-narrow-8}
\end{figure*}

\begin{figure*}[p]
\centering
\includegraphics[width=\textwidth]{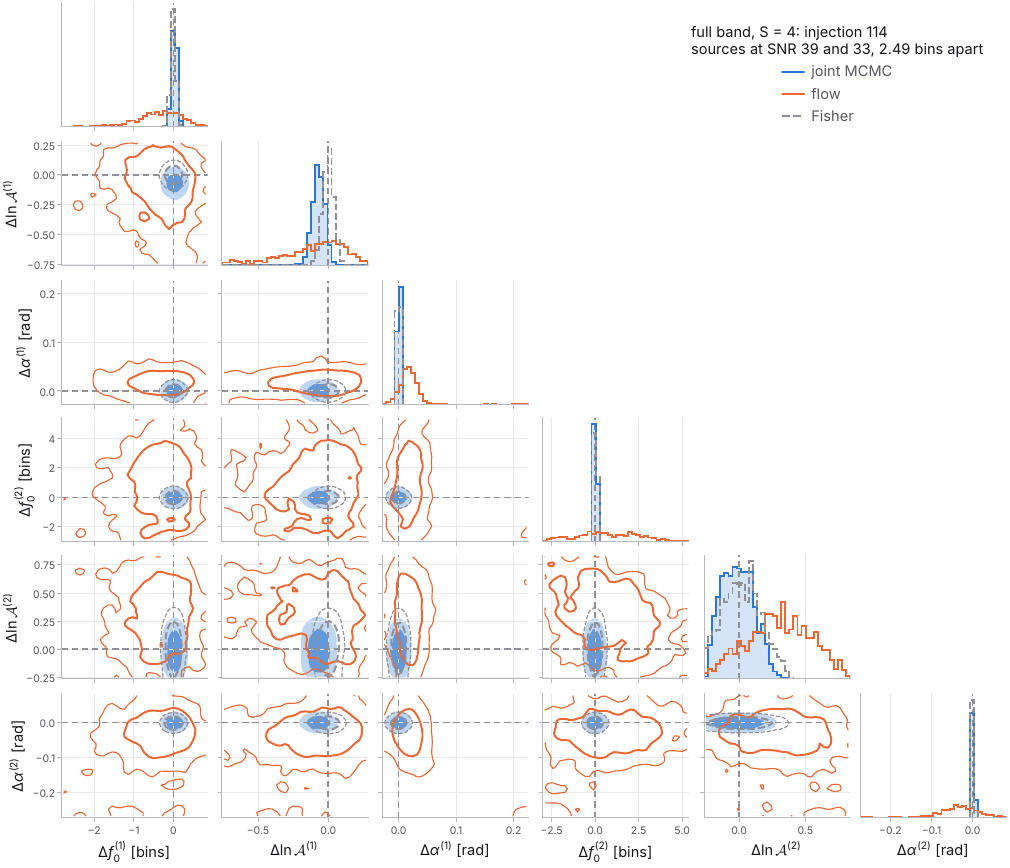}
\caption{Full band, $S=4$: for two loud sources a few bins apart the flow's $f_0$ of the louder one
peaks on the chains' with a tail of about a bin, that of the other spreads over several bins, its
amplitudes are skewed where the chains sit on the truth, and the sky has long tails. Contours as in
\cref{fig:corner-narrow-2}.}
\label{fig:corner-full-4}
\end{figure*}

\begin{figure*}[p]
\centering
\includegraphics[width=\textwidth]{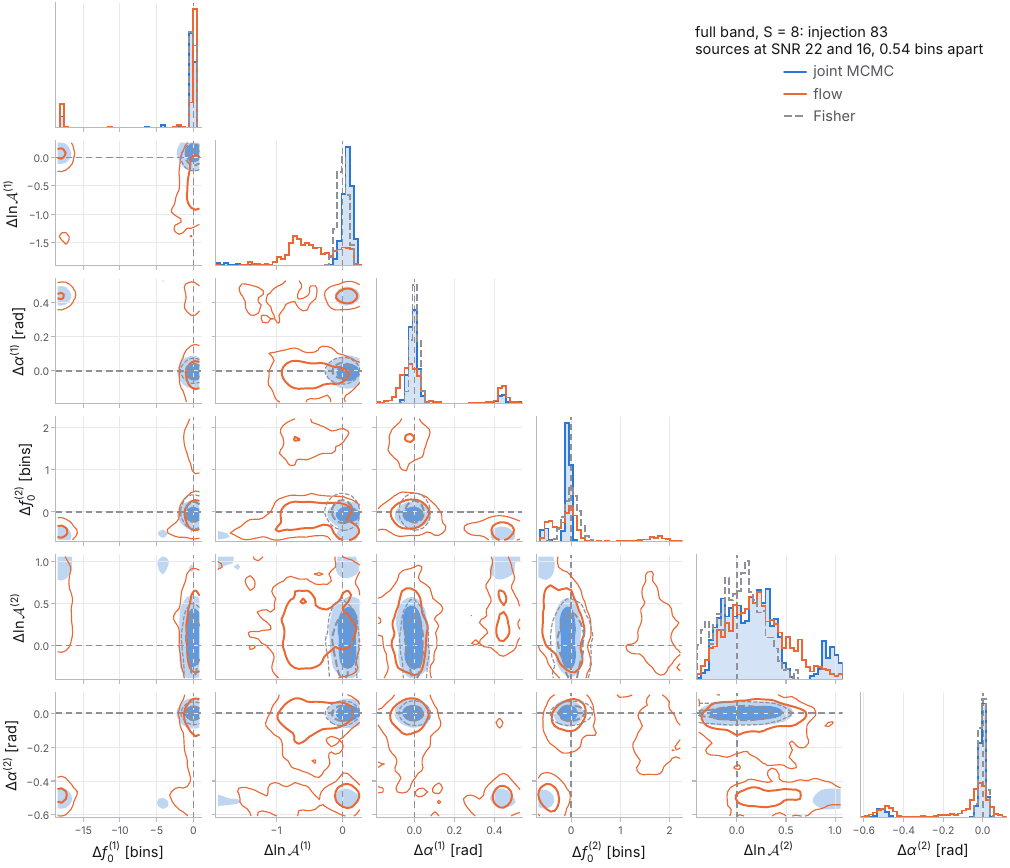}
\caption{Full band, $S=8$: the flow's posterior is multimodal where the chains have one mode.
Contours as in \cref{fig:corner-narrow-2}.}
\label{fig:corner-full-8}
\end{figure*}

\begin{figure*}[p]
\centering
\includegraphics[width=\textwidth]{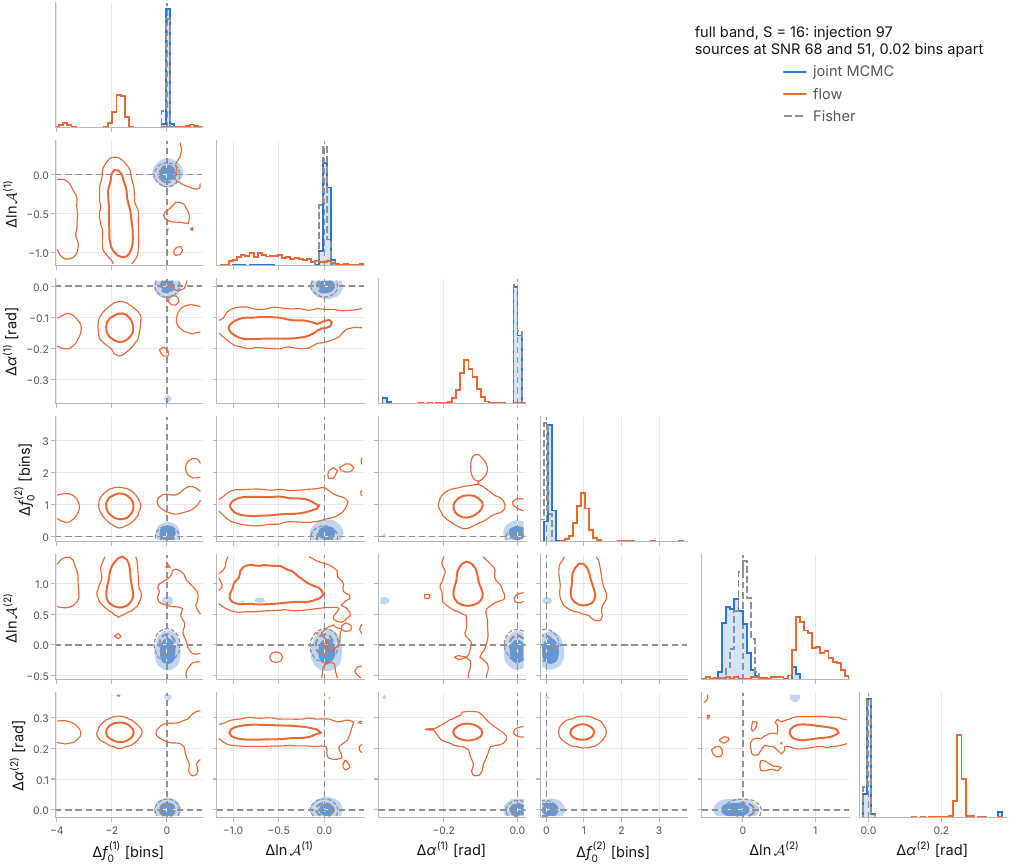}
\caption{Full band, $S=16$: for two very loud sources that nearly coincide in frequency the flow is
offset from the truth by one to two bins in $f_0$ and by a few tenths of a radian in the sky, where
the chains sit on it. Contours as in \cref{fig:corner-narrow-2}.}
\label{fig:corner-full-16}
\end{figure*}

\Cref{fig:corner-narrow-2,fig:corner-narrow-4-pair,fig:corner-narrow-8,fig:corner-full-4,fig:corner-full-8,fig:corner-full-16}
show one eval window for the model of each band and source count. The window is chosen by rule,
not by eye: the closest pair of sources at $\snr\ge15$ with \acp{snr} within a factor two (the
closest pair at $\snr\ge7$ if there is none). Each plot shows $f_0$, $\ln\Amp$ and right ascension
\rev{(equatorial, the simulator's frame, where \cref{fig:corner-narrow-4} uses ecliptic coordinates)}
of both sources as offsets from the truth, the sources in $f_0$ order and folded in $(\psi,\phi_0)$;
the axes span the central \qty{96}{\percent} of the flow and the chains together, so a few far-tail
draws fall outside. The legend gives the window, the two \acp{snr} and their separation.
The corner of the main text (\cref{fig:corner-narrow-4}) is chosen differently: all four sources of the
window, among those whose four sources are visible, where the flow overlaps the chains best.

\subsection{Longer Runs}
\label{app:results-forks}

\begin{table}[h]
\centering
\caption{Eval-catalogue loss / flow over joint-\ac{mcmc} $f_0$ width at $\snr=15$--40, for the forks that
start their cooldown at epochs 4 to 32, and for the main run, which starts it at the planned
\qty{80}{\percent} of its \ResultsEpochs{} epochs; in brackets the epoch evaluated. Full $S=4$'s main column is
its Student-t continuation from epoch 44, whose logged loss does not compare (n/c, \cref{app:fm-frame}).}
\label{tab:results-forks}
\small
\begin{tabular}{@{}lccccc@{}}
\toprule
 & e4 & e8 & e16 & e32 & main \\
\midrule
\ResultsForkRows
\bottomrule
\end{tabular}
\end{table}

The forks give the same configuration trained for fewer epochs (\cref{app:training-loop}), and their
eval-catalogue loss falls with the training length (\cref{tab:results-forks},
\cref{fig:results}c): by \ResultsNarrowTwoForkGain{} nats from epoch 4 to 32 on the narrow band with
$S=2$ and by \ResultsFullFourForkGain{} on the full band with $S=4$, and the $f_0$ widths shrink
toward the chains' with it. The cooldown matters most on the full band: the diverged run with
$S=4$, never cooled, has loud $f_0$ widths \ResultsFullFourMainLoud{} times the chains', against
\ResultsFullFourForkLoud{} for its fork from epoch 32 and \ResultsFullFourContLoud{} for its
continuation, cooled at the end. Beyond epoch 32 the main runs gain on two configurations and lose
on three (\cref{app:training-schedules}); with one seed per configuration, differences of a few
hundredths of a nat are within the run-to-run spread we have not measured (\cref{sec:followups}).

\clearpage
\section{Training Cost and Break-Even}
\label{app:amortisation}

Training is paid once per band, sampling once per window. This appendix shows how the fit depends on
the training time and after how many windows the training is repaid by the cheaper sampling, which backs
the cost claim of \cref{sec:experiments}.

\begin{figure}[t]
  \centering
  \includegraphics[width=\textwidth]{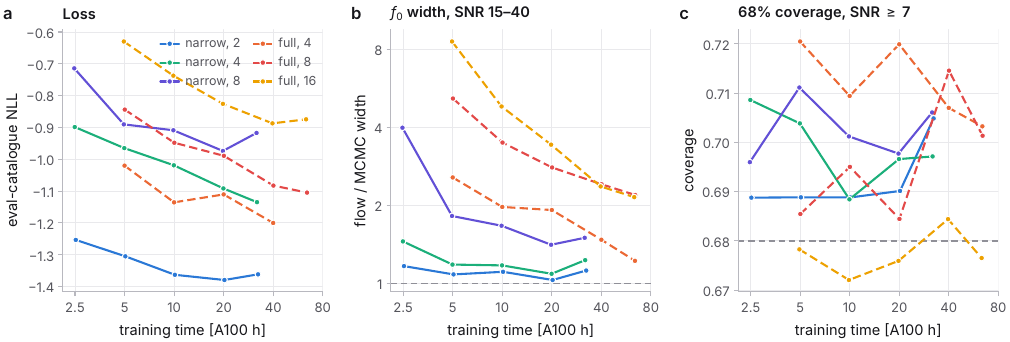}
  \caption{Fit against training time, for each configuration's cooldown forks (5, 10, 20 and 40
  epochs) and its \ResultsEpochs{}-epoch run (full $S=4$: its Student-t continuation, whose loss does
  not compare and is left out). Training time is the A100 hours of a run planned for that length.
  (a)~Eval-catalogue loss. (b)~Flow over joint-\ac{mcmc} $f_0$ width at $\snr=15$--40, median over the
  sources (\cref{app:results}). (c)~Share of true flow coordinates in the central
  \qty{68}{\percent} interval, pooled over the sources with $\snr\ge7$ and the eight coordinates.}
  \label{fig:amortisation-quality}
\end{figure}

\paragraph{Fit against training time.}
Each cooldown fork is the same run trained for a shorter schedule (\cref{app:training-loop}): the
forks from epochs 4, 8, 16 and 32 end after 5, 10, 20 and 40 epochs, and cost what a run planned
for that length would. \Cref{fig:amortisation-quality} places them, with the \ResultsEpochs{}-epoch
runs, at the A100 hours of that schedule. On the narrow band, \AmortisationNarrowShortEpochs{}
epochs, \AmortisationNarrowShortHours\,h, already give $f_0$ widths
\AmortisationNarrowShortRatio{} times the chains' at $\snr=15$--40, against
\AmortisationNarrowLongRatio{} after \AmortisationNarrowLongHours\,h; on the full band the same
widths fall from \AmortisationFullShortRatio{} after \AmortisationFullShortHours\,h to
\AmortisationFullLongRatio{} after \AmortisationFullLongHours\,h. The loss falls with training time on every
configuration up to 40 epochs, apart from one step on the full band with $S=4$, and the
\qty{68}{\percent} coverage stays between 0.67 and 0.72 at every training time
(\cref{fig:amortisation-quality}c), so shorter training gives wider posteriors whose coverage is
unchanged. The more sources a window holds,
the more training the same width costs. \Cref{tab:results-forks} lists the values.

\begin{figure}[t]
  \centering
  \includegraphics[width=0.5\textwidth]{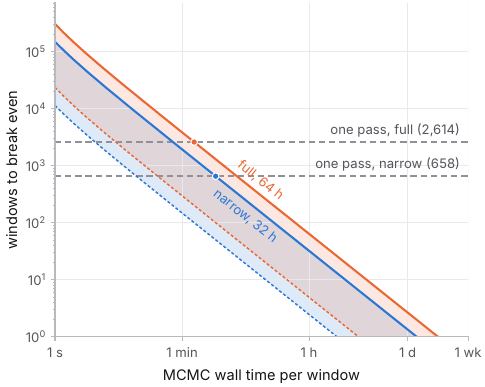}
  \caption{Windows after which the flow breaks even with \ac{mcmc}, \cref{eq:amortisation-break-even},
  against the \ac{mcmc} wall time per window. Solid: the \ResultsEpochs{}-epoch runs; dotted: the
  shortest fork (5 epochs); the shading lies between them. Dashed: one pass over each band; the
  dots mark the \ac{mcmc} time at which one pass repays the longest training.}
  \label{fig:amortisation-break-even}
\end{figure}

\paragraph{Break-even.}
Let $T$ be the training time, $t_\mathrm{flow}$ the flow's time per window and $t_\mathrm{mcmc}$ the
\ac{mcmc}'s. The flow is cheaper once it has inferred
\begin{equation}
  N^\star = \frac{T}{t_\mathrm{mcmc}-t_\mathrm{flow}}
  \label{eq:amortisation-break-even}
\end{equation}
windows. $t_\mathrm{flow}$ is measured: \AmortisationDraws{} draws take
\qty{\AmortisationNarrowFlowSeconds}{\second} on the narrow band and
\qty{\AmortisationFullFlowSeconds}{\second} on the full band (\cref{app:fm-sampling}). We do not fix
$t_\mathrm{mcmc}$. Our reference chains (\cref{app:mcmc}) were run as an evaluation reference, not tuned
for speed, and a production sampler with tuned proposals, more cores or a GPU likelihood would be faster by a margin
we have not measured. \Cref{fig:amortisation-break-even} therefore shows $N^\star$ against
$t_\mathrm{mcmc}$ from a second to a week. The time axis is wall time on whatever hardware runs the
chains; a different price per hour for the chains' hardware than for an A100 shifts the curves along
it.

As a scale for $N^\star$, one pass over a band infers each frequency once: its windows tiled by
their interiors of \AmortisationInteriorResponses$R$ bins, \num{\AmortisationNarrowPasses} windows
on the narrow band and \num{\AmortisationFullPasses} on the full band. A single pass repays the
training if a chain takes more than \qty{\AmortisationNarrowPassMinutes}{\minute} per window on the
narrow band, or \qty{\AmortisationFullPassMinutes}{\minute} on the full band. If a chain takes an
hour, the flow breaks even after \AmortisationNarrowHourWindows{} and
\AmortisationFullHourWindows{} windows. One pass understates the saving for a global fit, which
revisits each window at every iteration as the neighbouring signals change and starts again with each
data release, while the flow's training is reused as long as the noise model and the priors hold
(\cref{sec:followups}).

\end{document}